\documentclass[11pt]{article}
\usepackage{epsfig} 
\usepackage{graphics}
\newcommand{\sect}[1]{ \section{#1} \setcounter{equation}{0} } 

\newcommand{\half}{\mbox{\small{$\frac{1}{2}$}}} 
\newcommand{\third}{\mbox{\small{$\frac{1}{3}$}}} 
 
\newcommand{\threehalves}{\mbox{\small{$\frac{3}{2}$}}} 
\newcommand{\fivehalves}{\mbox{\small{$\frac{5}{2}$}}} 
\newcommand{\sevenhalves}{\mbox{\small{$\frac{7}{2}$}}} 
\newcommand{\ninehalves}{\mbox{\small{$\frac{9}{2}$}}} 
\newcommand{\twothirds}{\mbox{\small{$\frac{2}{3}$}}} 
\newcommand{\onefifth}{\mbox{\small{$\frac{1}{5}$}}} 
\newcommand{\twofifth}{\mbox{\small{$\frac{2}{5}$}}} 
 
\newcommand{\fourfifth}{\mbox{\small{$\frac{4}{5}$}}} 
\newcommand{\oneseventh}{\mbox{\small{$\frac{1}{7}$}}} 
\newcommand{\twoseventh}{\mbox{\small{$\frac{2}{7}$}}} 
\newcommand{\threeseventh}{\mbox{\small{$\frac{3}{7}$}}} 
 
\newcommand{\fiveseventh}{\mbox{\small{$\frac{5}{7}$}}} 
\newcommand{\sixseventh}{\mbox{\small{$\frac{6}{7}$}}}

\newcommand{\partialline}{\partial \! \! \! \backslash}
\newcommand{\pline}{p \! \! \! \backslash}

\begin{document}

\title{Renormalization of supersymmetric chiral theories in rational spacetime
dimensions}
\author{J.A. Gracey, \\ Theoretical Physics Division, \\ 
Department of Mathematical Sciences, \\ University of Liverpool, \\ P.O. Box 
147, \\ Liverpool, \\ L69 3BX, \\ United Kingdom.} 
\date{}

\maketitle 

\vspace{5cm} 
\noindent 
{\bf Abstract.} We renormalize models with scalar chiral superfields with an
odd superpotential to several orders in perturbation theory. These extensions 
of the cubic Wess-Zumino model are renormalizable in spacetime dimensions which
are rational. When endowed with an $O(N)$ symmetry it is shown that they share
the same property as their non-supersymmetric counterparts in that at a
particular fixed point there is an emergent $OSp(1|n-1)$ symmetry, where $n$ is
the power of the superpotential. This is shown at a loop order beyond that for 
which it was established in the parallel non-supersymmetric theory.

\vspace{-17.0cm}
\hspace{13.2cm}
{\bf LTH 1310}

\newpage 

\sect{Introduction.}

One of the more interesting developments in quantum field theory in recent
years has been that of emergent symmetries particularly in the case when a
model of bosons and fermions develops a configuration that possesses
supersymmetry, \cite{1,2,3}. Emergent properties derive from the critical point
analysis of the renormalization group functions of a multicoupling theory when 
treated in $d$-dimensions. Ordinarily in a single coupling theory the 
$\beta$-function has a Wilson-Fisher fixed point given by the first non-trivial
zero of the $d$-dimensional $\beta$-function. By contrast in the multicoupling 
case even with two coupling constants one can have a rich spectrum of fixed 
points in $d$-dimensions, \cite{2,3}. These can be stable in the ultraviolet 
limit or alternatively in the infrared if the running is in that direction,
in addition to the presence of saddle points. At each critical point the values
of critical exponents can be determined in the $\epsilon$ expansion where
$\epsilon$ is a measure of the difference between $d$ and the critical 
dimension of the theory. The concept of emergence then arises when a fixed 
point possesses an enlarged or extended symmetry over and above that of the 
fields in the original underlying Lagrangian. To illustrate the background to 
this, for instance, one well-studied case is that of the Gross-Neveu-Yukawa 
(GNY) system, \cite{4,5}, which is important for phase transitions in condensed 
matter systems. A comprehensive review can be found for instance in \cite{3}. 

In these GNY models one has several scalar fields coupled to a multiplet of 
fermions in a flavour symmetry group. It transpires that at one particular 
fixed point and a specific number of flavours the condition is met for the 
presence of supersymmetry, \cite{1,2,6,7}. By this we mean the critical point 
values of the two originally distinct coupling constants become equal. This is 
not sufficient for there to be supersymmetry alone. Instead it is also the 
observation that the field anomalous dimensions at this specific fixed point 
become equal. This occurs in the GNY related models of the chiral Ising and 
chiral XY models when the parameter $N$ takes the respective values of
$N$~$=$~$\frac{1}{4}$ and $N$~$=$~$\frac{1}{2}$, \cite{1,2,7} and has
subsequently been verified up to four loops, \cite{7,8,9,10}. In addition to 
the criteria for supersymmetry being satisfied at four loops at one
particular fixed point, the critical properties there have been connected, 
\cite{11,12}, for example, to those of the Wess-Zumino model, \cite{13}. This 
has been demonstrated to three loops, \cite{12}, and more recently at four 
loops, \cite{14}, using the explicit results of the renormalization group 
functions in the Wess-Zumino model available in \cite{13,15,16,17,18}. More 
recently the Wess-Zumino model has been renormalized to five loops in various 
schemes, \cite{14}, in preparation for verifying the emergence in the GNY 
system to the next order. In other words one can interpret the emergent 
supersymmetric theory of the GNY system as that of the Wess-Zumino model. This 
is important as it is believed that supersymmetry may be present in some 
condensed matter systems, like those on the boundaries of three dimensional 
topological insulators, \cite{6}, and so may be described by Wess-Zumino 
models. Interestingly the GNY model has a structure that is similar to the 
Standard Model of particle physics where the scalar field is analogous to the 
Higgs field. Therefore it has already been noted in, for instance, \cite{9}, 
that such emergence properties of the relatively simple GNY model could equally
hold in the Standard Model. If so there is the possibility that an emergent 
supersymmetry could be a route to an extension of the Standard Model.

It is worth stressing that emergent symmetries do not always lead to
supersymmetry. For instance, in a particular scalar cubic theory, 
\cite{20,21,22,23}, which is renormalizable in six dimensions, it was shown in
\cite{23}, that an emergent flavour symmetry is present. In particular the 
$O(3)$ symmetry of the original Lagrangian enhanced to an $SU(3)$ one at a 
particular critical point. A more recent example of such a flavour symmetry 
emergence was discussed in \cite{24}. In that work scalar field theories with 
an $O(N)$ symmetry and potentials with an odd power were studied. Although they
are renormalizable in rational spacetime dimensions, for specific values of $N$
there is a fixed point with an emergent $OSp(1|2M)$ symmetry, \cite{24}. The 
case of the quintic theory or Blume-Capel theory, \cite{25,26}, was of 
particular interest, \cite{27,28,29,30}, given that it is the next theory in 
the sequence after $\phi^3$ theory that underlies the Ising and Lee-Yang 
universality classes and has a rational critical dimension close to three 
dimensions. However, the underlying mechanism of the emergence in this instance
was that the anomalous dimensions of the fields in the $O(N)$ multiplet became 
equal to that of another scalar field in the theory. This field was analogous 
to the $\sigma$ field that arises in the $O(N)$ nonlinear sigma model. Indeed 
the sigma model is the first in the sequence of such odd power potentials for 
this $OSp(1|2M)$ emergence to arise. The next model in the sequence after the 
sigma model is the cubic theory akin to the one mentioned earlier. Indeed it is
structurally similar to the Wess-Zumino model in its superfield formulation 
with chiral superfields. Therefore given the parallel nature of the scalar 
cubic theory with the Wess-Zumino model a natural question to ask is whether 
there is an analogous sequence of supersymmetric models that is parallel to 
those considered in \cite{24} which have an emergent $OSp(1|2M)$ symmetry. 

This is the main aim of this article. It is possible to formulate these 
generalized Wess-Zumino theories given the superspace techniques that allowed 
the original component field formulation of the Wess-Zumino model, \cite{13}, 
to be rewritten in terms of chiral superfields, \cite{31}. One consequence was 
that the Wess-Zumino model was renormalized in an efficient way to very high 
loop order, \cite{14,16,18}. Therefore we will construct the relevant 
superspace actions for such a sequence of chirally supersymmetric theories and 
then renormalize them to second order which will be at an order beyond that 
considered in the scalar case of \cite{24}. This is primarily due to the chiral
property which rules out a substantial number of higher order graphs that would
ordinarily have to be determined for the wave function renormalization. 
Moreover the underlying supersymmetry Ward identity, \cite{1,2}, means that the 
$\beta$-functions will follow trivially from the field anomalous dimensions. 
One concern with following such a superspace approach here might be its 
relation with the associated component theory especially in light of the 
potential unequal boson and fermion degrees of freedom in a non-integer 
dimension. A similar issue arises when one regularizes a supersymmetric 
component Lagrangian. It is known that while canonical dimensional 
regularization does not preserve supersymmetry there is a way to circumvent the
degrees of freedom imbalance that is the underlying reason for this. Instead a 
modified regularization is used known as dimensional reduction and involves the
presence of additional fields termed $\epsilon$ scalars. They inhabitat the 
subspace of the regularizing spacetime that excludes the critical dimension 
spacetime. Such additional fields are absent in the critical dimension of the 
theory but their presence preserves the supersymmetry property of that physical
space. In the rational spacetime such fields will naturally also be necessary 
to preserve the degrees of freedom in the associated component theory. What 
would also be the case is that such a component theory will have a 
non-supersymmetric associate which has the same Lagrangian but each interaction
has a different coupling constant. Indeed it will be of a similar nature to the
three dimensional GNY systems that have an emergent supersymmetry where not 
only will there be a fixed point where all the critical couplings are equal but
the field anomalous dimensions will all be the same. In the three dimensional 
GNY case the underlying supersymmetric theory is the four dimensional 
Wess-Zumino model. Indeed it can be formulated in superspace and the $\epsilon$
expansion of its critical exponents agree precisely with the $\epsilon$ 
expansion of the exponents of the emergent supersymmetric fixed point of the 
related GNY system. In regard to the generalized Wess-Zumino theories we take 
a similar point of view that they in fact represent the emergent supersymmetric
fixed point of the associated non-supersymmetric partner theory. In studying 
the fixed point structures in the supersymmetric theories an $OSp(1|2M)$ 
emergent symmetry will be present but it arises in a subtle way compared to the
scalar case of \cite{24}. Aside from this main goal we will examine a more 
mundane aspect of the $\epsilon$ expansion in this class of theories with an 
odd power potential. For instance, the scalar quintic or Blume-Capel theory has
a critical dimension of $\frac{10}{3}$ which is close to the integer dimension 
of three. Therefore in $d$~$=$~$\frac{10}{3}$~$-$~$2\epsilon$ dimensions the 
value of $\epsilon$ needed to reach that integer dimension is relatively small 
compared to a theory with a critical dimension of four for example. In other 
words the convergence of the $\epsilon$ expansion in a quintic scalar theory 
should be quick.  Unfortunately with the inability to compute corrections 
beyond the leading order in that case due to difficult Feynman integrals, which
will be illustrated later, this convergence issue cannot be readily studied. In
the supersymmetric extension however we will be able to proceed to the next 
order as the corresponding difficult graphs are excluded by the chiral 
property. Thus we will examine convergence issues albeit in a simialar although
different class of theories. 

The paper is organized as follows. We devote Section $2$ to renormalizing the
basic chirally supersymmetric scalar theories with an odd potential to the 
first few orders. While we will concentrate on three specific theories some 
properties of critical exponents are provided for all models with odd 
potentials. To examine the emergent symmetry property we construct the $O(N)$ 
versions of the specific theories in Section $3$ before renormalizing them to 
allow us to analyse their fixed point properties in Section $4$. In Section $5$
we concentrate on establishing the $OSp(1|2M)$ enhancement at one particular 
critical point before summarizing our study in Section $6$. An appendix 
provides explicit expressions for the renormalization group functions of 
several of the $O(N)$ theories we focus on.

\sect{Background.}

First we consider the action of the most general superpotential with a chiral
superfield which is given by
\begin{equation}
S_{(n)} ~=~ \int d^d x \left[ \int d^2 \theta d^2 \bar{\theta} \,
\bar{\Phi}_{\mbox{\footnotesize{o}}} (x,\bar{\theta}) 
e^{-2 \theta {\partialline} \bar{\theta}}
\Phi_{\mbox{\footnotesize{o}}} (x,\theta) ~+~
\frac{g_{\mbox{\footnotesize{o}}}}{n!} \int d^2 \theta \, 
\Phi_{\mbox{\footnotesize{o}}}^n(x,\theta) ~+~
\frac{g_{\mbox{\footnotesize{o}}}}{n!} \int d^2 \bar{\theta} \, 
\bar{\Phi}_{\mbox{\footnotesize{o}}}^n(x,\bar{\theta}) \right]
\label{genact}
\end{equation}
where $\theta$ and $\bar{\theta}$ are anti-commuting superspace coordinates and
we use type I superfields with the subscript ${}_{\mbox{\footnotesize{o}}}$ 
denoting bare quantities and $g$ is the coupling constant. The kinetic term 
follows that used in the Wess-Zumino model, \cite{16,18,31}, where the 
$2$~$\times$~$2$ covariant Pauli matrices $\sigma^\mu$ play the role of the 
usual Dirac $\gamma$-matrices and satisfy the same Clifford algebra. We use a 
variation on the canonical notation by defining 
$\partialline$~$=$~$\sigma^\mu \partial_\mu$. At this stage we have not 
specified the canonical dimension of the action as $n$ is an arbitrary integer 
here. However it is a simple exercise to deduce that the critical dimension 
$D_n$ of (\ref{genact}) is
\begin{equation}
D_n ~=~ \frac{2(n-1)}{(n-2)} ~.
\label{critd}
\end{equation}
Clearly there are only two cases where $D_n$ is an integer which are
$D_3$~$=$~$4$ and $D_4$~$=$~$3$ with the former corresponding to the 
Wess-Zumino model. Subsequent potentials give $D_5$~$=$~$\frac{8}{3}$, 
$D_6$~$=$~$\frac{5}{2}$, $D_7$~$=$~$\frac{12}{5}$, $D_8$~$=$~$\frac{7}{3}$ and 
$D_9$~$=$~$\frac{16}{7}$ with $\lim_{n\to\infty} D_n$~$=$~$2$. It is worth 
contrasting (\ref{critd}) with the critical dimension of the corresponding 
non-supersymmetric theories which is, \cite{27,32,33},
\begin{equation}
D^{\mbox{\footnotesize{scalar}}}_n ~=~ \frac{2n}{(n-2)} ~.
\label{critdnsusy}
\end{equation}
In other words for each integer $n$~$\geq$~$3$ this is the dimension where the
coupling constant is dimensionless. The origin of the difference with $D_n$ is 
the integration measure over the dimensionful anticommuting spacetime 
coordinates in (\ref{genact}). The $n$~$=$~$5$ potential shares a similar 
property to its non-supersymmetric counterpart in that its critical dimension 
is close to three dimensions.  

The bare quantities in (\ref{genact}) are related to their renormalized
partners via  
\begin{equation}
\Phi_{\mbox{\footnotesize{o}}} ~=~ \sqrt{Z_\Phi} \Phi ~~,~~
\bar{\Phi}_{\mbox{\footnotesize{o}}} ~=~ \sqrt{Z_\Phi} \bar{\Phi} ~~,~~
g_{\mbox{\footnotesize{o}}} ~=~ \mu^\epsilon Z_g g
\end{equation}
where we will dimensionally regularize the superspace action in
$d$~$=$~$D_n$~$-$~$2\epsilon$ dimensions. The arbitrary mass scale $\mu$ being
introduced to ensure the coupling constant remains dimensionless in the
regularized theory. Like the Wess-Zumino model the suite of $n$ dependent
actions each satisfy a supersymmetry Ward identity which follows simply by
generalizing the argument given in \cite{13,15,31}. This means that there is 
only one independent renormalization constant since the Ward identity implies 
\begin{equation}
Z_g Z_\Phi^{{\mbox{\small{$\frac{n}{2}$}}}} ~=~ 1 ~.
\label{susywi}
\end{equation}
This provides a simple strategy to determine the $\beta$-function of
(\ref{genact}) since $Z_g$ can be deduced from $Z_\Phi$ which means we only
need to renormalize the $2$-point function. In other words vertex functions are
finite and so do not need to be evaluated. A further simplification comes from 
the use of superspace techniques. From the action (\ref{genact}) the propagator
in momentum superspace is, \cite{18}, 
\begin{equation}
\langle \Phi(p,\theta) \bar{\Phi}(-p,\bar{\theta}) \rangle ~=~
\frac{\exp{(2 \theta \pline \bar{\theta}})}{p^2}
\end{equation}
which means that prior to carrying out the integration over the loop momenta 
the $\theta$ coordinate integration has to be performed. As these variables are 
anti-commuting the exponential associated with each propagator will truncate 
after a finite number of terms. Once this has been implemented the 
$\theta$-integration is carried out. As this effectively equates to 
differentiating with respect to the internal anticommuting variables, and is
equivalent to the so-called $D$-algebra, it results in simple traces over the 
covariant Pauli matrices. This procedure is based on the approach used in the
four loop renormalization of the Wess-Zumino model, \cite{18}, and more 
recently at five loops, \cite{14}. In the latter case the $\theta$ coordinate
integration for each graph was carried out automatically through a routine
written in the symbolic manipulation language {\sc Form}, \cite{34,35}. We have
used that same procedure for each of the three cases we focus on here. These 
will be the $n$~$=$~$5$, $7$ and $9$ potentials. Once the $\theta$ integration 
has been carried out the integration over the loop momenta remains. For 
(\ref{genact}) this is possible for both the first two orders of graphs that 
contribute.

{\begin{figure}[ht]
\begin{center}
\includegraphics[width=13.0cm,height=2.5cm]{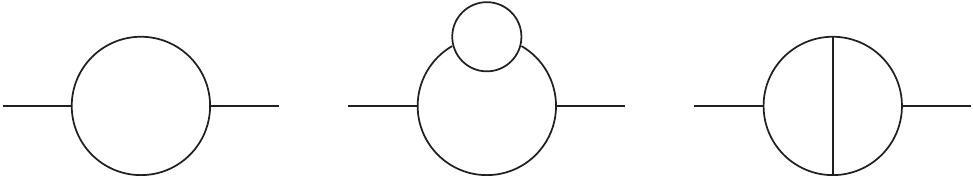}
\end{center}
%\vspace{0.5cm}
\caption{Basic one and two loop topologies for a $2$-point function in a scalar
cubic theory.}
\label{figsca}
\end{figure}}

To appreciate this for theories with higher order potentials it is instructive
to focus for the moment on the basic one and two loop topologies that can arise
in a scalar $\phi^3$ theory. These are illustrated in Figure \ref{figsca}. For 
the Wess-Zumino model, which has a cubic interaction, these are in principle
the only topologies that would determine the $\beta$-function. However the 
Wess-Zumino model is the $n$~$=$~$3$ version of (\ref{genact}) and has a chiral
symmetry. This implies that the propagators are directed and in a Feynman 
diagram have an arrow on each line. Moreover the chirality means that at a 
vertex the arrows all point towards the interaction location or away from
it. Simple reasoning indicates that this ordering excludes any topology where 
there is a subgraph with an odd number of propagators. So in Figure 
\ref{figsca} the second two loop graph is excluded. The relevance of this to 
(\ref{genact}) for odd values of $n$~$>$~$3$ is that for these higher order 
potentials the $2$-point function graphs will have the same underlying 
topological structure. This can be observed at leading order for (\ref{genact})
where the only contributing graph is given in Figure \ref{figpn1}. The number 
beside ellipses between propagators will always indicate the number of 
propagators between and including the bounding propagators. In this and 
subsequent figures lines will be directed with arrows reflecting the underlying
chirality. The relation of the graph of Figure \ref{figpn1} to the first 
topology of Figure \ref{figsca} can be seen by notionally deleting the number 
of internal lines connecting each vertex to leave vertices with only three 
lines. By way of example this observation with the core topologies of Figure 
\ref{figsca} at next order can be viewed in the $n$~$=$~$5$ case where the 
graphs are shown in Figure \ref{figp52}. These and the graphs for all the other
theories have been generated with the {\sc Qgraf} package, \cite{36}. It is 
evident that each of the three graphs of Figure \ref{figp52} are extensions of 
the middle topology of Figure \ref{figsca} where propagators are added to each 
vertex in such a way that five propagators intersect there.

{\begin{figure}[ht]
\begin{center}
\includegraphics[width=5.0cm,height=2.0cm]{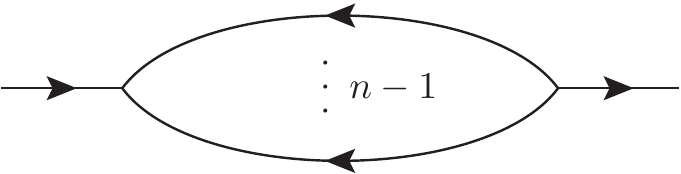}
\end{center}
%\vspace{0.5cm}
\caption{Leading order $(n-2)$ loop graph for $\Phi^n$ $2$-point function.} 
\label{figpn1}
\end{figure}}

As the structure of the leading two orders of $2$-point function graphs is
relatively simple the implementation of the $D$-algebra resulting from the
$\theta$ integration is straightforward. This is in part due to the simple
bubble graphs that comprise each $2$-point function for (\ref{genact}) when $n$
is odd. For each of the topologies beyond leading order the only minor 
complication is that the loop integrals of each central bubble in the three 
bubble sequence has a contraction of two internal loop momenta. This is not a 
hindrance to evaluating a graph as one simply makes use of the momentum 
conservation to rewrite the scalar product in terms of the squares of the 
momenta of related propagators. In other words the effect of the $D$-algebra at
this order is the removal of a propagator from the original topology similar to
what was observed in the Wess-Zumino model, \cite{18}. The consequence of the 
$D$-algebra is that all the Feynman integrals at the leading two orders are 
quickly reduced to simple scalar bubble integrals which are elementary to 
evaluate.  

{\begin{figure}[ht]
\begin{center}
\includegraphics[width=14cm,height=8cm]{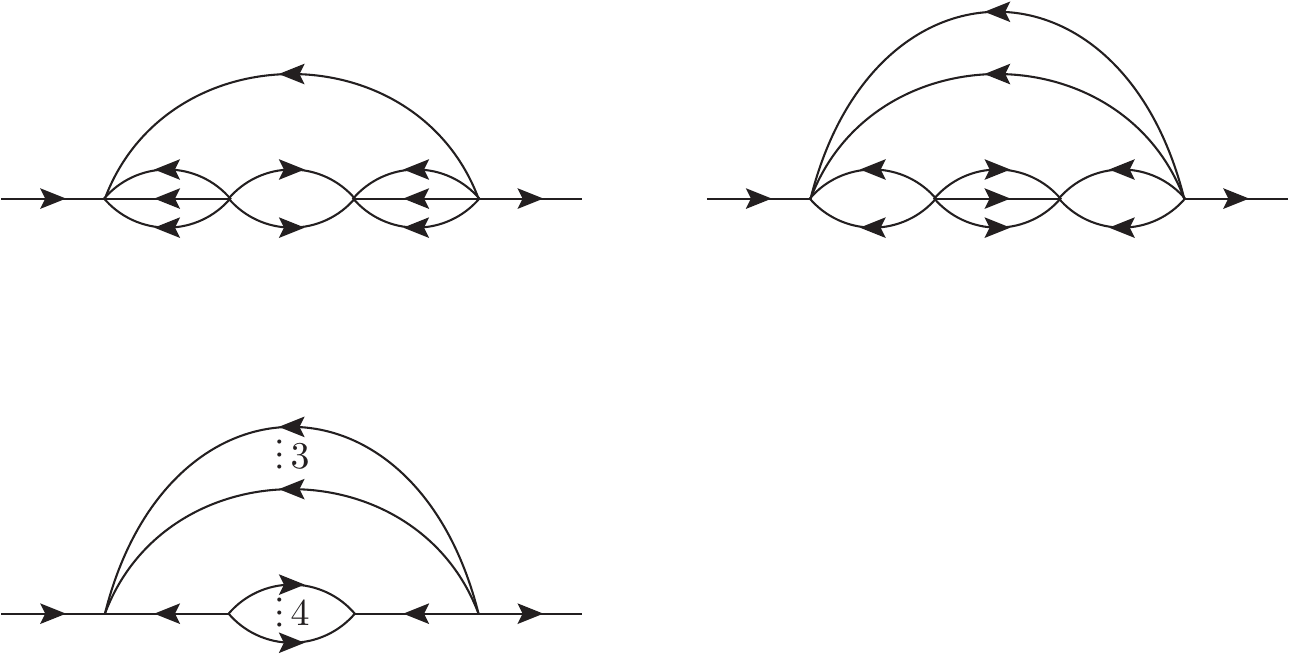}
\end{center}
%\vspace{0.5cm}
\caption{Six loop graphs for $\Phi^5$ theory $2$-point function.}
\label{figp52}
\end{figure}}

If we focus for the moment on the case of $n$~$=$~$5$ applying the algorithm
to the $\Phi^5$ theory we find that the anomalous dimension is  
\begin{equation}
\gamma^{\Phi^5}(a) ~=~
\frac{\sqrt{3} \pi^3 a}{9 \Gamma^3(\twothirds) } ~-~
\left[ 40 \sqrt{3} \pi^3 + 81 \Gamma^3(\twothirds) \right]
\frac{4 \pi^6 a^2}{729 \Gamma^9(\twothirds) } ~+~ O(a^3)
\label{gamma5}
\end{equation}
where here and elsewhere the factor arising from the surface area of the
$d$-dimensional unit sphere is absorbed in the combination
\begin{equation}
a ~=~ \frac{g^2}{(4\pi)^{{\mbox{\small{$\frac{D_n}{2}$}}}}} ~.
\end{equation}
In (\ref{gamma5}) we have applied the identity
\begin{equation}
\Gamma(\third) ~=~ \frac{2\pi}{\sqrt{3} \Gamma(\twothirds)}
\end{equation}
to simplify the expression. While there are three higher order graphs there are
only two terms at $O(a^2)$. The second of these two terms arises from the final
graph of Figure \ref{figp52} and this graph is the insertion of Figure 
\ref{figpn1} on one of the internal lines of the graph itself when $n$~$=$~$5$.
The remaining two graphs correspond to vertex corrections arising from the
graph of Figure \ref{figv51}. As it is clearly finite this means that the first
two graphs of Figure \ref{figp52} are primitives.

{\begin{figure}[ht]
\begin{center}
\includegraphics[width=2.2cm,height=6cm]{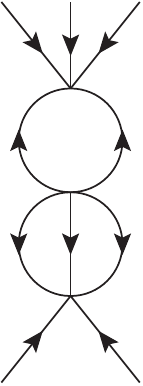}
\end{center}
%\vspace{0.5cm}
\caption{Leading order vertex correction for $\Phi^5$ theory.}
\label{figv51}
\end{figure}}

{\begin{figure}[ht]
\begin{center}
\includegraphics[width=14cm,height=12.0cm]{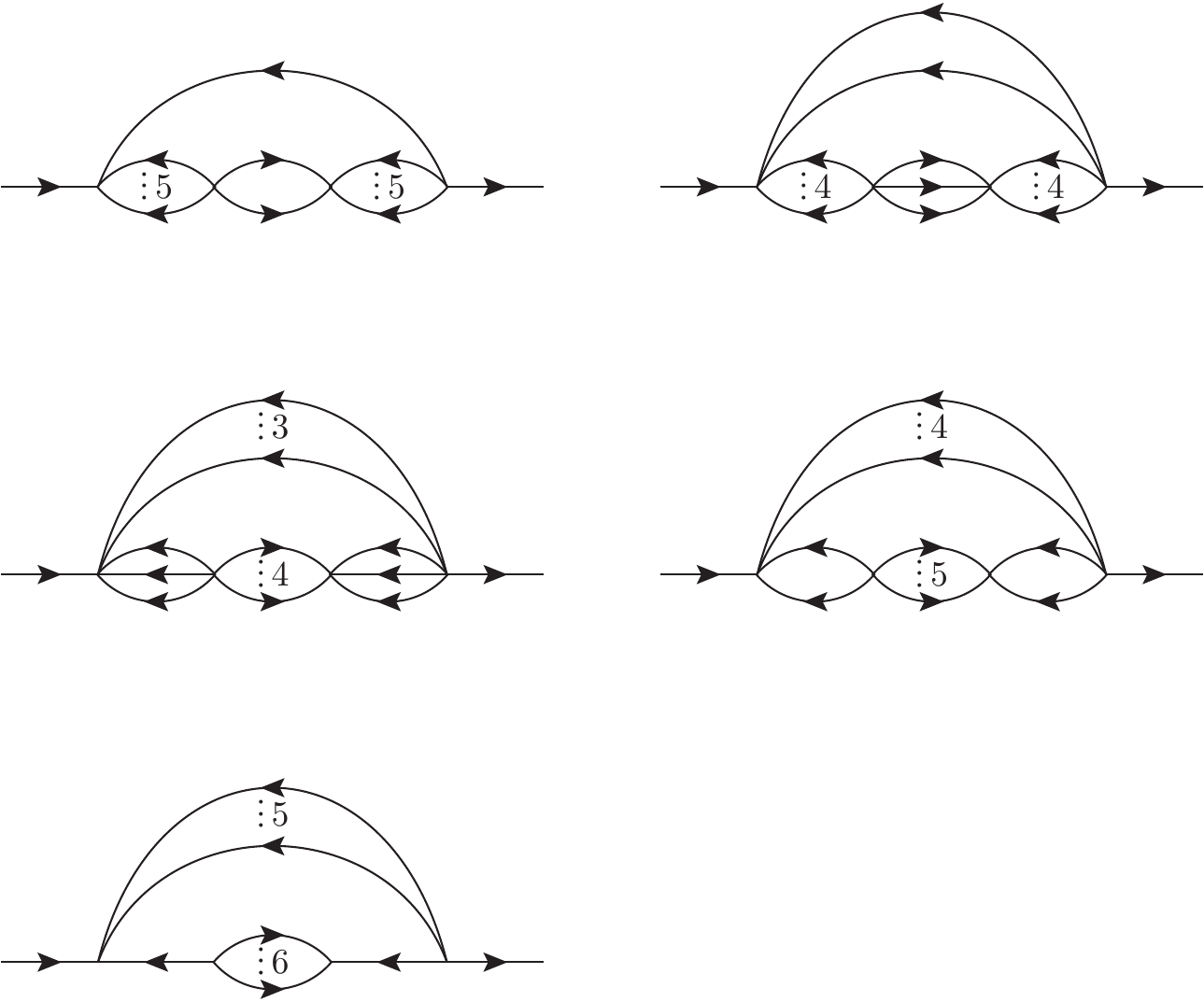}
\end{center}
%\vspace{0.5cm}
\caption{Ten loop graphs for $\Phi^7$ theory $2$-point function.}
\label{figp72}
\end{figure}}

Having discussed the $n$~$=$~$5$ case in detail the procedure to renormalize
the other two cases we consider here, $n$~$=$~$7$ and $9$, is completely 
parallel. The main differences, however, rest in the increase in the number of
graphs for each theory which are illustrated respectively in Figures 
\ref{figp72} and \ref{figp92}. Again the final graph of each figure corresponds
to the self-energy correction on a propagator of the leading order $2$-point 
function. This means the remaining graphs are all primitives as they contain 
vertex subgraph corrections and the leading order vertex graph is finite. The 
resulting anomalous dimensions for both theories are
\begin{eqnarray}
\gamma^{\Phi^7}(a) &=&
\frac{\Gamma^5(\onefifth) a}{144} \nonumber \\
&& 
-~ \left[
63 \Gamma^2(\fourfifth) \Gamma^3(\onefifth)
+ 150 \Gamma(\fourfifth) \Gamma(\twofifth)
+ 175 \Gamma^2(\twofifth) \Gamma^2(\onefifth)
\right]
\frac{\Gamma^{10}(\onefifth) a^2}
{103680 \Gamma(\fourfifth) \Gamma(\twofifth)} \nonumber \\
&& +~ O(a^3)
\end{eqnarray}
and
\begin{eqnarray}
\gamma^{\Phi^9}(a) &=&
\frac{\Gamma^7(\oneseventh) a}{5760} \nonumber \\
&& -~ 
\left[
36 \Gamma^2(\sixseventh) \Gamma(\fiveseventh) \Gamma(\threeseventh)
\Gamma^3(\oneseventh)
+ 98 \Gamma(\sixseventh) \Gamma(\fiveseventh) \Gamma(\threeseventh)
\Gamma(\twoseventh)
+ 441 \Gamma(\sixseventh) \Gamma^2(\threeseventh) \Gamma(\twoseventh) 
\Gamma^2(\oneseventh)
\right. \nonumber \\
&& \left. ~~~~
+ 196 \Gamma^2(\fiveseventh) \Gamma^2(\twoseventh) \Gamma^2(\oneseventh)
\right] \frac{\Gamma^{14}(\oneseventh) a^2}
{58060800 \Gamma(\sixseventh) \Gamma(\fiveseventh) \Gamma(\threeseventh) 
\Gamma(\twoseventh)} ~+~ O(a^3) ~.
\end{eqnarray}
The appearance of factors of the form $\Gamma(p/(n-2))$ where 
$1$~$\leq$~$p$~$\leq$~$(n-3)$ may seem at odds with expectations but arises
from the basic loop bubble integrals. For instance, denoting the value of the
leading order graph of Figure \ref{figsca} by $\Gamma_{(2)}^{\Phi^n}$ then
\begin{equation}
\frac{\Gamma^{n-1}\left(\frac{1}{(n-2)}-\epsilon\right) 
\Gamma((n-2)\epsilon)}
{\Gamma\left(\frac{(n-1)}{(n-2)}-(n-1)\epsilon\right)}
\end{equation}
in $d$-dimensions. The divergence clearly arises from the second numerator
factor while the other numerator one and that in the denominator lead to a 
final factor of $\Gamma^{n-2}\left(\frac{1}{(n-2)}\right)$ in each anomalous
dimension at leading order. Clearly for the Wess-Zumino model, which is cubic,
no $\Gamma$-functions appear in the wave function renormalization at low loop
orders for this reason. 

{\begin{figure}[ht]
\begin{center}
\includegraphics[width=14cm,height=16.0cm]{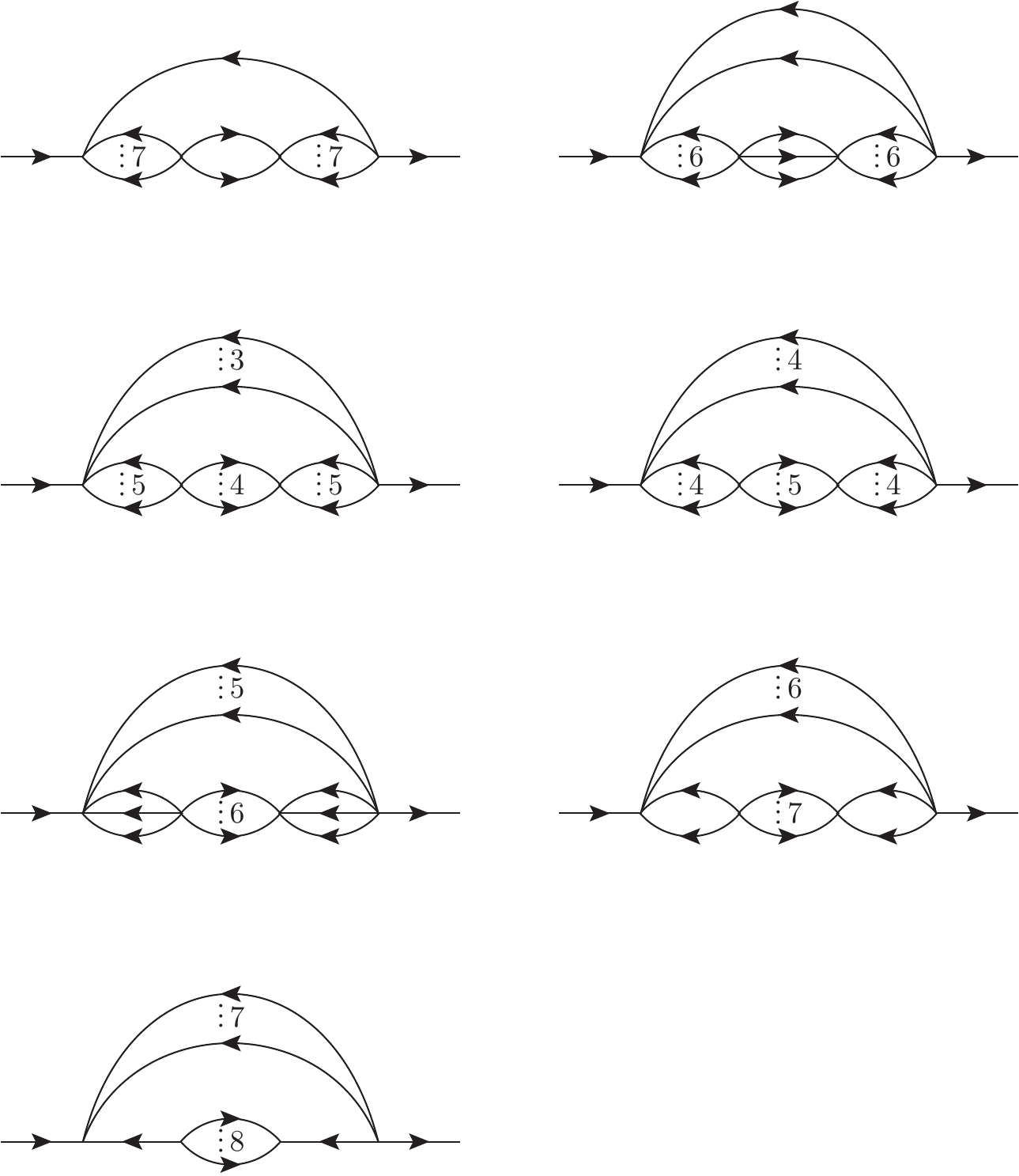}
\end{center}
%\vspace{0.5cm}
\caption{Fourteen loop graphs for $\Phi^9$ theory $2$-point function.}
\label{figp92}
\end{figure}}

With the graphs for both the $n$~$=$~$7$ and $9$ cases available as well as the
explicit anomalous dimensions for the leading two orders we note that there is 
one more graph than there are terms at $O(a^2)$ as was the case for 
$n$~$=$~$5$. This is because two graphs for each theory evaluate to the same 
$\Gamma$-function structure. These are the first two graphs in Figure 
\ref{figp52}, the first and fourth in Figure \ref{figp72} and the first and
sixth graph of Figure \ref{figp92}. The reason why these graphs have the same
structure derives from the underlying $D$-algebra. The consequence of rewriting
the resulting scalar products between loop momenta of the fully internal bubble 
after enacting the $\theta$ integration is to remove or delete a propagator 
from one of the bubbles immediately adjoining it. Applying this observation to 
these specific graphs in the figure produces a pair of graphs with bubbles 
which have the same number of propagators in each or a single propagator. Since
all the bubble integrals are scalar integrals they will each evaluate to the 
same $d$-dimensional expression and hence have the same $\epsilon$ expansion. 
As a final part of the renormalization it is worth providing the numerical 
values for the anomalous dimensions. We have
\begin{eqnarray}
\gamma^{\Phi^5}(a) &=& 2.403246 a ~-~ 809.582836 a^2 ~+~ O(a^3) \nonumber \\
\gamma^{\Phi^7}(a) &=& 14.161200 a ~-~ 416179.106979 a^2 ~+~ O(a^3) 
\nonumber \\
\gamma^{\Phi^9}(a) &=& 89.612261 a ~-~ 225108066.08 a^2 ~+~ O(a^3) ~.
\end{eqnarray}
The large coefficients are not to be regarded as indicating a lack of
convergence. For instance, absorbing the factor of $\Gamma^7(\oneseventh)$ into
$a$ for the $n$~$=$~$9$ case the respective one and two loop coefficients
become $0.000173611$ and $0.000844912$. These are of the same order in much
the same way as for four dimensional theories. Of course in that case the 
corresponding factor would involve powers of $\Gamma(1)$ which have no
consequence.

Equipped with the anomalous dimensions and the $\beta$-functions through the
supersymmetry Ward identities we can determine the critical exponents of each
theory at the Wilson-Fisher fixed point. That associated with the field 
anomalous dimension, $\eta^{\Phi^n}$~$=$~$\gamma^{\Phi^n}(a^\ast)$, where
$a^\ast$ is the critical coupling, can be determined {\em exactly} to all
orders in perturbation as
\begin{equation}
\eta^{\Phi^n} ~=~ \frac{(n-2)}{n} \epsilon 
\label{etaexact}
\end{equation}
for each value of $n$ odd with $n$~$>$~$1$. This follows trivially from 
(\ref{critd}) and (\ref{susywi}). In the case of $n$~$=$~$3$ the four 
dimensional result of \cite{9,11} emerges. For the other integer dimensions of 
interest we find
\begin{equation}
\left. \frac{}{} \eta^{\Phi^n} \right|_{d=2} ~=~ \frac{1}{n} ~~~,~~~
\left. \frac{}{} \eta^{\Phi^n} \right|_{d=3} ~=~ -~ \frac{(n-4)}{n}
\end{equation}
if one assumes a negative value of $\epsilon$ is valid when $D_n$~$<$~$3$. As
$n$~$\to$~$\infty$ the former vanishes while the latter tends to $(-1)$. The
situation with the other exponent, which is the $\beta$-function slope at
criticality, is different in that there is no exact expression for any value of
$n$. Defining $\omega^{\Phi^n}$~$=$~$2 {\beta^{\Phi^n}}^\prime (a^\ast)$ we
have
\begin{eqnarray}
\omega^{\Phi^5} &=& 6 \epsilon ~-~ 
\left[ 40 \sqrt{3} \pi^3 + 81 \Gamma^3(\twothirds) \right] 
\frac{8\epsilon^2}{15\Gamma^3(\twothirds)} ~+~ O(\epsilon^3) \nonumber \\
\omega^{\Phi^7} &=& 10 \epsilon ~-~ 
\left[ 63 \Gamma^2(\fourfifth) \Gamma^3(\onefifth) 
+ 150 \Gamma(\fourfifth) \Gamma(\twofifth) 
+ 175 \Gamma^2(\twofifth) \Gamma^2(\onefifth) \right]
\frac{10\epsilon^2}{7\Gamma(\fourfifth) \Gamma(\twofifth)} ~+~ O(\epsilon^3) 
\nonumber \\
\omega^{\Phi^9} &=& 14 \epsilon \nonumber \\
&& -~ \left[ 
36 \Gamma^2(\sixseventh) \Gamma(\fiveseventh) \Gamma(\threeseventh) 
\Gamma^3(\oneseventh) 
+ 98 \Gamma(\sixseventh) \Gamma(\fiveseventh) \Gamma(\threeseventh) 
\Gamma(\twoseventh) 
+ 441 \Gamma(\sixseventh) \Gamma^2(\threeseventh) \Gamma(\twoseventh) 
\Gamma^2(\oneseventh) 
\right. \nonumber \\
&& ~~~~ \left.
+ 196 \Gamma^2(\fiveseventh) \Gamma^2(\twoseventh) \Gamma^2(\oneseventh)
\right]
\frac{56\epsilon^2}{9 \Gamma(\sixseventh) \Gamma(\fiveseventh) 
\Gamma( \threeseventh) \Gamma(\twoseventh)} ~+~ O(\epsilon^3)
\end{eqnarray}
or
\begin{eqnarray}
\omega^{\Phi^5} &=& 6 \epsilon ~-~ 504.623267 \epsilon^2 ~+~ O(\epsilon^3)
\nonumber \\
\omega^{\Phi^7} &=& 10 \epsilon ~-~ 14823.547215 \epsilon^2 ~+~ O(\epsilon^3)
\nonumber \\
\omega^{\Phi^9} &=& 14 \epsilon ~-~ 305238.813694 \epsilon^2 ~+~ O(\epsilon^3)
\end{eqnarray}
numerically. Clearly there are large corrections for each theory which would
suggest that it is not possible to extract anything meaningful by naively
substituting even a small value of $\epsilon$. However, if we use a $[1,1]$ 
Pad\'{e} approximant we find
\begin{eqnarray}
\left. \frac{}{} \omega^{\Phi^5} \right|_{d=2} &=& 0.0688833 \nonumber \\
\left. \frac{}{} \omega^{\Phi^7} \right|_{d=2} &=& 0.00672335 \nonumber \\
\left. \frac{}{} \omega^{\Phi^9} \right|_{d=2} &=& 0.000642017 
\end{eqnarray}
for instance in two dimensions which appear credible. These are significantly 
smaller than the canonical term which is $2$ for all odd $n$. Under the same 
assumptions as before we deduce
\begin{eqnarray}
\left. \frac{}{} \omega^{\Phi^5} \right|_{d=3} &=& 0.0768208 \nonumber \\
\left. \frac{}{} \omega^{\Phi^7} \right|_{d=3} &=& 0.00676123 \nonumber \\
\left. \frac{}{} \omega^{\Phi^9} \right|_{d=3} &=& 0.000642258
\end{eqnarray}
for the extension to three dimensions.

\sect{$O(N)$ symmetric theories.}

Having considered the renormalization of the core higher order potentials we
consider their $O(N)$ symmetric counterparts in this section. This requires two
distinct superfields $\Phi^i(x,\theta)$ and $\sigma(x,\theta)$ together with 
their chiral partners. The former field takes values in $O(N)$ where 
$1$~$\leq$~$i$~$\leq$~$N$. The presence of two sets of superfields means that 
the action for each core potential is more involved and moreover the number of 
interactions increases with the order of the potential. For instance, when
$n$~$=$~$5$ we have  
\begin{eqnarray}
S^{O(N)}_{(5)} &=& \int d^4 x \left[ \int d^2 \theta d^2 \bar{\theta} \left[
\bar{\Phi}_{\mbox{\footnotesize{o}}}^i (x,\bar{\theta}) 
e^{-2 \theta {\partialline} \bar{\theta}}
\Phi_{\mbox{\footnotesize{o}}}^i (x,\theta) ~+~
\bar{\sigma}_{\mbox{\footnotesize{o}}} (x,\bar{\theta}) 
e^{-2 \theta {\partialline} \bar{\theta}}
\sigma_{\mbox{\footnotesize{o}}} (x,\theta) \right] \right. \nonumber \\
&& \left. ~~~~~~~~~+~
\frac{\tilde{g_1}_{\mbox{\footnotesize{o}}}}{24} \int d^2 \theta \,
\sigma_{\mbox{\footnotesize{o}}} \left( \Phi_{\mbox{\footnotesize{o}}}^i 
\Phi_{\mbox{\footnotesize{o}}}^i \right)^2 ~+~
\frac{\tilde{g_1}_{\mbox{\footnotesize{o}}}}{24} \int d^2 \bar{\theta} \, 
\bar{\sigma}_{\mbox{\footnotesize{o}}}
\left( \bar{\Phi}_{\mbox{\footnotesize{o}}}^i 
\bar{\Phi}_{\mbox{\footnotesize{o}}}^i \right)^2 \right. \nonumber \\
&& \left. ~~~~~~~~~+~
\frac{\tilde{g_2}_{\mbox{\footnotesize{o}}}}{12} \int d^2 \theta \,
\sigma_{\mbox{\footnotesize{o}}}^3 \Phi_{\mbox{\footnotesize{o}}}^i 
\Phi_{\mbox{\footnotesize{o}}}^i ~+~
\frac{\tilde{g_2}_{\mbox{\footnotesize{o}}}}{12} \int d^2 \bar{\theta} \, 
\bar{\sigma}_{\mbox{\footnotesize{o}}}^3
\bar{\Phi}_{\mbox{\footnotesize{o}}}^i 
\bar{\Phi}_{\mbox{\footnotesize{o}}}^i \right. \nonumber \\
&& \left. ~~~~~~~~~+~
\frac{\tilde{g_3}_{\mbox{\footnotesize{o}}}}{120} \int d^2 \theta \, 
\sigma_{\mbox{\footnotesize{o}}}^5 ~+~
\frac{\tilde{g_3}_{\mbox{\footnotesize{o}}}}{120} \int d^2 \bar{\theta} \, 
\bar{\sigma}_{\mbox{\footnotesize{o}}}^5 \right]
\label{act5on}
\end{eqnarray}
for the action in terms of bare quantities where 
$\tilde{g}_i$~$=$~$(4\pi)^{{\mbox{\small{$\frac{D_n}{4}$}}}}g_i$ here and 
throughout. Setting both $\Phi^i(x,\theta)$ and $\bar{\Phi}^i(x,\bar{\theta})$ 
formally to zero recovers the $n$~$=$~$5$ case of (\ref{genact}). An equivalent
way of producing this is to put $g_1$~$=$~$g_2$~$=$~$0$ whence the $O(N)$
multiplet decouples. For the next two theories in the sequence of odd 
potentials the respective actions are
\begin{eqnarray}
S^{O(N)}_{(7)} &=& \int d^4 x \left[ \int d^2 \theta d^2 \bar{\theta} \left[
\bar{\Phi}_{\mbox{\footnotesize{o}}}^i (x,\bar{\theta}) 
e^{-2 \theta {\partialline} \bar{\theta}}
\Phi_{\mbox{\footnotesize{o}}}^i (x,\theta) ~+~
\bar{\sigma}_{\mbox{\footnotesize{o}}} (x,\bar{\theta}) 
e^{-2 \theta {\partialline} \bar{\theta}}
\sigma_{\mbox{\footnotesize{o}}} (x,\theta) \right] \right. \nonumber \\
&& \left. ~~~~~~~~~+~
\frac{\tilde{g_1}_{\mbox{\footnotesize{o}}}}{720} \int d^2 \theta \,
\sigma_{\mbox{\footnotesize{o}}} \left( \Phi_{\mbox{\footnotesize{o}}}^i 
\Phi_{\mbox{\footnotesize{o}}}^i \right)^3 ~+~
\frac{\tilde{g_1}_{\mbox{\footnotesize{o}}}}{720} \int d^2 \bar{\theta} \, 
\bar{\sigma}_{\mbox{\footnotesize{o}}}
\left( \bar{\Phi}_{\mbox{\footnotesize{o}}}^i 
\bar{\Phi}_{\mbox{\footnotesize{o}}}^i \right)^3 \right. \nonumber \\
&& \left. ~~~~~~~~~+~
\frac{\tilde{g_2}_{\mbox{\footnotesize{o}}}}{144} \int d^2 \theta \,
\sigma^3_{\mbox{\footnotesize{o}}} \left( \Phi_{\mbox{\footnotesize{o}}}^i 
\Phi_{\mbox{\footnotesize{o}}}^i \right)^2 ~+~
\frac{\tilde{g_2}_{\mbox{\footnotesize{o}}}}{144} \int d^2 \bar{\theta} \, 
\bar{\sigma}^3_{\mbox{\footnotesize{o}}}
\left( \bar{\Phi}_{\mbox{\footnotesize{o}}}^i 
\bar{\Phi}_{\mbox{\footnotesize{o}}}^i \right)^2 \right. \nonumber \\
&& \left. ~~~~~~~~~+~
\frac{\tilde{g_3}_{\mbox{\footnotesize{o}}}}{240} \int d^2 \theta \,
\sigma^5_{\mbox{\footnotesize{o}}} \Phi_{\mbox{\footnotesize{o}}}^i 
\Phi_{\mbox{\footnotesize{o}}}^i ~+~
\frac{\tilde{g_3}_{\mbox{\footnotesize{o}}}}{240} \int d^2 \bar{\theta} \, 
\bar{\sigma}^5_{\mbox{\footnotesize{o}}}
\bar{\Phi}_{\mbox{\footnotesize{o}}}^i 
\bar{\Phi}_{\mbox{\footnotesize{o}}}^i \right. \nonumber \\
&& \left. ~~~~~~~~~+~
\frac{\tilde{g_4}_{\mbox{\footnotesize{o}}}}{5040} \int d^2 \theta \,
\sigma^7_{\mbox{\footnotesize{o}}} ~+~
\frac{\tilde{g_4}_{\mbox{\footnotesize{o}}}}{5040} \int d^2 \bar{\theta} \, 
\bar{\sigma}^7_{\mbox{\footnotesize{o}}} \right]
\label{act7on}
\end{eqnarray}
and
\begin{eqnarray}
S^{O(N)}_{(9)} &=& \int d^4 x \left[ \int d^2 \theta d^2 \bar{\theta} \left[
\bar{\Phi}_{\mbox{\footnotesize{o}}}^i (x,\bar{\theta}) 
e^{-2 \theta {\partialline} \bar{\theta}}
\Phi_{\mbox{\footnotesize{o}}}^i (x,\theta) ~+~
\bar{\sigma}_{\mbox{\footnotesize{o}}} (x,\bar{\theta}) 
e^{-2 \theta {\partialline} \bar{\theta}}
\sigma_{\mbox{\footnotesize{o}}} (x,\theta) \right] \right. \nonumber \\
&& \left. ~~~~~~~~~+~
\frac{\tilde{g_1}_{\mbox{\footnotesize{o}}}}{40320} \int d^2 \theta \,
\sigma_{\mbox{\footnotesize{o}}} \left( \Phi_{\mbox{\footnotesize{o}}}^i 
\Phi_{\mbox{\footnotesize{o}}}^i \right)^4 ~+~
\frac{\tilde{g_1}_{\mbox{\footnotesize{o}}}}{40320} \int d^2 \bar{\theta} \, 
\bar{\sigma}_{\mbox{\footnotesize{o}}}
\left( \bar{\Phi}_{\mbox{\footnotesize{o}}}^i 
\bar{\Phi}_{\mbox{\footnotesize{o}}}^i \right)^4 \right. \nonumber \\
&& \left. ~~~~~~~~~+~
\frac{\tilde{g_2}_{\mbox{\footnotesize{o}}}}{4320} \int d^2 \theta \,
\sigma^3_{\mbox{\footnotesize{o}}} \left( \Phi_{\mbox{\footnotesize{o}}}^i 
\Phi_{\mbox{\footnotesize{o}}}^i \right)^3 ~+~
\frac{\tilde{g_2}_{\mbox{\footnotesize{o}}}}{4320} \int d^2 \bar{\theta} \, 
\bar{\sigma}^3_{\mbox{\footnotesize{o}}}
\left( \bar{\Phi}_{\mbox{\footnotesize{o}}}^i 
\bar{\Phi}_{\mbox{\footnotesize{o}}}^i \right)^3 \right. \nonumber \\
&& \left. ~~~~~~~~~+~
\frac{\tilde{g_3}_{\mbox{\footnotesize{o}}}}{2880} \int d^2 \theta \,
\sigma^5_{\mbox{\footnotesize{o}}} \left( \Phi_{\mbox{\footnotesize{o}}}^i 
\Phi_{\mbox{\footnotesize{o}}}^i \right)^2 ~+~
\frac{\tilde{g_3}_{\mbox{\footnotesize{o}}}}{2880} \int d^2 \bar{\theta} \, 
\bar{\sigma}^5_{\mbox{\footnotesize{o}}}
\left( \bar{\Phi}_{\mbox{\footnotesize{o}}}^i 
\bar{\Phi}_{\mbox{\footnotesize{o}}}^i \right)^2 \right. \nonumber \\
&& \left. ~~~~~~~~~+~
\frac{\tilde{g_4}_{\mbox{\footnotesize{o}}}}{10080} \int d^2 \theta \,
\sigma^7_{\mbox{\footnotesize{o}}} \Phi_{\mbox{\footnotesize{o}}}^i 
\Phi_{\mbox{\footnotesize{o}}}^i ~+~
\frac{\tilde{g_4}_{\mbox{\footnotesize{o}}}}{10080} \int d^2 \bar{\theta} \, 
\bar{\sigma}^7_{\mbox{\footnotesize{o}}}
\bar{\Phi}_{\mbox{\footnotesize{o}}}^i 
\bar{\Phi}_{\mbox{\footnotesize{o}}}^i \right. \nonumber \\
&& \left. ~~~~~~~~~+~
\frac{\tilde{g_5}_{\mbox{\footnotesize{o}}}}{362880} \int d^2 \theta \, 
\sigma_{\mbox{\footnotesize{o}}}^9 ~+~
\frac{\tilde{g_5}_{\mbox{\footnotesize{o}}}}{362880} \int d^2 \bar{\theta} \, 
\bar{\sigma}_{\mbox{\footnotesize{o}}}^9 \right] 
\label{act9on}
\end{eqnarray}
which illustrate the increase in number of interactions with $n$. Consequently 
a larger number of Feynman graphs have to be computed to extract the 
renormalization group functions. The precise numbers are given in Table 
\ref{tabgraphson} for both sets of $2$-point functions. Like previously the 
$\beta$-functions of the respective coupling constants are determined by a 
generalization of the supersymmetry Ward identities. For $n$~$=$~$5$ these are  
\begin{equation}
Z_{g_1} Z_\Phi^2 Z_\sigma^{\half} ~=~ 
Z_{g_2} Z_\Phi Z_\sigma^{\threehalves} ~=~  
Z_{g_3} Z_\sigma^{\fivehalves} ~=~ 1 
\end{equation}
with
\begin{equation}
Z_{g_1} Z_\Phi^3 Z_\sigma^{\half} ~=~ 
Z_{g_2} Z_\Phi^2 Z_\sigma^{\threehalves} ~=~  
Z_{g_3} Z_\Phi Z_\sigma^{\fivehalves} ~=~ 
Z_{g_4} Z_\sigma^{\sevenhalves} ~=~ 1 
\end{equation}
for $n$~$=$~$7$. Finally 
\begin{equation}
Z_{g_1} Z_\Phi^4 Z_\sigma^{\half} ~=~ 
Z_{g_2} Z_\Phi^3 Z_\sigma^{\threehalves} ~=~  
Z_{g_3} Z_\Phi^2 Z_\sigma^{\fivehalves} ~=~ 
Z_{g_4} Z_\Phi Z_\sigma^{\sevenhalves} ~=~ 
Z_{g_5} Z_\sigma^{\ninehalves} ~=~ 1 
\end{equation}
for (\ref{act9on}) by extending (\ref{susywi}) in the same way.

{\begin{table}[ht]
\begin{center}
\begin{tabular}{|c|r||r|r|r|}
\hline
\rule{0pt}{12pt}
$n$ & $L$ & $\langle \Phi^i \bar{\Phi}^j \rangle$ & 
$\langle \sigma \bar{\sigma} \rangle$ & Total \\
\hline
$5$ & $3$ & $2$ & $3$ & $5$ \\
    & $6$ & $34$ & $40$ & $74$ \\
\hline
$7$ & $5$ & $3$ & $4$ & $7$ \\
    & $10$ & $155$ & $174$ & $329$ \\
\hline
$9$ & $7$ & $4$ & $5$ & $9$ \\
    & $14$ & $480$ & $521$ & $1001$ \\
\hline
\end{tabular}
\end{center}
\begin{center}
\caption{Number of graphs at each loop order $L$ required to renormalize the 
$\Phi^i$ and $\sigma$ $2$-point functions in the $O(N)$ theories.}
\label{tabgraphson}
\end{center}
\end{table}}

For the remainder of this section we focus on the $n$~$=$~$5$ case as an 
example. The procedure to renormalize (\ref{act5on}) follows the same as that
used for (\ref{genact}) with respect to applying the $D$-algebra and the
evaluation of the $79$ $2$-point graphs. The resulting anomalous dimensions are
\begin{eqnarray}
\gamma_\Phi^{\Phi^5}(g_i) &=&
\left[ 4 N g_1^2
+ 8 g_1^2 
+ 12 g_2^2 \right]
\frac{\sqrt{3}\pi^3}{27\Gamma^3(\twothirds)}
\nonumber \\
&& -~
\left[
\left[
128 N^3 g_1^4
+ 1536 N^2 g_1^4
+ 6144 N g_1^4
+ 7168 g_1^4
+ 2304 N^2 g_1^2 g_2^2
+ 16128 N g_1^2 g_2^2
\right. \right. \nonumber \\
&& \left. \left. ~~~~~
+ 23040 g_1^2 g_2^2
+ 2304 N g_1 g_2^2 g_3
+ 4608 g_1 g_2^2 g_3
+ 5760 N g_2^4
+ 20736 g_2^4
\right. \right. \nonumber \\
&& \left. \left. ~~~~~
+ 2304 g_2^2 g_3^2
\right]
\sqrt{3} \pi^9
\right. \nonumber \\
&& \left. ~~~~~+
\left[
324 N^3 g_1^4
+ 5184 N^2 g_1^4
+ 16848 N g_1^4
+ 15552 g_1^4
+ 8748 N^2 g_1^2 g_2^2
\right. \right. \nonumber \\
&& \left. \left. ~~~~~~~~~
+ 33048 N g_1^2 g_2^2
+ 31104 g_1^2 g_2^2
+ 972 N g_1^2 g_3^2
+ 1944 g_1^2 g_3^2
+ 52488 N g_2^4
\right. \right. \nonumber \\
&& \left. \left. ~~~~~~~~~
+ 11664 g_2^4
+ 8748 g_2^2 g_3^2
\right]
\Gamma^3(\twothirds) \pi^6
\right]
\frac{1}{6561\Gamma^9(\twothirds)} ~+~ O( g_i^7) 
\end{eqnarray}
and
\begin{eqnarray}
\gamma_\sigma^{\Phi^5}(g_i) &=&
\left[
N^2 g_1^2
+ 2 N g_1^2
+ 18 N g_2^2
+ 3 g_3^2
\right]
\frac{\sqrt{3} \pi^3}{27\Gamma^3(\twothirds)}
\nonumber \\
&& -~
\left[
\left[
32 N^4 g_1^4
+ 384 N^3 g_1^4
+ 1536 N^2 g_1^4
+ 1792 N g_1^4
+ 1536 N^3 g_1^2 g_2^2
+ 10752 N^2 g_1^2 g_2^2
\right. \right. \nonumber \\
&& \left. \left. ~~~~~
+ 15360 N g_1^2 g_2^2
+ 3456 N^2 g_1 g_2^2 g_3
+ 6912 N g_1 g_2^2 g_3
+ 8640 N^2 g_2^4
+ 31104 N g_2^4
\right. \right. \nonumber \\
&& \left. \left. ~~~~~
+ 9216 N g_2^2 g_3^2
+ 1440 g_3^4
\right]
\sqrt{3} \pi^9 
\right. \nonumber \\
&& \left. ~~~~~+
\left[
1296 N^3 g_1^4
+ 5184 N^2 g_1^4
+ 5184 N g_1^4
+ 2916 N^3 g_1^2 g_2^2
+ 21384 N^2 g_1^2 g_2^2
\right. \right. \nonumber \\
&& \left. \left. ~~~~~~~~~
+ 31104 N g_1^2 g_2^2
+ 972 N^2 g_1^2 g_3^2
+ 1944 N g_1^2 g_3^2
+ 52488 N^2 g_2^4
+ 34992 N g_2^4
\right. \right. \nonumber \\
&& \left. \left. ~~~~~~~~~
+ 26244 N g_2^2 g_3^2
+ 2916 g_3^4
\right]
\Gamma(\twothirds)^3 \pi^6
\right]
\frac{1}{6561\Gamma^9(\twothirds)} ~+~ O( g_i^7) ~. 
\end{eqnarray}
As a trivial check setting $g_1$~$=$~$g_2$~$=$~$0$ in
$\gamma_\sigma^{\Phi^5}(g_i)$ reproduces (\ref{gamma5}). Consequently using the
supersymmetry Ward identities we can deduce the $\beta$-functions which are
\begin{eqnarray}
\beta_1^{\Phi^5}(g_i) &=&
\left[ N^2 g_1^3
+ 18 N g_1^3
+ 32 g_1^3
+ 18 N g_1 g_2^2
+ 48 g_1 g_2^2
+ 3 g_1 g_3^2
\right]
\frac{\sqrt{3}\pi^3}{27\Gamma^3(\twothirds)}
\nonumber \\
&&
-~ \left[
\left[
32 N^4 g_1^5
+ 896 N^3 g_1^5
+ 7680 N^2 g_1^5
+ 26368 N g_1^5
+ 28672 g_1^5
+ 1536 N^3 g_1^3 g_2^2
\right. \right. \nonumber \\
&& \left. \left. ~~~~~
+ 19968 N^2 g_1^3 g_2^2
+ 79872 N g_1^3 g_2^2
+ 92160 g_1^3 g_2^2
+ 3456 N^2 g_1^2 g_2^2 g_3
\right. \right. \nonumber \\
&& \left. \left. ~~~~~
+ 16128 N g_1^2 g_2^2 g_3
+ 18432 g_1^2 g_2^2 g_3
+ 8640 N^2 g_1 g_2^4
+ 54144 N g_1 g_2^4
+ 82944 g_1 g_2^4
\right. \right. \nonumber \\
&& \left. \left. ~~~~~
+ 9216 N g_1 g_2^2 g_3^2
+ 9216 g_1 g_2^2 g_3^2
+ 1440 g_1 g_3^4
\right]
\sqrt{3} \pi^9 
\right.
\nonumber \\
&& \left. ~~~~~
+ \left[
2592 N^3 g_1^5
+ 25920 N^2 g_1^5
+ 72576 N g_1^5
+ 62208 g_1^5
+ 2916 N^3 g_1^3 g_2^2
\right. \right. \nonumber \\
&& \left. \left. ~~~~~~~~~
+ 56376 N^2 g_1^3 g_2^2
+ 163296 N g_1^3 g_2^2
+ 124416 g_1^3 g_2^2
+ 972 N^2 g_1^3 g_3^2
\right. \right. \nonumber \\
&& \left. \left. ~~~~~~~~~
+ 5832 N g_1^3 g_3^2
+ 7776 g_1^3 g_3^2
+ 52488 N^2 g_1 g_2^4
+ 244944 N g_1 g_2^4
+ 46656 g_1 g_2^4
\right. \right. \nonumber \\
&& \left. \left. ~~~~~~~~~
+ 26244 N g_1 g_2^2 g_3^2
+ 34992 g_1 g_2^2 g_3^2
+ 2916 g_1 g_3^4 \right]
\Gamma^3(\twothirds) \pi^6 
\right]
\frac{1}{6561\Gamma^9(\twothirds)} \nonumber \\
&& +~ O( g_i^7) \nonumber \\
\beta_2^{\Phi^5}(g_i) &=&
\frac{\sqrt{3} \pi^3}{27\Gamma^3(\twothirds)}
\left[ 3 N^2 g_1^2 g_2
+ 14 N g_1^2 g_2
+ 16 g_1^2 g_2
+ 54 N g_2^3
+ 24 g_2^3
+ 9 g_2 g_3^2 \right] \nonumber \\
&& -~ \left[
\left[ 96 N^4 g_1^4 g_2
+ 1408 N^3 g_1^4 g_2
+ 7680 N^2 g_1^4 g_2
+ 17664 N g_1^4 g_2
+ 14336 g_1^4 g_2
\right. \right. \nonumber \\
&& \left. \left. ~~~~~
+ 4608 N^3 g_1^2 g_2^3
+ 36864 N^2 g_1^2 g_2^3
+ 78336 N g_1^2 g_2^3
+ 46080 g_1^2 g_2^3
\right. \right. \nonumber \\
&& \left. \left. ~~~~~
+ 10368 N^2 g_1 g_2^3 g_3
+ 25344 N g_1 g_2^3 g_3
+ 9216 g_1 g_2^3 g_3 
+ 25920 N^2 g_2^5 
\right. \right. \nonumber \\
&& \left. \left. ~~~~~
+ 104832 N g_2^5 
+ 41472 g_2^5 
+ 27648 N g_2^3 g_3^2 
+ 4608 g_2^3 g_3^2 
+ 4320 g_2 g_3^4 \right]
\sqrt{3}\pi^9
\right. \nonumber \\
&& \left. ~~~~~
+ \left[
4536 N^3 g_1^4 g_2   
+ 25920 N^2 g_1^4 g_2 
+ 49248 N g_1^4 g_2 
+ 31104 g_1^4 g_2 
+ 8748 N^3 g_1^2 g_2^3 
\right. \right. \nonumber \\
&& \left. \left. ~~~~~~~~~
+ 81648 N^2 g_1^2 g_2^3 
+ 159408 N g_1^2 g_2^3 
+ 62208 g_1^2 g_2^3 
+ 2916 N^2 g_1^2 g_2 g_3^2 
\right. \right. \nonumber \\
&& \left. \left. ~~~~~~~~~
+ 7776 N g_1^2 g_2 g_3^2 
+ 3888 g_1^2 g_2 g_3^2 
+ 157464 N^2 g_2^5 
+ 209952 N g_2^5 
\right. \right. \nonumber \\
&& \left. \left. ~~~~~~~~~
+ 23328 g_2^5 
+ 78732 N g_2^3 g_3^2 
+ 17496 g_2^3 g_3^2 
+ 8748 g_2 g_3^4 
\right]
\Gamma^3(\twothirds) \pi^6
\right]
\frac{1}{6561\Gamma^9(\twothirds)} \nonumber \\
&& +~ O( g_i^7) \nonumber \\
\beta_3^{\Phi^5}(g_i) &=&
\left[ N^2 g_1^2 g_3
+ 10 N g_1^2 g_3
+ 90 N g_2^2 g_3
+ 15 g_3^3 \right]
\frac{\sqrt{3}\pi^3}{27\Gamma^3(\twothirds)}
\nonumber \\
&& -~
\left[
\left[ 160 N^4 g_1^4 g_3
+ 1920 N^3 g_1^4 g_3
+ 7680 N^2 g_1^4 g_3
+ 8960 N g_1^4 g_3
+ 7680 N^3 g_1^2 g_2^2 g_3
\right. \right. \nonumber \\
&& \left. \left. ~~~~~
+ 53760 N^2 g_1^2 g_2^2 g_3
+ 76800 N g_1^2 g_2^2 g_3
+ 17280 N^2 g_1 g_2^2 g_3^2
+ 34560 N g_1 g_2^2 g_3^2
\right. \right. \nonumber \\
&& \left. \left. ~~~~~
+ 43200 N^2 g_2^4 g_3
+ 155520 N g_2^4 g_3
+ 46080 N g_2^2 g_3^3
+ 7200 g_3^5
\right]
\sqrt{3} \pi^9
\right. \nonumber \\
&& \left. ~~~~~
+ \left[
6480 N^3 g_1^4 g_3
+ 25920 N^2 g_1^4 g_3
+ 25920 N g_1^4 g_3
+ 14580 N^3 g_1^2 g_2^2 g_3
\right. \right. \nonumber \\
&& \left. \left. ~~~~~~~~~
+ 106920 N^2 g_1^2 g_2^2 g_3
+ 155520 N g_1^2 g_2^2 g_3
+ 4860 N^2 g_1^2 g_3^3
+ 9720 N g_1^2 g_3^3
\right. \right. \nonumber \\
&& \left. \left. ~~~~~~~~~
+ 262440 N^2 g_2^4 g_3
+ 174960 N g_2^4 g_3
+ 131220 N g_2^2 g_3^3
\right. \right. \nonumber \\
&& \left. \left. ~~~~~~~~~
+ 14580 g_3^5
\right]
\Gamma^3(\twothirds) \pi^6
\right]
\frac{1}{6561\Gamma^9(\twothirds)} ~+~ O( g_i^7) ~.
\end{eqnarray}
Clearly $\beta_1^{\Phi^5}(g_i)$ and $\beta_2^{\Phi^5}(g_i)$ vanish when 
$g_1$~$=$~$g_2$~$=$~$0$ leaving $\beta_3^{\Phi^5}(g_i)$ as five times 
$\gamma_\sigma^{\Phi^5}(g_i)$ under the same condition. This is consistent with
the Ward identity of (\ref{genact}) at $n$~$=$~$5$. Renormalization group 
functions for $n$~$=$~$7$ and $9$ are recorded in the Appendices. Expressions 
for the renormalization group functions for each of the three theories are 
provided in electronic format in the associated data file.

\sect{Fixed point analysis.}

Having established the renormalization group functions we now examine the fixed
point properties of the theories. In the first instance we focus on the
$n$~$=$~$5$ case for arbitrary $N$ and consider the Wilson-Fisher fixed point. 
Setting
\begin{equation}
g_i ~=~ \frac{x_i \sqrt{\epsilon}}
{\left(\Gamma\left(\frac{1}{(n-2)}\right)\right)^{n-2}}
\end{equation}
in general we find that there is a large set of solutions. A significant number
are merely various coupling constant reflections $g_i$~$\to$~$-$~$g_i$ of a 
core subset. Therefore we only record the independent ones for $n$~$=$~$5$ and 
other cases in the region of coupling constant space where $g_i$~$\geq$~$0$. 
The location of those where there is one nonzero critical coupling are
\begin{eqnarray}
x_1^{(1)} &=& \left[ 6 
+ \left[ 12 (N+16) (N^2+10 N+28) \Gamma^3(\third) + 864 (N+2) (N+6) \right]
\frac{\epsilon}{(N+2)(N+16)^2} \right. \nonumber \\
&& \left. ~ + O(\epsilon^2) \right]
\sqrt{\frac{2}{(N+2)(N+16)}} ~~,~~
x_2^{(1)} ~=~ 0 ~~,~~
x_3^{(1)} ~=~ 0 ~; \nonumber \\
x_1^{(2)} &=& 0 \nonumber \\
x_2^{(2)} &=& \left[ 2 
+ \left[ 6 (9N+4) (5 N+18) \Gamma^3(\third) + 54 (27 N^2+36 N+4) \right]
\frac{\epsilon}{3(9N+4)^2} \right. \nonumber \\
&& \left. ~ + O(\epsilon^2) \right] \sqrt{\frac{3}{(9N+4)}} ~~,~~
x_3^{(2)} ~=~ 0 ~; \nonumber \\
x_1^{(3)} &=& 0 ~~,~~
x_2^{(3)} ~=~ 0 ~~,~~ 
x_3^{(3)} ~=~ \frac{2}{5} \sqrt{30} ~+~ \left[ 5 \Gamma^3(\third) + 9 \right]
\frac{4\sqrt{30}}{25} \epsilon ~+~ O(\epsilon^2)
\end{eqnarray}
with associated anomalous dimensions
\begin{eqnarray}
\eta^{\Phi^5}_{\Phi \, (1)} &=& \frac{12\epsilon}{(N+16)} ~-~
\frac{108N(N-4)\epsilon^2}{(N+16)^3} ~+~ O(\epsilon^3) \nonumber \\
\eta^{\Phi^5}_{\sigma \, (1)} &=& \frac{3N\epsilon}{(N+16)} ~-~
\frac{432N(N-4)\epsilon^2}{(N+16)^3} ~+~ O(\epsilon^3) ~; \nonumber \\
\eta^{\Phi^5}_{\Phi \, (2)} &=& \frac{6\epsilon}{(9N+4)} ~-~
\frac{486N(3N-2)\epsilon^2}{(9N+4)^3} ~+~ O(\epsilon^3) \nonumber \\
\eta^{\Phi^5}_{\sigma \, (2)} &=& \frac{9N\epsilon}{(9N+4)} ~-~
\frac{324N(3N-2)\epsilon^2}{(9N+4)^3} ~+~ O(\epsilon^3) ~; \nonumber \\
\eta^{\Phi^5}_{\Phi \, (3)} &=& O(\epsilon^3) ~~,~~ 
\eta^{\Phi^5}_{\sigma \, (3)} ~=~ \frac{3}{5} \epsilon ~+~ O(\epsilon^3) ~.
\end{eqnarray}
One interesting feature is that for both solutions $1$ and $2$ is that
$\eta^{\Phi^5}_\Phi$ and $\eta^{\Phi^5}_\sigma$ are equal for a specific but 
different value of $N$. For solution $1$ this is $N$~$=$~$4$ while it is
$N$~$=$~$\frac{3}{2}$ for solution $2$. The latter case is formal in the sense
that $N$ is non-integer. However in both instances the value of the exponent
is $\frac{3}{5} \epsilon$. The final solution labelled $3$ corresponds to 
(\ref{etaexact}). The next scenario is when only one of the couplings vanishes 
at criticality. Again there are three cases with the critical couplings given 
by
\begin{eqnarray}
x_1^{(12)} &=& \left[ \frac{6}{5}
+ \left[ (N+6) (N^2+16 N+4) \Gamma^3(\third) + 90 N (N+2) \right]
\frac{6\epsilon}{125N(N+2)} \right. \nonumber \\
&& \left. ~+ O(\epsilon^2) \right]
\sqrt{\frac{(3N-2)}{N(N+2)}} \nonumber \\
x_2^{(12)} &=& \left[ \frac{1}{5}
+ \left[ (7 N^2+54 N+32) \Gamma^3(\third) + 180 N \right]
\frac{\epsilon}{250N} ~+~ O(\epsilon^2) \right] \sqrt{\frac{6(4-N)}{N}}
\nonumber \\ 
x_3^{(12)} &=& 0 ~; \nonumber \\
x_1^{(13)} &=& \left[ \frac{3}{5}
+ \left[ (N^2+10 N+28) \Gamma^3(\third) + 36 (N+2) \right]
\frac{3\epsilon}{50(N+2)} ~+~ O(\epsilon^2) \right]
\sqrt{\frac{10}{(N+2)}} \nonumber \\
x_2^{(13)} &=& 0 \nonumber \\
x_3^{(13)} &=& \left[ \frac{1}{5}
- \left[ 5 (N-4) \Gamma^3(\third) - 36 \right]
\frac{\epsilon}{50} ~+~ O(\epsilon^2) \right] \sqrt{30(4-N)} ~;
\nonumber \\ 
x_1^{(23)} &=& 0 \nonumber \\
x_2^{(23)} &=& \left[ \frac{1}{5}
- \left[ (7 N-26) \Gamma^3(\third) - 36  \right]
\frac{\epsilon}{50} ~+~ O(\epsilon^2) \right] \sqrt{30}
\nonumber \\ 
x_3^{(23)} &=& \left[ \frac{2}{5}
- \left[ 5 (N-2) \Gamma^3(\third) - 18 \right]
\frac{2\epsilon}{25} ~+~ O(\epsilon^2) \right] \sqrt{15(2-3N)} ~.
\end{eqnarray}
In each case the anomalous dimensions are all the same since
\begin{eqnarray}
\eta^{\Phi^5}_{\Phi \, (12)} &=& 
\eta^{\Phi^5}_{\sigma \, (12)} ~=~ \frac{3}{5} \epsilon ~+~ O(\epsilon^3) 
\nonumber \\
\eta^{\Phi^5}_{\Phi \, (13)} &=& 
\eta^{\Phi^5}_{\sigma \, (13)} ~=~ \frac{3}{5} \epsilon ~+~ O(\epsilon^3) 
\nonumber \\
\eta^{\Phi^5}_{\Phi \, (23)} &=& 
\eta^{\Phi^5}_{\sigma \, (23)} ~=~ \frac{3}{5} \epsilon ~+~ O(\epsilon^3) ~.
\end{eqnarray}
As a check on these fixed point solutions we note that
\begin{eqnarray}
\lim_{N\to 4} x_1^{(12)} &=& \lim_{N\to 4} x_1^{(1)} ~~~,~~~
\lim_{N\to \frac{2}{3}} x_2^{(12)} ~=~ \lim_{N\to \frac{2}{3}} x_2^{(2)} 
\nonumber \\
\lim_{N\to 4} x_1^{(13)} &=& \lim_{N\to 4} x_1^{(1)}  ~~~,~~~
\lim_{N\to \frac{2}{3}} x_2^{(23)} ~=~ \lim_{N\to \frac{2}{3}} x_2^{(2)} 
\end{eqnarray}
and for these cases the anomalous dimensions all equate to 
$\frac{3}{5} \epsilon$. These particular values of $N$ point to a deeper aspect
of the latter set of fixed point solutions. For instance for solutions $12$ and
$13$ one critical coupling of the pair becomes complex for $N$~$>$~$4$ with a 
similar observation for solutions $12$ and $23$ when $N$~$>$~$\frac{2}{3}$. In 
this case there is then no real solution for any positive integer $N$. So it 
appears that the $N$~$=$~$4$ represents a watershed in terms of the set of 
possible real fixed point solutions. This is especially the case since for that
value the solution $1$ $\eta^{\Phi^5}_\Phi$ and $\eta^{\Phi^5}_\sigma$ are 
equal but there is only one pair of interaction terms at criticality with 
$\sigma$ and $\bar{\sigma}$ appearing linearly in (\ref{act5on}). The remaining
single coupling solutions equally identify one pair of interactions but with 
$\sigma$ and its partner occuring nonlinearly. The final case is when none of
the critical couplings vanish at the Wilson-Fisher fixed point. This will be 
considered in the next section as a special case.

For the other two theories we focus on, the properties of the critical points 
is completely parallel. By this we mean that there are fixed points both for 
only one non-zero critical coupling as well as a set for pairs. To illustrate 
this we record the explicit forms of the field critical anomalous dimensions. 
For $n$~$=$~$7$ we have 
\begin{eqnarray}
\eta^{\Phi^7}_{\Phi \, (1)} &=& \frac{30\epsilon}{(N+36)} ~-~
\frac{750N(N-6)\epsilon^2}{(N+36)^3} ~+~ O(\epsilon^3) \nonumber \\
\eta^{\Phi^7}_{\sigma \, (1)} &=& \frac{5N\epsilon}{(N+36)} ~+~
\frac{4500N(N-6)\epsilon^2}{(N+36)^3} ~+~ O(\epsilon^3) ~; \nonumber \\
\eta^{\Phi^7}_{\Phi \, (2)} &=& \frac{20\epsilon}{(9N+16)} ~-~
\frac{4500N(3N-4)\epsilon^2}{(9N+16)^3} ~+~ O(\epsilon^3) \nonumber \\
\eta^{\Phi^7}_{\sigma \, (2)} &=& \frac{15N\epsilon}{(9N+16)} ~+~
\frac{6000N(3N-4)\epsilon^2}{(9N+16)^3} ~+~ O(\epsilon^3) ~; \nonumber \\
\eta^{\Phi^7}_{\Phi \, (3)} &=& \frac{10\epsilon}{(25N+4)} ~-~
\frac{6250N(5N-2)\epsilon^2}{(25N+4)^3} ~+~ O(\epsilon^3) \nonumber \\
\eta^{\Phi^7}_{\sigma \, (3)} &=& \frac{25N\epsilon}{(25N+4)} ~+~
\frac{2500N(5N-2)\epsilon^2}{(25N+4)^3} ~+~ O(\epsilon^3) ~; \nonumber \\
\eta^{\Phi^7}_{\Phi \, (4)} &=& O(\epsilon^3) ~~,~~ 
\eta^{\Phi^7}_{\sigma \, (4)} ~=~ \frac{5}{7} \epsilon ~+~ O(\epsilon^3) ~; 
\nonumber \\
\eta^{\Phi^7}_{\Phi \, (ij)} &=& 
\eta^{\Phi^7}_{\sigma \, (ij)} ~=~ \frac{5}{7} \epsilon ~+~ O(\epsilon^3) 
\end{eqnarray}
for $1$~$\geq$~$i$~$>$~$j$~$\geq$~$5$. While for $n$~$=$~$9$ we find 
\begin{eqnarray}
\eta^{\Phi^9}_{\Phi \, (1)} &=& \frac{56\epsilon}{(N+64)} ~-~
\frac{2744N(N-8)\epsilon^2}{(N+64)^3} ~+~ O(\epsilon^3) \nonumber \\
\eta^{\Phi^9}_{\sigma \, (1)} &=& \frac{7N\epsilon}{(N+64)} ~+~
\frac{21952N(N-8)\epsilon^2}{(N+64)^3} ~+~ O(\epsilon^3) ~; \nonumber \\
\eta^{\Phi^9}_{\Phi \, (2)} &=& \frac{14\epsilon}{3(N+4)} ~-~
\frac{686N(N-2)\epsilon^2}{9(N+4)^3} ~+~ O(\epsilon^3) \nonumber \\
\eta^{\Phi^9}_{\sigma \, (2)} &=& \frac{7N\epsilon}{(N+4)} ~+~
\frac{1372N(N-2)\epsilon^2}{9(N+4)^3} ~+~ O(\epsilon^3) ~; \nonumber \\
\eta^{\Phi^9}_{\Phi \, (3)} &=& \frac{28\epsilon}{(25N+16)} ~-~
\frac{34300N(5N-4)\epsilon^2}{(25N+16)^3} ~+~ O(\epsilon^3) \nonumber \\
\eta^{\Phi^9}_{\sigma \, (3)} &=& \frac{35N\epsilon}{(25N+16)} ~+~
\frac{27440N(5N-4)\epsilon^2}{(25N+16)^3} ~+~ O(\epsilon^3) ~; \nonumber \\
\eta^{\Phi^9}_{\Phi \, (4)} &=& \frac{14\epsilon}{(49N+4)} ~-~
\frac{33614N(7N-2)\epsilon^2}{(49N+4)^3} ~+~ O(\epsilon^3) \nonumber \\
\eta^{\Phi^9}_{\sigma \, (4)} &=& \frac{49N\epsilon}{(49N+4)} ~+~
\frac{9604N(7N-2)\epsilon^2}{(49N+4)^3} ~+~ O(\epsilon^3) ~; \nonumber \\
\eta^{\Phi^9}_{\Phi \, (5)} &=& O(\epsilon^3) ~~,~~ 
\eta^{\Phi^9}_{\sigma \, (5)} ~=~ \frac{7}{9} \epsilon ~+~ O(\epsilon^3) ~; 
\nonumber \\
\eta^{\Phi^9}_{\Phi \, (ij)} &=& 
\eta^{\Phi^9}_{\sigma \, (ij)} ~=~ \frac{7}{9} \epsilon ~+~ O(\epsilon^3) 
\end{eqnarray}
for $1$~$\geq$~$i$~$>$~$j$~$\geq$~$5$. From these it is equally clear that for
special values of $N$ the $\Phi^i$ and $\sigma$ exponents equate. Moreover
they follow a general pattern which is
\begin{equation}
\eta^{\Phi^n}_{\Phi \, (r)} ~=~ \eta^{\Phi^n}_{\sigma \, (r)} 
\end{equation}
when 
\begin{equation}
N ~=~ \frac{(n-2r+1)}{(2r-1)}
\label{Nvalue}
\end{equation}
for each fixed point labelled by $r$ in the range 
$1$~$\leq$~$r$~$\leq$~$\half (n-1)$. The final single coupling fixed point 
denoted by solution $5$ corresponds to the single field case of the previous 
section.

\sect{$OSp(1|2M)$ enhancement.}

We now turn to a special case of when all critical coupling are non-zero and
either real or complex. This is motivated by the observation in the 
non-supersymmetric case, \cite{24}, that there is a symmetry enhancement for
a specific value of $N$ for each $n$. Briefly for each group $O(N)$ the 
enhancement is to the group $OSp(1|2M)$, where $n$~$=$~$(2M+1)$. In particular 
the value for $N$ when this occurs is $N$~$=$~$-$~$2M$, \cite{24}. While
this was for the case of the non-supersymmetric model the property should also
hold for (\ref{act5on}), (\ref{act7on}) and (\ref{act9on}). To make this
manifest in the Lagrangian formulation will involve the superfields $\sigma$ 
and $\Theta^i$ and their chiral partners. Unlike $\Phi^i$ of previous sections 
$\Theta^i$ is a Grassmann field in order to realize the symplectic aspect of 
the group. Similar to \cite{24} this allows one to express the superpotential 
as a function of both sets of fields. In particular the $OSp(1|2M)$ action is
\begin{eqnarray}
S^{OSp(1|2M)} &=& \int d^4 x \left[ \int d^2 \theta d^2 \bar{\theta} \left[
\bar{\Phi}_{\mbox{\footnotesize{o}}}^i (x,\bar{\theta}) 
e^{-2 \theta {\partialline} \bar{\theta}}
\Phi_{\mbox{\footnotesize{o}}}^i (x,\theta) ~+~
\bar{\sigma}_{\mbox{\footnotesize{o}}} (x,\bar{\theta}) 
e^{-2 \theta {\partialline} \bar{\theta}}
\sigma_{\mbox{\footnotesize{o}}} (x,\theta) \right] \right. \nonumber \\
&& \left. ~~~~~~~~~+~
\tilde{g}_{\mbox{\footnotesize{o}}} \int d^2 \theta \,
\left( \sigma^2_{\mbox{\footnotesize{o}}} ~+~
\Theta_{\mbox{\footnotesize{o}}}^i 
\Theta_{\mbox{\footnotesize{o}}}^i \right)^{{\half}(2M+1)} \right. \nonumber \\
&& \left. ~~~~~~~~~+~
\tilde{g}_{\mbox{\footnotesize{o}}} \int d^2 \bar{\theta} \,
\left( \bar{\sigma}^2_{\mbox{\footnotesize{o}}} ~+~
\bar{\Theta}_{\mbox{\footnotesize{o}}}^i 
\bar{\Theta}_{\mbox{\footnotesize{o}}}^i \right)^{{\half}(2M+1)} \right] 
\label{actosp}
\end{eqnarray}
where the subscript again indicates bare objects. If we define the 
superpotential by 
\begin{equation}
V_M(\sigma,\Theta) ~=~ \left( \sigma^2 ~+~ \Theta^i \Theta^i 
\right)^{{\half}(2M+1)}
\end{equation}
motivated by the construction of \cite{24} then the first few cases are 
\begin{eqnarray}
V_2(\sigma,\Theta) &=& 
\frac{15}{8} \sigma \left( \Theta^i \Theta^i \right)^2 ~+~
\frac{5}{2} \sigma^3 \Theta^i \Theta^i ~+~
\sigma^5 \nonumber \\
V_3(\sigma,\Theta) &=& 
\frac{35}{16} \sigma \left( \Theta^i \Theta^i \right)^3 ~+~
\frac{35}{8} \sigma^3 \left( \Theta^i \Theta^i \right)^2 ~+~
\frac{7}{2} \sigma^5 \Theta^i \Theta^i ~+~
\sigma^7 \nonumber \\
V_4(\sigma,\Theta) &=& 
\frac{315}{128} \sigma \left( \Theta^i \Theta^i \right)^4 ~+~
\frac{105}{16} \sigma^3 \left( \Theta^i \Theta^i \right)^3 ~+~
\frac{63}{8} \sigma^5 \left( \Theta^i \Theta^i \right)^2 \nonumber \\
&& +~ \frac{9}{2} \sigma^7 \Theta^i \Theta^i ~+~ \sigma^9 
\label{superpot}
\end{eqnarray}
due to the Grassmann property of $\Theta^i$. When $M$~$=$~$1$ the $OSp(1|1)$
version of the Wess-Zumino model results. The relative coefficients of the
terms in each of the superpotentials of (\ref{superpot}) are instrumental in
deducing the emergent $OSp(1|2M)$ symmetry for various values of $N$. These
will be in the same ratio as discovered in the non-supersymmetric case of
\cite{24}. In particular the vector of critical couplings to the first two
orders are 
\begin{eqnarray}
\left. \frac{}{} \left( g_1^\ast, g_2^\ast, g_3^\ast \right) 
\right|_{N=-4}^{n=5} &=&
\left[ 1 ~+~ \left[ \frac{18}{5} - 3 \Gamma^3(\third) \right] \epsilon ~+~
O(\epsilon^2) \right]
i \sqrt{\frac{3\epsilon}{5\Gamma^3(\third)}} \left( 3, 2, 8 \right) 
\nonumber \\
\left. \frac{}{} \left( g_1^\ast, g_2^\ast, g_3^\ast, g_4^\ast \right) 
\right|_{N=-6}^{n=7} &=&
\left[ 1 ~+~ \left[ 30 
- \frac{7 \Gamma(\fourfifth) \Gamma^3(\onefifth)}{\Gamma(\twofifth)} 
+ \frac{35 \Gamma(\twofifth) \Gamma^2(\onefifth)}{\Gamma(\fourfifth)} \right]
\frac{5\epsilon}{14} 
\right. \nonumber \\
&& \left. ~~~~
+~ O(\epsilon^2) \right]
\sqrt{\frac{35\epsilon}{7\Gamma^5(\onefifth)}}
\left( 15, 6, 8, 48 \right) \nonumber \\ 
\left. \frac{}{} \left( g_1^\ast, g_2^\ast, g_3^\ast, g_4^\ast, g_5^\ast 
\right) \right|_{N=-8}^{n=9} &=&
\left[ 1 ~+~ 
\left[
980 \Gamma(\sixseventh) \Gamma(\fiveseventh) \Gamma(\threeseventh) 
\Gamma(\twoseventh) 
- 126 \Gamma^2(\sixseventh) \Gamma(\fiveseventh) \Gamma(\threeseventh) 
\Gamma^3(\oneseventh) 
\right. \right. \nonumber \\
&& \left. \left. ~~~~~~~~~
- 2205 \Gamma(\sixseventh) \Gamma^2(\threeseventh) \Gamma(\twoseventh) 
\Gamma^2(\oneseventh) 
\right. \right. \nonumber \\
&& \left. \left. ~~~~~~~~~
+ 490 \Gamma^2(\fiveseventh) \Gamma^2(\twoseventh) \Gamma^2(\oneseventh)
\right] 
\frac{\epsilon}{45 \Gamma(\sixseventh) \Gamma(\fiveseventh) 
\Gamma(\threeseventh) \Gamma(\twoseventh)}
\right. \nonumber \\
&& \left. ~~~~
+~ O(\epsilon^2) \right]
i \sqrt{\frac{7\epsilon}{\Gamma^7(\oneseventh)}}
\left( 35, 10, 8, 16, 128 \right) ~.
\label{emegeccospn}
\end{eqnarray}
That this emergent symmetry holds for the supersymmetric case is not too
surprising given that it occurs in the non-supersymmetric equivalent theories.
However the observation is subtle here in that the specific value of 
$N$~$=$~$(1-n)$ for the emergence has connections with the non-Grassmann
$O(N)$ partner theory if one sets $r$~$=$~$1$ in (\ref{Nvalue}). It is known
that properties of the $Sp(N)$ group can be related to those of an orthogonal
group $O(N)$ if one maps $N$~$\to$~$-$~$N$. What is the case for $N$ not equal 
to the emergent value value of $(1-n)$ is that the field anomalous dimensions 
are not equal. It is only for each value of $N$~$=$~$(1-n)$ that 
\begin{equation}
\eta^{\Phi^n}_\Phi ~=~ \eta^{\Phi^n}_\sigma 
\end{equation}
for the critical couplings (\ref{emegeccospn}) whence the emergent $OSp(1|2M)$
symmetry is realized in the supersymmetric theory. 

As we are able to go to a higher order in the $\epsilon$ expansion compared to 
the non-supersymmetric cases it is instructive to determine the critical 
$\beta$-function slope for the emergent $OSp(1|2M)$ theories. In particular we
have 
\begin{eqnarray}
\left. \omega^{\Phi^5} \right|_{N=-4} &=& 6 \epsilon ~+~ 
\left[ 180 \Gamma^3(\third) - 216 \right] \frac{\epsilon^2}{5} ~+~
O(\epsilon^3) \nonumber \\
\left. \omega^{\Phi^7} \right|_{N=-6} &=& 10 \epsilon \nonumber \\
&& +~ \left[ 350 \Gamma^2(\fourfifth) \Gamma^3(\onefifth) 
- 1500 \Gamma(\fourfifth) \Gamma(\twofifth) 
- 1750 \Gamma^2(\twofifth) \Gamma^2(\onefifth) \right]
\frac{\epsilon^2}{7 \Gamma(\fourfifth) \Gamma(\twofifth)} ~+~ O(\epsilon^3) 
\nonumber \\
\left. \omega^{\Phi^9} \right|_{N=-8} &=& 14 \epsilon \nonumber \\
&& +~ \left[ 18 \Gamma^2(\sixseventh) \Gamma(\fiveseventh) 
\Gamma(\threeseventh) \Gamma^3(\oneseventh) 
- 140 \Gamma(\sixseventh) \Gamma(\fiveseventh) \Gamma(\threeseventh) 
\Gamma(\twoseventh) 
+ 315 \Gamma(\sixseventh) \Gamma^2(\threeseventh) 
\Gamma(\twoseventh) \Gamma^2(\oneseventh) 
\right. \nonumber \\
&& \left. ~~~~
- 70 \Gamma^2(\fiveseventh) \Gamma^2(\twoseventh) \Gamma^2(\oneseventh) \right]
\frac{196 \epsilon^2}{45 \Gamma(\sixseventh) \Gamma(\fiveseventh) 
\Gamma(\threeseventh) \Gamma(\twoseventh)} ~+~ O(\epsilon^3)
\end{eqnarray}
analytically which equates to
\begin{eqnarray}
\left. \omega^{\Phi^5} \right|_{N=-4} &=& 6 \epsilon ~+~ 
648.934900 \epsilon^2 ~+~ O(\epsilon^3) \nonumber \\
\left. \omega^{\Phi^7} \right|_{N=-6} &=& 10 \epsilon ~-~ 
7713.844209 \epsilon^2 ~+~ O(\epsilon^3) \nonumber \\
\left. \omega^{\Phi^9} \right|_{N=-8} &=& 14 \epsilon ~+~ 
79461.413957 \epsilon^2 ~+~ O(\epsilon^3)
\end{eqnarray}
numerically. Clearly the coefficient of the $O(\epsilon^2)$ term is
significantly large and that increases with $n$. However this needs to be
tempered by the fact that the limit of $D_n$ is $2$ as $n$ increases. Indeed
with $d$~$=$~$D_n$~$-$~$2\epsilon$ then setting 
$\epsilon$~$=$~$\frac{1}{(n-2)}$ produces $d$~$=$~$2$. However even with this 
choice of $\epsilon$ the value of $\omega$ for the respective theories carries 
no meaning. One option would be to improve the convergence by using a Pad\'{e} 
approximant to estimate $\omega$ in $d$~$=$~$2$. For $n$~$=$~$5$ and $9$, 
however, the Pad\'{e} approximant is singular in the range 
$2$~$<$~$d$~$<$~$D_n$ since the correction term is positive. This is not the 
case for $n$~$=$~$7$ when a $[1,1]$ Pad\'{e} approximant gives 
$\left. \omega^{\Phi^7} \right|_{N=-6}$~$=$~$0.012880$ which is significantly
lower than the canonical value. What remains to be clarified is the effect of 
the as yet uncalculated subsequent $\epsilon$ term would be to this estimate. 
Indeed a value of the $O(\epsilon^2)$ term could produce a non-singular 
Pad\'{e} approximant for the other two theories.

{\begin{figure}[ht]
\begin{center}
\includegraphics[width=7.0cm,height=7.0cm]{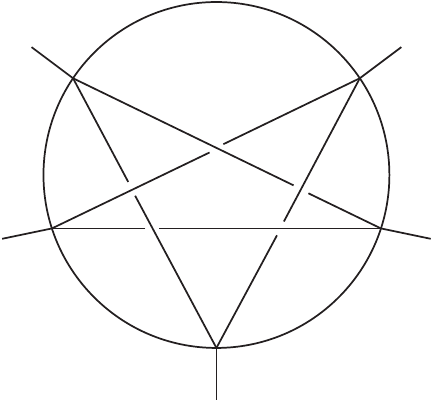}
\end{center}
%\vspace{0.5cm}
\caption{Primitive graph contributing to second order $\beta$-function in 
non-supersymmetric $\phi^5$ theory.}
\label{figv52s}
\end{figure}}

\sect{Discussion.}

The main interest in exploring the supersymmetric extension of theories with a
potential with an odd number of fields was to ascertain whether the $OSp(1|2M)$
emergence of the non-supersymm\-etric case, \cite{24}, was maintained. It was
not surprising that this is indeed the case, which we expect to be manifest
beyond the three cases studies in depth here, but there are subtle aspects to 
the analysis. For instance the lowest order potential with $n$~$=$~$3$ has been
extensively studied as it corresponds to the Wess-Zumino model, \cite{13}. In
that theory it was known that as a consequence of the supersymmetry Ward
identities the critical exponents of the basic fields of the theory can be
deduced {\em exactly} in the $\epsilon$ expansion near the model's critical
dimension. For the extension to $n$~$>$~$3$ with $n$ odd none of these theories
have an integer critical dimension. While this may indicate limited physical
interest $D_n$ is relatively close to an integer dimension which is either two 
or three. Therefore the convergence of critical exponent estimates for the
variety of fixed points we examined in the $O(N)$ theory should be relatively
quick. This was an important exercise for this class of theories with 
non-integer dimensions. Aside from \cite{24} there have been other studies of
the non-supersymmetric non-integer critical dimension theories, 
\cite{27,32,33}, with that of the Blume-Capel model being just above three 
dimensions. In that case only the leading order renormalization group functions
are known since the underlying Feynman graphs are straightforward to evaluate. 
However the corrections to the coupling constant renormalization involve a
significantly large number of graphs. One of these is illustrated in Figure 
\ref{figv52s}. It is clearly non-planar as well as being a primitive and has 
yet to be evaluated. It is likely to have to be treated in the same way as the 
analogous graphs of $\phi^6$ theory in the third order determination of its 
$\beta$-function, \cite{37}. Clearly the graph is absent in the supersymmetric 
extension due to the chiral property of the interaction which simplified the 
analysis of this article. Consequently it has not been possible to ascertain 
whether the $\epsilon$ expansion of critical exponents in the Blume-Capel case 
improves let alone obtain more accurate estimates. It is in this context that 
our supersymmetric analysis has provided some insight. Even in this case,
however, we expect there to be a calculational hurdle to overcome at the next
order to determine the $\beta$-function of the supersymmetric theories which
will have an intricacy akin to that of Figure \ref{figv52s}.

\vspace{0.5cm}
\noindent
{\bf Acknowledgements.} The author thanks I.R. Klebanov for valuable 
discussions. This work was carried out with the support of STFC through the 
Consolidated Grant ST/T000988/1 and partly with the support of a DFG Mercator 
Fellowship. The diagrams were prepared with the {\sc Axodraw} package, 
\cite{38}. For the purpose of open access, the author has applied a Creative 
Commons Attribution (CC-BY) licence to any Author Accepted Manuscript version 
arising. The data representing the main results here are accessible in 
electronic form from the arXiv ancillary directory associated with the article.

\appendix

\sect{Results for the $O(N)$ $\Phi^7$ theory.}

This appendix records the renormalization group functions for the $O(N)$ 
symmetric theory based on an $n$~$=$~$7$ potential. These results and those for
the other two $O(N)$ theories are available in electronic form in the 
associated data file. First the anomalous dimensions for the fields are
\begin{eqnarray}
\gamma_\Phi^{\Phi^7}(g_i) &=&
\left[ 
3 N^2 g_1^2
+ 18 N g_1^2
+ 24 g_1^2
+ 50 N g_2^2
+ 100 g_2^2
+ 45 g_3^2
\right]
\frac{\Gamma^5(\onefifth)}{1080} \nonumber \\
&& -~
\left[ 
\left[ 
27 N^5 g_1^4
+ 648 N^4 g_1^4
+ 6588 N^3 g_1^4
+ 32400 N^2 g_1^4
+ 74304 N g_1^4
+ 62208 g_1^4
\right. \right. \nonumber \\
&& \left. \left. ~~~~~~
+ 1875 N^4 g_1^2 g_2^2
+ 33000 N^3 g_1^2 g_2^2
+ 199500 N^2 g_1^2 g_2^2
+ 498000 N g_1^2 g_2^2
\right. \right. \nonumber \\
&& \left. \left. ~~~~~~
+ 432000 g_1^2 g_2^2
+ 540 N^3 g_1^2 g_3^2
+ 16200 N^2 g_1^2 g_3^2
+ 82080 N g_1^2 g_3^2
\right. \right. \nonumber \\
&& \left. \left. ~~~~~~
+ 103680 g_1^2 g_3^2
+ 18000 N^3 g_1 g_2^2 g_3
+ 144000 N^2 g_1 g_2^2 g_3
+ 360000 N g_1 g_2^2 g_3
\right. \right. \nonumber \\
&& \left. \left. ~~~~~~
+ 288000 g_1 g_2^2 g_3
+ 4050 N^2 g_1 g_2 g_3 g_4
+ 24300 N g_1 g_2 g_3 g_4
+ 32400 g_1 g_2 g_3 g_4
\right. \right. \nonumber \\
&& \left. \left. ~~~~~~
+ 16250 N^3 g_2^4
+ 265000 N^2 g_2^4
+ 925000 N g_2^4
+ 920000 g_2^4
\right. \right. \nonumber \\
&& \left. \left. ~~~~~~
+ 175500 N^2 g_2^2 g_3^2
+ 904500 N g_2^2 g_3^2
+ 1107000 g_2^2 g_3^2
+ 6750 N g_2^2 g_4^2
\right. \right. \nonumber \\
&& \left. \left. ~~~~~~
+ 13500 g_2^2 g_4^2
+ 27000 N g_2 g_3^2 g_4
+ 54000 g_2 g_3^2 g_4
+ 204525 N g_3^4
\right. \right. \nonumber \\
&& \left. \left. ~~~~~~
+ 202500 g_3^4
+ 22275 g_3^2 g_4^2
\right]
\Gamma^2(\fourfifth) \Gamma^{13} (\onefifth)
\right. \nonumber \\
&& \left. ~~~~~~+
\left[ 
75 N^5 g_1^4
+ 3150 N^4 g_1^4
+ 30900 N^3 g_1^4
+ 124200 N^2 g_1^4
\right. \right. \nonumber \\
&& \left. \left. ~~~~~~~~~~
+ 220800 N g_1^4
+ 144000 g_1^4
+ 9375 N^4 g_1^2 g_2^2
+ 135000 N^3 g_1^2 g_2^2
\right. \right. \nonumber \\
&& \left. \left. ~~~~~~~~~~
+ 667500 N^2 g_1^2 g_2^2
+ 1350000 N g_1^2 g_2^2
+ 960000 g_1^2 g_2^2
+ 22500 N^3 g_1^2 g_3^2
\right. \right. \nonumber \\
&& \left. \left. ~~~~~~~~~~
+ 175500 N^2 g_1^2 g_3^2
+ 423000 N g_1^2 g_3^2
+ 324000 g_1^2 g_3^2
+ 1125 N^2 g_1^2 g_4^2
\right. \right. \nonumber \\
&& \left. \left. ~~~~~~~~~~
+ 6750 N g_1^2 g_4^2
+ 9000 g_1^2 g_4^2
+ 281250 N^3 g_2^4
+ 1500000 N^2 g_2^4
\right. \right. \nonumber \\
&& \left. \left. ~~~~~~~~~~
+ 2625000 N g_2^4
+ 1500000 g_2^4
+ 1265625 N^2 g_2^2 g_3^2
+ 2981250 N g_2^2 g_3^2
\right. \right. \nonumber \\
&& \left. \left. ~~~~~~~~~~
+ 900000 g_2^2 g_3^2
+ 56250 N g_2^2 g_4^2
+ 112500 g_2^2 g_4^2
+ 1265625 N g_3^4
\right. \right. \nonumber \\
&& \left. \left. ~~~~~~~~~~
+ 101250 g_3^4
+ 84375 g_3^2 g_4^2
\right]
\Gamma(\fourfifth) \Gamma(\twofifth) \Gamma^{10}(\onefifth)
\right. \nonumber \\
&& \left. ~~~~~~+
\left[ 
27 N^5 g_1^4
+ 972 N^4 g_1^4
+ 12204 N^3 g_1^4
+ 69408 N^2 g_1^4
\right. \right. \nonumber \\
&& \left. \left. ~~~~~~~~~~
+ 177984 N g_1^4
+ 161280 g_1^4
+ 2625 N^4 g_1^2 g_2^2
+ 60000 N^3 g_1^2 g_2^2
\right. \right. \nonumber \\
&& \left. \left. ~~~~~~~~~~
+ 466500 N^2 g_1^2 g_2^2
+ 1434000 N g_1^2 g_2^2
+ 1440000 g_1^2 g_2^2
+ 2700 N^3 g_1^2 g_3^2
\right. \right. \nonumber \\
&& \left. \left. ~~~~~~~~~~
+ 27000 N^2 g_1^2 g_3^2
+ 86400 N g_1^2 g_3^2
+ 86400 g_1^2 g_3^2
+ 18000 N^3 g_1 g_2^2 g_3
\right. \right. \nonumber \\
&& \left. \left. ~~~~~~~~~~
+ 288000 N^2 g_1 g_2^2 g_3
+ 1224000 N g_1 g_2^2 g_3
+ 1440000 g_1 g_2^2 g_3
\right. \right. \nonumber \\
&& \left. \left. ~~~~~~~~~~
+ 6750 N^2 g_1 g_2 g_3 g_4
+ 40500 N g_1 g_2 g_3 g_4
+ 54000 g_1 g_2 g_3 g_4
\right. \right. \nonumber \\
&& \left. \left. ~~~~~~~~~~
+ 46250 N^3 g_2^4
+ 565000 N^2 g_2^4
+ 2245000 N g_2^4
\right. \right. \nonumber \\
&& \left. \left. ~~~~~~~~~~
+ 2600000 g_2^4
+ 337500 N^2 g_2^2 g_3^2
+ 2362500 N g_2^2 g_3^2
\right. \right. \nonumber \\
&& \left. \left. ~~~~~~~~~~
+ 3375000 g_2^2 g_3^2
+ 3750 N g_2^2 g_4^2
+ 7500 g_2^2 g_4^2
+ 135000 N g_2 g_3^2 g_4
\right. \right. \nonumber \\
&& \left. \left. ~~~~~~~~~~
+ 270000 g_2 g_3^2 g_4
+ 253125 N g_3^4
+ 810000 g_3^4
\right. \right. \nonumber \\
&& \left. \left. ~~~~~~~~~~
+ 50625 g_3^2 g_4^2
\right]
\Gamma^2(\twofifth) \Gamma^{12}(\onefifth)
\right]
\frac{1}{11664000\Gamma(\fourfifth)\Gamma(\twofifth)} ~+~ O(g_i^6)
\end{eqnarray}
and
\begin{eqnarray}
\gamma_\sigma^{\Phi^7}(g_i) &=&
\left[ 
N^3 g_1^2
+ 6 N^2 g_1^2
+ 8 N g_1^2
+ 75 N^2 g_2^2
+ 150 N g_2^2
+ 225 N g_3^2
+ 15 g_4^2 
\right]
\frac{\Gamma^5(\onefifth)}{2160} \nonumber \\
&& -~
\left[ 
\left[ 
3 N^6 g_1^4
+ 72 N^5 g_1^4
+ 732 N^4 g_1^4
+ 3600 N^3 g_1^4
+ 8256 N^2 g_1^4
+ 6912 N g_1^4
\right. \right. \nonumber \\
&& \left. \left. ~~~~~~
+ 500 N^5 g_1^2 g_2^2
+ 8800 N^4 g_1^2 g_2^2
+ 53200 N^3 g_1^2 g_2^2
+ 132800 N^2 g_1^2 g_2^2
\right. \right. \nonumber \\
&& \left. \left. ~~~~~~
+ 115200 N g_1^2 g_2^2
+ 270 N^4 g_1^2 g_3^2
+ 8100 N^3 g_1^2 g_3^2
+ 41040 N^2 g_1^2 g_3^2
\right. \right. \nonumber \\
&& \left. \left. ~~~~~~
+ 51840 N g_1^2 g_3^2
+ 9000 N^4 g_1 g_2^2 g_3
+ 72000 N^3 g_1 g_2^2 g_3
+ 180000 N^2 g_1 g_2^2 g_3
\right. \right. \nonumber \\
&& \left. \left. ~~~~~~
+ 144000 N g_1 g_2^2 g_3
+ 3600 N^3 g_1 g_2 g_3 g_4
+ 21600 N^2 g_1 g_2 g_3 g_4
\right. \right. \nonumber \\
&& \left. \left. ~~~~~~
+ 28800 N g_1 g_2 g_3 g_4
+ 8125 N^4 g_2^4
+ 132500 N^3 g_2^4
+ 462500 N^2 g_2^4
+ 460000 N g_2^4
\right. \right. \nonumber \\
&& \left. \left. ~~~~~~
+ 156000 N^3 g_2^2 g_3^2
+ 804000 N^2 g_2^2 g_3^2
+ 984000 N g_2^2 g_3^2
+ 11250 N^2 g_2^2 g_4^2
\right. \right. \nonumber \\
&& \left. \left. ~~~~~~
+ 22500 N g_2^2 g_4^2
+ 45000 N^2 g_2 g_3^2 g_4
+ 90000 N g_2 g_3^2 g_4
+ 340875 N^2 g_3^4
\right. \right. \nonumber \\
&& \left. \left. ~~~~~~
+ 337500 N g_3^4
+ 89100 N g_3^2 g_4^2
+ 4725 g_4^4
\right]
\Gamma^2(\fourfifth) \Gamma^{13}(\onefifth)
\right. \nonumber \\
&& \left. ~~~~~~+
\left[ 
300 N^5 g_1^4
+ 3600 N^4 g_1^4
+ 15600 N^3 g_1^4
+ 28800 N^2 g_1^4
\right. \right. \nonumber \\
&& \left. \left. ~~~~~~~~~~
+ 19200 N g_1^4
+ 1250 N^5 g_1^2 g_2^2
+ 30000 N^4 g_1^2 g_2^2
+ 185000 N^3 g_1^2 g_2^2
\right. \right. \nonumber \\
&& \left. \left. ~~~~~~~~~~
+ 420000 N^2 g_1^2 g_2^2
+ 320000 N g_1^2 g_2^2
+ 7500 N^4 g_1^2 g_3^2
+ 72000 N^3 g_1^2 g_3^2
\right. \right. \nonumber \\
&& \left. \left. ~~~~~~~~~~
+ 222000 N^2 g_1^2 g_3^2
+ 216000 N g_1^2 g_3^2
+ 750 N^3 g_1^2 g_4^2
+ 4500 N^2 g_1^2 g_4^2
\right. \right. \nonumber \\
&& \left. \left. ~~~~~~~~~~
+ 6000 N g_1^2 g_4^2
+ 93750 N^4 g_2^4
+ 625000 N^3 g_2^4
+ 1375000 N^2 g_2^4
\right. \right. \nonumber \\
&& \left. \left. ~~~~~~~~~~
+ 1000000 N g_2^4
+ 843750 N^3 g_2^2 g_3^2
+ 2287500 N^2 g_2^2 g_3^2
+ 1200000 N g_2^2 g_3^2
\right. \right. \nonumber \\
&& \left. \left. ~~~~~~~~~~
+ 75000 N^2 g_2^2 g_4^2
+ 150000 N g_2^2 g_4^2
+ 1687500 N^2 g_3^4
+ 337500 N g_3^4
\right. \right. \nonumber \\
&& \left. \left. ~~~~~~~~~~
+ 281250 N g_3^2 g_4^2
+ 11250 g_4^4
\right]
\Gamma(\fourfifth) \Gamma(\twofifth) \Gamma^{10}(\onefifth)
\right. \nonumber \\
&& \left. ~~~~~~+
\left[ 
3 N^6 g_1^4
+ 108 N^5 g_1^4
+ 1356 N^4 g_1^4
+ 7712 N^3 g_1^4
\right. \right. \nonumber \\
&& \left. \left. ~~~~~~~~~~
+ 19776 N^2 g_1^4
+ 17920 N g_1^4
+ 700 N^5 g_1^2 g_2^2
+ 16000 N^4 g_1^2 g_2^2
\right. \right. \nonumber \\
&& \left. \left. ~~~~~~~~~~
+ 124400 N^3 g_1^2 g_2^2
+ 382400 N^2 g_1^2 g_2^2
+ 384000 N g_1^2 g_2^2
+ 1350 N^4 g_1^2 g_3^2
\right. \right. \nonumber \\
&& \left. \left. ~~~~~~~~~~
+ 13500 N^3 g_1^2 g_3^2
+ 43200 N^2 g_1^2 g_3^2
+ 43200 N g_1^2 g_3^2
+ 9000 N^4 g_1 g_2^2 g_3
\right. \right. \nonumber \\
&& \left. \left. ~~~~~~~~~~
+ 144000 N^3 g_1 g_2^2 g_3
+ 612000 N^2 g_1 g_2^2 g_3
+ 720000 N g_1 g_2^2 g_3
\right. \right. \nonumber \\
&& \left. \left. ~~~~~~~~~~
+ 6000 N^3 g_1 g_2 g_3 g_4
+ 36000 N^2 g_1 g_2 g_3 g_4
+ 48000 N g_1 g_2 g_3 g_4
+ 23125 N^4 g_2^4
\right. \right. \nonumber \\
&& \left. \left. ~~~~~~~~~~
+ 282500 N^3 g_2^4
+ 1122500 N^2 g_2^4
+ 1300000 N g_2^4
+ 300000 N^3 g_2^2 g_3^2
\right. \right. \nonumber \\
&& \left. \left. ~~~~~~~~~~
+ 2100000 N^2 g_2^2 g_3^2
+ 3000000 N g_2^2 g_3^2
+ 6250 N^2 g_2^2 g_4^2
+ 12500 N g_2^2 g_4^2
\right. \right. \nonumber \\
&& \left. \left. ~~~~~~~~~~
+ 225000 N^2 g_2 g_3^2 g_4
+ 450000 N g_2 g_3^2 g_4
+ 421875 N^2 g_3^4
+ 1350000 N g_3^4
\right. \right. \nonumber \\
&& \left. \left. ~~~~~~~~~~
+ 202500 N g_3^2 g_4^2
+ 13125 g_4^4
\right]
\Gamma^2(\twofifth) \Gamma^{12}(\onefifth)
\right]
\frac{1}{7776000\Gamma(\fourfifth)\Gamma(\twofifth)} \nonumber \\
&& +~ O(g_i^6) ~.
\end{eqnarray}
Consequently the supersymmetry Ward identities determine the four
$\beta$-functions as 
\begin{eqnarray}
\beta_1^{\Phi^7}(g_i) &=&
\left[ 
N^3 g_1^3 
+ 42 N^2 g_1^3
+ 224 N g_1^3
+ 288 g_1^3
+ 75 N^2 g_1 g_2^2
+ 750 N g_1 g_2^2
+ 1200 g_1 g_2^2
\right. \nonumber \\
&& \left. ~
+ 225 N g_1 g_3^2
+ 540 g_1 g_3^2
+ 15 g_1 g_4^2 
\right]
\frac{\Gamma^5(\onefifth)^5}{2160}
\nonumber \\
&& -~
\left[ 
\left[ 
3 N^6 g_1^5
+ 180 N^5 g_1^5
+ 3324 N^4 g_1^5
+ 29952 N^3 g_1^5
+ 137856 N^2 g_1^5
+ 304128 N g_1^5
\right. \right. \nonumber \\
&& \left. \left. ~~~~~~
+ 248832 g_1^5
+ 500 N^5 g_1^3 g_2^2
+ 16300 N^4 g_1^3 g_2^2
+ 185200 N^3 g_1^3 g_2^2
\right. \right. \nonumber \\
&& \left. \left. ~~~~~~
+ 930800 N^2 g_1^3 g_2^2
+ 2107200 N g_1^3 g_2^2
+ 1728000 g_1^3 g_2^2
+ 270 N^4 g_1^3 g_3^2
\right. \right. \nonumber \\
&& \left. \left. ~~~~~~
+ 10260 N^3 g_1^3 g_3^2
+ 105840 N^2 g_1^3 g_3^2
+ 380160 N g_1^3 g_3^2
+ 414720 g_1^3 g_3^2
\right. \right. \nonumber \\
&& \left. \left. ~~~~~~
+ 9000 N^4 g_1^2 g_2^2 g_3
+ 144000 N^3 g_1^2 g_2^2 g_3
+ 756000 N^2 g_1^2 g_2^2 g_3
+ 1584000 N g_1^2 g_2^2 g_3
\right. \right. \nonumber \\
&& \left. \left. ~~~~~~
+ 1152000 g_1^2 g_2^2 g_3
+ 3600 N^3 g_1^2 g_2 g_3 g_4
+ 37800 N^2 g_1^2 g_2 g_3 g_4
\right. \right. \nonumber \\
&& \left. \left. ~~~~~~
+ 126000 N g_1^2 g_2 g_3 g_4
+ 129600 g_1^2 g_2 g_3 g_4
+ 8125 N^4 g_1 g_2^4
+ 197500 N^3 g_1 g_2^4
\right. \right. \nonumber \\
&& \left. \left. ~~~~~~
+ 1522500 N^2 g_1 g_2^4
+ 4160000 N g_1 g_2^4
+ 3680000 g_1 g_2^4
+ 156000 N^3 g_1 g_2^2 g_3^2
\right. \right. \nonumber \\
&& \left. \left. ~~~~~~
+ 1506000 N^2 g_1 g_2^2 g_3^2
+ 4602000 N g_1 g_2^2 g_3^2
+ 4428000 g_1 g_2^2 g_3^2
+ 11250 N^2 g_1 g_2^2 g_4^2
\right. \right. \nonumber \\
&& \left. \left. ~~~~~~
+ 49500 N g_1 g_2^2 g_4^2
+ 54000 g_1 g_2^2 g_4^2
+ 45000 N^2 g_1 g_2 g_3^2 g_4
+ 198000 N g_1 g_2 g_3^2 g_4
\right. \right. \nonumber \\
&& \left. \left. ~~~~~~
+ 216000 g_1 g_2 g_3^2 g_4
+ 340875 N^2 g_1 g_3^4
+ 1155600 N g_1 g_3^4
+ 810000 g_1 g_3^4
\right. \right. \nonumber \\
&& \left. \left. ~~~~~~
+ 89100 N g_1 g_3^2 g_4^2
+ 89100 g_1 g_3^2 g_4^2
+ 4725 g_1 g_4^4
\right]
\Gamma^2(\fourfifth) \Gamma^{13}(\onefifth) 
\right. \nonumber \\
&& \left. ~~~~~~+
\left[ 
600 N^5 g_1^5
+ 16200 N^4 g_1^5
+ 139200 N^3 g_1^5
+ 525600 N^2 g_1^5
+ 902400 N g_1^5
\right. \right. \nonumber \\
&& \left. \left. ~~~~~~~~~~
+ 576000 g_1^5
+ 1250 N^5 g_1^3 g_2^2
+ 67500 N^4 g_1^3 g_2^2
+ 725000 N^3 g_1^3 g_2^2
\right. \right. \nonumber \\
&& \left. \left. ~~~~~~~~~~
+ 3090000 N^2 g_1^3 g_2^2
+ 5720000 N g_1^3 g_2^2
+ 3840000 g_1^3 g_2^2
+ 7500 N^4 g_1^3 g_3^2
\right. \right. \nonumber \\
&& \left. \left. ~~~~~~~~~~
+ 162000 N^3 g_1^3 g_3^2
+ 924000 N^2 g_1^3 g_3^2
+ 1908000 N g_1^3 g_3^2
+ 1296000 g_1^3 g_3^2
\right. \right. \nonumber \\
&& \left. \left. ~~~~~~~~~~
+ 750 N^3 g_1^3 g_4^2
+ 9000 N^2 g_1^3 g_4^2
+ 33000 N g_1^3 g_4^2
+ 36000 g_1^3 g_4^2
\right. \right. \nonumber \\
&& \left. \left. ~~~~~~~~~~
+ 93750 N^4 g_1 g_2^4
+ 1750000 N^3 g_1 g_2^4
+ 7375000 N^2 g_1 g_2^4
+ 11500000 N g_1 g_2^4
\right. \right. \nonumber \\
&& \left. \left. ~~~~~~~~~~
+ 6000000 g_1 g_2^4
+ 843750 N^3 g_1 g_2^2 g_3^2
+ 7350000 N^2 g_1 g_2^2 g_3^2
\right. \right. \nonumber \\
&& \left. \left. ~~~~~~~~~~
+ 13125000 N g_1 g_2^2 g_3^2
+ 3600000 g_1 g_2^2 g_3^2
+ 75000 N^2 g_1 g_2^2 g_4^2
\right. \right. \nonumber \\
&& \left. \left. ~~~~~~~~~~
+ 375000 N g_1 g_2^2 g_4^2
+ 450000 g_1 g_2^2 g_4^2
+ 1687500 N^2 g_1 g_3^4
\right. \right. \nonumber \\
&& \left. \left. ~~~~~~~~~~
+ 5400000 N g_1 g_3^4
+ 405000 g_1 g_3^4
+ 281250 N g_1 g_3^2 g_4^2
\right. \right. \nonumber \\
&& \left. \left. ~~~~~~~~~~
+ 337500 g_1 g_3^2 g_4^2
+ 11250 g_1 g_4^4
\right]
\Gamma(\fourfifth) \Gamma(\twofifth) \Gamma^{10}(\onefifth) 
\right. \nonumber \\
&& \left. ~~~~~~+
\left[ 
3 N^6 g_1^5
+ 216 N^5 g_1^5
+ 5244 N^4 g_1^5
+ 56528 N^3 g_1^5
+ 297408 N^2 g_1^5
\right. \right. \nonumber \\
&& \left. \left. ~~~~~~~~~~
+ 729856 N g_1^5
+ 645120 g_1^5
+ 700 N^5 g_1^3 g_2^2
+ 26500 N^4 g_1^3 g_2^2
\right. \right. \nonumber \\
&& \left. \left. ~~~~~~~~~~
+ 364400 N^3 g_1^3 g_2^2
+ 2248400 N^2 g_1^3 g_2^2
+ 6120000 N g_1^3 g_2^2
+ 5760000 g_1^3 g_2^2
\right. \right. \nonumber \\
&& \left. \left. ~~~~~~~~~~
+ 1350 N^4 g_1^3 g_3^2
+ 24300 N^3 g_1^3 g_3^2
+ 151200 N^2 g_1^3 g_3^2
+ 388800 N g_1^3 g_3^2
\right. \right. \nonumber \\
&& \left. \left. ~~~~~~~~~~
+ 345600 g_1^3 g_3^2
+ 9000 N^4 g_1^2 g_2^2 g_3
+ 216000 N^3 g_1^2 g_2^2 g_3
+ 1764000 N^2 g_1^2 g_2^2 g_3
\right. \right. \nonumber \\
&& \left. \left. ~~~~~~~~~~
+ 5616000 N g_1^2 g_2^2 g_3
+ 5760000 g_1^2 g_2^2 g_3
+ 6000 N^3 g_1^2 g_2 g_3 g_4
\right. \right. \nonumber \\
&& \left. \left. ~~~~~~~~~~
+ 63000 N^2 g_1^2 g_2 g_3 g_4
+ 210000 N g_1^2 g_2 g_3 g_4
+ 216000 g_1^2 g_2 g_3 g_4
\right. \right. \nonumber \\
&& \left. \left. ~~~~~~~~~~
+ 23125 N^4 g_1 g_2^4
+ 467500 N^3 g_1 g_2^4
+ 3382500 N^2 g_1 g_2^4
\right. \right. \nonumber \\
&& \left. \left. ~~~~~~~~~~
+ 10280000 N g_1 g_2^4
+ 10400000 g_1 g_2^4
+ 300000 N^3 g_1 g_2^2 g_3^2
\right. \right. \nonumber \\
&& \left. \left. ~~~~~~~~~~
+ 3450000 N^2 g_1 g_2^2 g_3^2
+ 12450000 N g_1 g_2^2 g_3^2
+ 13500000 g_1 g_2^2 g_3^2
\right. \right. \nonumber \\
&& \left. \left. ~~~~~~~~~~
+ 6250 N^2 g_1 g_2^2 g_4^2
+ 27500 N g_1 g_2^2 g_4^2
+ 30000 g_1 g_2^2 g_4^2
+ 225000 N^2 g_1 g_2 g_3^2 g_4
\right. \right. \nonumber \\
&& \left. \left. ~~~~~~~~~~
+ 990000 N g_1 g_2 g_3^2 g_4
+ 1080000 g_1 g_2 g_3^2 g_4
+ 421875 N^2 g_1 g_3^4
\right. \right. \nonumber \\
&& \left. \left. ~~~~~~~~~~
+ 2362500 N g_1 g_3^4
+ 3240000 g_1 g_3^4
+ 202500 N g_1 g_3^2 g_4^2
\right. \right. \nonumber \\
&& \left. \left. ~~~~~~~~~~
+ 202500 g_1 g_3^2 g_4^2
+ 13125 g_1 g_4^4 
\right]
\Gamma^2(\twofifth) \Gamma^{12}(\onefifth) 
\right]
\frac{1}{7776000\Gamma(\fourfifth)\Gamma(\twofifth)} \nonumber \\
&& +~ O(g_i^7) \\
\beta_2^{\Phi^7}(g_i) &=&
\left[ 
3 N^3 g_1^2 g_2
+ 42 N^2 g_1^2 g_2
+ 168 N g_1^2 g_2
+ 192 g_1^2 g_2
+ 225 N^2 g_2^3
+ 850 N g_2^3
+ 800 g_2^3
\right. \nonumber \\
&& \left. ~
+ 675 N g_2 g_3^2
+ 360 g_2 g_3^2
+ 45 g_2 g_4^2 
\right]
\frac{\Gamma^5(\onefifth)}{2160} \nonumber \\
&& -~ 
\left[ 
\left[ 
27 N^6 g_1^4 g_2
+ 864 N^5 g_1^4 g_2
+ 11772 N^4 g_1^4 g_2
+ 85104 N^3 g_1^4 g_2
+ 333504 N^2 g_1^4 g_2
\right. \right. \nonumber \\
&& \left. \left. ~~~~~~
+ 656640 N g_1^4 g_2
+ 497664 g_1^4 g_2
+ 4500 N^5 g_1^2 g_2^3
+ 94200 N^4 g_1^2 g_2^3
\right. \right. \nonumber \\
&& \left. \left. ~~~~~~
+ 742800 N^3 g_1^2 g_2^3
+ 2791200 N^2 g_1^2 g_2^3
+ 5020800 N g_1^2 g_2^3
+ 3456000 g_1^2 g_2^3
\right. \right. \nonumber \\
&& \left. \left. ~~~~~~
+ 2430 N^4 g_1^2 g_2 g_3^2
+ 77220 N^3 g_1^2 g_2 g_3^2
+ 498960 N^2 g_1^2 g_2 g_3^2
+ 1123200 N g_1^2 g_2 g_3^2
\right. \right. \nonumber \\
&& \left. \left. ~~~~~~
+ 829440 g_1^2 g_2 g_3^2
+ 81000 N^4 g_1 g_2^3 g_3
+ 792000 N^3 g_1 g_2^3 g_3
+ 2772000 N^2 g_1 g_2^3 g_3
\right. \right. \nonumber \\
&& \left. \left. ~~~~~~
+ 4176000 N g_1 g_2^3 g_3
+ 2304000 g_1 g_2^3 g_3
+ 32400 N^3 g_1 g_2^2 g_3 g_4
\right. \right. \nonumber \\
&& \left. \left. ~~~~~~
+ 226800 N^2 g_1 g_2^2 g_3 g_4
+ 453600 N g_1 g_2^2 g_3 g_4
+ 259200 g_1 g_2^2 g_3 g_4
\right. \right. \nonumber \\
&& \left. \left. ~~~~~~
+ 73125 N^4 g_2^5
+ 1322500 N^3 g_2^5
+ 6282500 N^2 g_2^5
+ 11540000 N g_2^5
\right. \right. \nonumber \\
&& \left. \left. ~~~~~~
+ 7360000 g_2^5
+ 1404000 N^3 g_2^3 g_3^2
+ 8640000 N^2 g_2^3 g_3^2
+ 16092000 N g_2^3 g_3^2
\right. \right. \nonumber \\
&& \left. \left. ~~~~~~
+ 8856000 g_2^3 g_3^2
+ 101250 N^2 g_2^3 g_4^2
+ 256500 N g_2^3 g_4^2
+ 108000 g_2^3 g_4^2
\right. \right. \nonumber \\
&& \left. \left. ~~~~~~
+ 405000 N^2 g_2^2 g_3^2 g_4
+ 1026000 N g_2^2 g_3^2 g_4
+ 432000 g_2^2 g_3^2 g_4
+ 3067875 N^2 g_2 g_3^4
\right. \right. \nonumber \\
&& \left. \left. ~~~~~~
+ 4673700 N g_2 g_3^4
+ 1620000 g_2 g_3^4
+ 801900 N g_2 g_3^2 g_4^2
+ 178200 g_2 g_3^2 g_4^2
\right. \right. \nonumber \\
&& \left. \left. ~~~~~~
+ 42525 g_2 g_4^4
\right]
\Gamma^2(\fourfifth) \Gamma^{13}(\onefifth)
\right. \nonumber \\
&& \left. ~~~~~~+
\left[ 
3300 N^5 g_1^4 g_2
+ 57600 N^4 g_1^4 g_2
+ 387600 N^3 g_1^4 g_2
+ 1252800 N^2 g_1^4 g_2
\right. \right. \nonumber \\
&& \left. \left. ~~~~~~~~~~
+ 1939200 N g_1^4 g_2
+ 1152000 g_1^4 g_2
+ 11250 N^5 g_1^2 g_2^3
+ 345000 N^4 g_1^2 g_2^3
\right. \right. \nonumber \\
&& \left. \left. ~~~~~~~~~~
+ 2745000 N^3 g_1^2 g_2^3
+ 9120000 N^2 g_1^2 g_2^3
+ 13680000 N g_1^2 g_2^3
+ 7680000 g_1^2 g_2^3
\right. \right. \nonumber \\
&& \left. \left. ~~~~~~~~~~
+ 67500 N^4 g_1^2 g_2 g_3^2
+ 828000 N^3 g_1^2 g_2 g_3^2
+ 3402000 N^2 g_1^2 g_2 g_3^2
\right. \right. \nonumber \\
&& \left. \left. ~~~~~~~~~~
+ 5328000 N g_1^2 g_2 g_3^2
+ 2592000 g_1^2 g_2 g_3^2
+ 6750 N^3 g_1^2 g_2 g_4^2
+ 49500 N^2 g_1^2 g_2 g_4^2
\right. \right. \nonumber \\
&& \left. \left. ~~~~~~~~~~
+ 108000 N g_1^2 g_2 g_4^2
+ 72000 g_1^2 g_2 g_4^2
+ 843750 N^4 g_2^5
+ 7875000 N^3 g_2^5
\right. \right. \nonumber \\
&& \left. \left. ~~~~~~~~~~
+ 24375000 N^2 g_2^5
+ 30000000 N g_2^5
+ 12000000 g_2^5
+ 7593750 N^3 g_2^3 g_3^2
\right. \right. \nonumber \\
&& \left. \left. ~~~~~~~~~~
+ 30712500 N^2 g_2^3 g_3^2
+ 34650000 N g_2^3 g_3^2
+ 7200000 g_2^3 g_3^2
+ 675000 N^2 g_2^3 g_4^2
\right. \right. \nonumber \\
&& \left. \left. ~~~~~~~~~~
+ 1800000 N g_2^3 g_4^2
+ 900000 g_2^3 g_4^2
+ 15187500 N^2 g_2 g_3^4
+ 13162500 N g_2 g_3^4
\right. \right. \nonumber \\
&& \left. \left. ~~~~~~~~~~
+ 810000 g_2 g_3^4
+ 2531250 N g_2 g_3^2 g_4^2
+ 675000 g_2 g_3^2 g_4^2
\right. \right. \nonumber \\
&& \left. \left. ~~~~~~~~~~
+ 101250 g_2 g_4^4
\right]
\Gamma(\fourfifth) \Gamma(\twofifth) \Gamma^{10}(\onefifth)
\right. \nonumber \\
&& \left. ~~~~~~+
\left[ 
27 N^6 g_1^4 g_2
+ 1188 N^5 g_1^4 g_2
+ 19980 N^4 g_1^4 g_2
+ 167040 N^3 g_1^4 g_2
\right. \right. \nonumber \\
&& \left. \left. ~~~~~~~~~~
+ 733248 N^2 g_1^4 g_2
+ 1585152 N g_1^4 g_2
+ 1290240 g_1^4 g_2
+ 6300 N^5 g_1^2 g_2^3
\right. \right. \nonumber \\
&& \left. \left. ~~~~~~~~~~
+ 165000 N^4 g_1^2 g_2^3
+ 1599600 N^3 g_1^2 g_2^3
+ 7173600 N^2 g_1^2 g_2^3
\right. \right. \nonumber \\
&& \left. \left. ~~~~~~~~~~
+ 14928000 N g_1^2 g_2^3
+ 11520000 g_1^2 g_2^3
+ 12150 N^4 g_1^2 g_2 g_3^2
\right. \right. \nonumber \\
&& \left. \left. ~~~~~~~~~~
+ 143100 N^3 g_1^2 g_2 g_3^2
+ 604800 N^2 g_1^2 g_2 g_3^2
+ 1080000 N g_1^2 g_2 g_3^2
\right. \right. \nonumber \\
&& \left. \left. ~~~~~~~~~~
+ 691200 g_1^2 g_2 g_3^2
+ 81000 N^4 g_1 g_2^3 g_3
+ 1440000 N^3 g_1 g_2^3 g_3
\right. \right. \nonumber \\
&& \left. \left. ~~~~~~~~~~
+ 7812000 N^2 g_1 g_2^3 g_3
+ 16272000 N g_1 g_2^3 g_3
+ 11520000 g_1 g_2^3 g_3
\right. \right. \nonumber \\
&& \left. \left. ~~~~~~~~~~
+ 54000 N^3 g_1 g_2^2 g_3 g_4
+ 378000 N^2 g_1 g_2^2 g_3 g_4
+ 756000 N g_1 g_2^2 g_3 g_4
\right. \right. \nonumber \\
&& \left. \left. ~~~~~~~~~~
+ 432000 g_1 g_2^2 g_3 g_4
+ 208125 N^4 g_2^5
+ 2912500 N^3 g_2^5
+ 14622500 N^2 g_2^5
\right. \right. \nonumber \\
&& \left. \left. ~~~~~~~~~~
+ 29660000 N g_2^5
+ 20800000 g_2^5
+ 2700000 N^3 g_2^3 g_3^2
+ 21600000 N^2 g_2^3 g_3^2
\right. \right. \nonumber \\
&& \left. \left. ~~~~~~~~~~
+ 45900000 N g_2^3 g_3^2
+ 27000000 g_2^3 g_3^2
+ 56250 N^2 g_2^3 g_4^2
+ 142500 N g_2^3 g_4^2
\right. \right. \nonumber \\
&& \left. \left. ~~~~~~~~~~
+ 60000 g_2^3 g_4^2
+ 2025000 N^2 g_2^2 g_3^2 g_4
+ 5130000 N g_2^2 g_3^2 g_4
+ 2160000 g_2^2 g_3^2 g_4
\right. \right. \nonumber \\
&& \left. \left. ~~~~~~~~~~
+ 3796875 N^2 g_2 g_3^4
+ 14175000 N g_2 g_3^4
+ 6480000 g_2 g_3^4
+ 1822500 N g_2 g_3^2 g_4^2
\right. \right. \nonumber \\
&& \left. \left. ~~~~~~~~~~
+ 405000 g_2 g_3^2 g_4^2
+ 118125 g_2 g_4^4 
\right]
\Gamma^2(\twofifth) \Gamma^{12}(\onefifth)
\right]
\frac{1}{23328000\Gamma(\fourfifth)\Gamma(\twofifth)} \nonumber \\
&& +~ O(g_i^7) \\
\beta_3^{\Phi^7}(g_i) &=&
\left[ 
5 N^3 g_1^2 g_3
+ 42 N^2 g_1^2 g_3
+ 112 N g_1^2 g_3
+ 96 g_1^2 g_3
+ 375 N^2 g_2^2 g_3
+ 950 N g_2^2 g_3
\right. \nonumber \\
&& \left. ~
+ 400 g_2^2 g_3
+ 1125 N g_3^3
+ 180 g_3^3
+ 75 g_3 g_4^2
\right]
\frac{\Gamma^5(\onefifth)}{2160} \nonumber \\
&& -~
\left[ 
\left[ 
45 N^6 g_1^4 g_3
+ 1188 N^5 g_1^4 g_3
+ 13572 N^4 g_1^4 g_3
+ 80352 N^3 g_1^4 g_3
+ 253440 N^2 g_1^4 g_3
\right. \right. \nonumber \\
&& \left. \left. ~~~~~~
+ 400896 N g_1^4 g_3
+ 248832 g_1^4 g_3
+ 7500 N^5 g_1^2 g_2^2 g_3
+ 139500 N^4 g_1^2 g_2^2 g_3
\right. \right. \nonumber \\
&& \left. \left. ~~~~~~
+ 930000 N^3 g_1^2 g_2^2 g_3
+ 2790000 N^2 g_1^2 g_2^2 g_3
+ 3720000 N g_1^2 g_2^2 g_3
\right. \right. \nonumber \\
&& \left. \left. ~~~~~~
+ 1728000 g_1^2 g_2^2 g_3
+ 4050 N^4 g_1^2 g_3^3
+ 123660 N^3 g_1^2 g_3^3
+ 680400 N^2 g_1^2 g_3^3
\right. \right. \nonumber \\
&& \left. \left. ~~~~~~
+ 1105920 N g_1^2 g_3^3
+ 414720 g_1^2 g_3^3
+ 135000 N^4 g_1 g_2^2 g_3^2
\right. \right. \nonumber \\
&& \left. \left. ~~~~~~
+ 1152000 N^3 g_1 g_2^2 g_3^2
+ 3276000 N^2 g_1 g_2^2 g_3^2
+ 3600000 N g_1 g_2^2 g_3^2
\right. \right. \nonumber \\
&& \left. \left. ~~~~~~
+ 1152000 g_1 g_2^2 g_3^2
+ 54000 N^3 g_1 g_2 g_3^2 g_4
+ 340200 N^2 g_1 g_2 g_3^2 g_4
\right. \right. \nonumber \\
&& \left. \left. ~~~~~~
+ 529200 N g_1 g_2 g_3^2 g_4
+ 129600 g_1 g_2 g_3^2 g_4
+ 121875 N^4 g_2^4 g_3
+ 2052500 N^3 g_2^4 g_3
\right. \right. \nonumber \\
&& \left. \left. ~~~~~~
+ 7997500 N^2 g_2^4 g_3
+ 10600000 N g_2^4 g_3
+ 3680000 g_2^4 g_3
+ 2340000 N^3 g_2^2 g_3^3
\right. \right. \nonumber \\
&& \left. \left. ~~~~~~
+ 12762000 N^2 g_2^2 g_3^3
+ 18378000 N g_2^2 g_3^3
+ 4428000 g_2^2 g_3^3
+ 168750 N^2 g_2^2 g_3 g_4^2
\right. \right. \nonumber \\
&& \left. \left. ~~~~~~
+ 364500 N g_2^2 g_3 g_4^2
+ 54000 g_2^2 g_3 g_4^2
+ 675000 N^2 g_2 g_3^3 g_4
+ 1458000 N g_2 g_3^3 g_4
\right. \right. \nonumber \\
&& \left. \left. ~~~~~~
+ 216000 g_2 g_3^3 g_4
+ 5113125 N^2 g_3^5
+ 5880600 N g_3^5
+ 810000 g_3^5
\right. \right. \nonumber \\
&& \left. \left. ~~~~~~
+ 1336500 N g_3^3 g_4^2
+ 89100 g_3^3 g_4^2
+ 70875 g_3 g_4^4
\right]
\Gamma^2(\fourfifth) \Gamma^{13}(\onefifth)
\right. \nonumber \\
&& \left. ~~~~~~+
\left[ 
4800 N^5 g_1^4 g_3
+ 66600 N^4 g_1^4 g_3
+ 357600 N^3 g_1^4 g_3
+ 928800 N^2 g_1^4 g_3
\right. \right. \nonumber \\
&& \left. \left. ~~~~~~~~~~
+ 1171200 N g_1^4 g_3
+ 576000 g_1^4 g_3
+ 18750 N^5 g_1^2 g_2^2 g_3
+ 487500 N^4 g_1^2 g_2^2 g_3
\right. \right. \nonumber \\
&& \left. \left. ~~~~~~~~~~
+ 3315000 N^3 g_1^2 g_2^2 g_3
+ 8970000 N^2 g_1^2 g_2^2 g_3
+ 10200000 N g_1^2 g_2^2 g_3
\right. \right. \nonumber \\
&& \left. \left. ~~~~~~~~~~
+ 3840000 g_1^2 g_2^2 g_3
+ 112500 N^4 g_1^2 g_3^3
+ 1170000 N^3 g_1^2 g_3^3
\right. \right. \nonumber \\
&& \left. \left. ~~~~~~~~~~
+ 4032000 N^2 g_1^2 g_3^3
+ 4932000 N g_1^2 g_3^3
+ 1296000 g_1^2 g_3^3
\right. \right. \nonumber \\
&& \left. \left. ~~~~~~~~~~
+ 11250 N^3 g_1^2 g_3 g_4^2
+ 72000 N^2 g_1^2 g_3 g_4^2
+ 117000 N g_1^2 g_3 g_4^2
\right. \right. \nonumber \\
&& \left. \left. ~~~~~~~~~~
+ 36000 g_1^2 g_3 g_4^2
+ 1406250 N^4 g_2^4 g_3
+ 10500000 N^3 g_2^4 g_3
\right. \right. \nonumber \\
&& \left. \left. ~~~~~~~~~~
+ 26625000 N^2 g_2^4 g_3
+ 25500000 N g_2^4 g_3
+ 6000000 g_2^4 g_3
\right. \right. \nonumber \\
&& \left. \left. ~~~~~~~~~~
+ 12656250 N^3 g_2^2 g_3^3
+ 39375000 N^2 g_2^2 g_3^3
+ 29925000 N g_2^2 g_3^3
\right. \right. \nonumber \\
&& \left. \left. ~~~~~~~~~~
+ 3600000 g_2^2 g_3^3
+ 1125000 N^2 g_2^2 g_3 g_4^2
+ 2475000 N g_2^2 g_3 g_4^2
\right. \right. \nonumber \\
&& \left. \left. ~~~~~~~~~~
+ 450000 g_2^2 g_3 g_4^2
+ 25312500 N^2 g_3^5
+ 10125000 N g_3^5
+ 405000 g_3^5
\right. \right. \nonumber \\
&& \left. \left. ~~~~~~~~~~
+ 4218750 N g_3^3 g_4^2
+ 337500 g_3^3 g_4^2
+ 168750 g_3 g_4^4
\right]
\Gamma(\fourfifth) \Gamma(\twofifth) \Gamma^{10} (\onefifth)
\right. \nonumber \\
&& \left. ~~~~~~+
\left[ 
45 N^6 g_1^4 g_3
+ 1728 N^5 g_1^4 g_3
+ 24228 N^4 g_1^4 g_3
+ 164496 N^3 g_1^4 g_3
\right. \right. \nonumber \\
&& \left. \left. ~~~~~~~~~~
+ 574272 N^2 g_1^4 g_3
+ 980736 N g_1^4 g_3
+ 645120 g_1^4 g_3
+ 10500 N^5 g_1^2 g_2^2 g_3
\right. \right. \nonumber \\
&& \left. \left. ~~~~~~~~~~
+ 250500 N^4 g_1^2 g_2^2 g_3
+ 2106000 N^3 g_1^2 g_2^2 g_3
+ 7602000 N^2 g_1^2 g_2^2 g_3
\right. \right. \nonumber \\
&& \left. \left. ~~~~~~~~~~
+ 11496000 N g_1^2 g_2^2 g_3
+ 5760000 g_1^2 g_2^2 g_3
+ 20250 N^4 g_1^2 g_3^3
\right. \right. \nonumber \\
&& \left. \left. ~~~~~~~~~~
+ 213300 N^3 g_1^2 g_3^3
+ 756000 N^2 g_1^2 g_3^3
+ 993600 N g_1^2 g_3^3
+ 345600 g_1^2 g_3^3
\right. \right. \nonumber \\
&& \left. \left. ~~~~~~~~~~
+ 135000 N^4 g_1 g_2^2 g_3^2
+ 2232000 N^3 g_1 g_2^2 g_3^2
+ 10332000 N^2 g_1 g_2^2 g_3^2
\right. \right. \nonumber \\
&& \left. \left. ~~~~~~~~~~
+ 15696000 N g_1 g_2^2 g_3^2
+ 5760000 g_1 g_2^2 g_3^2
+ 90000 N^3 g_1 g_2 g_3^2 g_4
\right. \right. \nonumber \\
&& \left. \left. ~~~~~~~~~~
+ 567000 N^2 g_1 g_2 g_3^2 g_4
+ 882000 N g_1 g_2 g_3^2 g_4
+ 216000 g_1 g_2 g_3^2 g_4
\right. \right. \nonumber \\
&& \left. \left. ~~~~~~~~~~
+ 346875 N^4 g_2^4 g_3
+ 4422500 N^3 g_2^4 g_3
+ 19097500 N^2 g_2^4 g_3
\right. \right. \nonumber \\
&& \left. \left. ~~~~~~~~~~
+ 28480000 N g_2^4 g_3
+ 10400000 g_2^4 g_3
+ 4500000 N^3 g_2^2 g_3^3
+ 32850000 N^2 g_2^2 g_3^3
\right. \right. \nonumber \\
&& \left. \left. ~~~~~~~~~~
+ 54450000 N g_2^2 g_3^3
+ 13500000 g_2^2 g_3^3
+ 93750 N^2 g_2^2 g_3 g_4^2
+ 202500 N g_2^2 g_3 g_4^2
\right. \right. \nonumber \\
&& \left. \left. ~~~~~~~~~~
+ 30000 g_2^2 g_3 g_4^2
+ 3375000 N^2 g_2 g_3^3 g_4
+ 7290000 N g_2 g_3^3 g_4
+ 1080000 g_2 g_3^3 g_4
\right. \right. \nonumber \\
&& \left. \left. ~~~~~~~~~~
+ 6328125 N^2 g_3^5
+ 21262500 N g_3^5
+ 3240000 g_3^5
+ 3037500 N g_3^3 g_4^2
\right. \right. \nonumber \\
&& \left. \left. ~~~~~~~~~~
+ 202500 g_3^3 g_4^2
+ 196875 g_3 g_4^4 
\right]
\Gamma^2(\twofifth) \Gamma^{12}(\onefifth)
\right]
\frac{1}{23328000\Gamma(\fourfifth)\Gamma(\twofifth)} \nonumber \\ 
&& +~ O(g_i^7)
\end{eqnarray}
and
\begin{eqnarray}
\beta_4^{\Phi^7}(g_i) &=&
\left[ 
7 N^3 g_1^2 g_4
+ 42 N^2 g_1^2 g_4
+ 56 N g_1^2 g_4
+ 525 N^2 g_2^2 g_4
+ 1050 N g_2^2 g_4
\right. \nonumber \\
&& \left. ~
+ 1575 N g_3^2 g_4
+ 105 g_4^3 
\right]
\frac{\Gamma^5(\onefifth)}{2160} \nonumber \\
&& -~
\left[ 
\left[ 
21 N^6 g_1^4 g_4
+ 504 N^5 g_1^4 g_4
+ 5124 N^4 g_1^4 g_4
+ 25200 N^3 g_1^4 g_4
+ 57792 N^2 g_1^4 g_4
\right. \right. \nonumber \\
&& \left. \left. ~~~~~~
+ 48384 N g_1^4 g_4
+ 3500 N^5 g_1^2 g_2^2 g_4
+ 61600 N^4 g_1^2 g_2^2 g_4
+ 372400 N^3 g_1^2 g_2^2 g_4
\right. \right. \nonumber \\
&& \left. \left. ~~~~~~
+ 929600 N^2 g_1^2 g_2^2 g_4
+ 806400 N g_1^2 g_2^2 g_4
+ 1890 N^4 g_1^2 g_3^2 g_4
+ 56700 N^3 g_1^2 g_3^2 g_4
\right. \right. \nonumber \\
&& \left. \left. ~~~~~~
+ 287280 N^2 g_1^2 g_3^2 g_4
+ 362880 N g_1^2 g_3^2 g_4
+ 63000 N^4 g_1 g_2^2 g_3 g_4
\right. \right. \nonumber \\
&& \left. \left. ~~~~~~
+ 504000 N^3 g_1 g_2^2 g_3 g_4
+ 1260000 N^2 g_1 g_2^2 g_3 g_4
+ 1008000 N g_1 g_2^2 g_3 g_4
\right. \right. \nonumber \\
&& \left. \left. ~~~~~~
+ 25200 N^3 g_1 g_2 g_3 g_4^2
+ 151200 N^2 g_1 g_2 g_3 g_4^2
+ 201600 N g_1 g_2 g_3 g_4^2
\right. \right. \nonumber \\
&& \left. \left. ~~~~~~
+ 56875 N^4 g_2^4 g_4
+ 927500 N^3 g_2^4 g_4
+ 3237500 N^2 g_2^4 g_4
+ 3220000 N g_2^4 g_4
\right. \right. \nonumber \\
&& \left. \left. ~~~~~~
+ 1092000 N^3 g_2^2 g_3^2 g_4
+ 5628000 N^2 g_2^2 g_3^2 g_4
+ 6888000 N g_2^2 g_3^2 g_4
+ 78750 N^2 g_2^2 g_4^3
\right. \right. \nonumber \\
&& \left. \left. ~~~~~~
+ 157500 N g_2^2 g_4^3
+ 315000 N^2 g_2 g_3^2 g_4^2
+ 630000 N g_2 g_3^2 g_4^2
+ 2386125 N^2 g_3^4 g_4
\right. \right. \nonumber \\
&& \left. \left. ~~~~~~
+ 2362500 N g_3^4 g_4
+ 623700 N g_3^2 g_4^3
+ 33075 g_4^5
\right]
\Gamma^2(\fourfifth) \Gamma^{13}(\onefifth)
\right. \nonumber \\
&& \left. ~~~~~~+
\left[ 
2100 N^5 g_1^4 g_4
+ 25200 N^4 g_1^4 g_4
+ 109200 N^3 g_1^4 g_4
+ 201600 N^2 g_1^4 g_4
\right. \right. \nonumber \\
&& \left. \left. ~~~~~~~~~~
+ 134400 N g_1^4 g_4
+ 8750 N^5 g_1^2 g_2^2 g_4
+ 210000 N^4 g_1^2 g_2^2 g_4
\right. \right. \nonumber \\
&& \left. \left. ~~~~~~~~~~
+ 1295000 N^3 g_1^2 g_2^2 g_4
+ 2940000 N^2 g_1^2 g_2^2 g_4
+ 2240000 N g_1^2 g_2^2 g_4
\right. \right. \nonumber \\
&& \left. \left. ~~~~~~~~~~
+ 52500 N^4 g_1^2 g_3^2 g_4
+ 504000 N^3 g_1^2 g_3^2 g_4
+ 1554000 N^2 g_1^2 g_3^2 g_4
\right. \right. \nonumber \\
&& \left. \left. ~~~~~~~~~~
+ 1512000 N g_1^2 g_3^2 g_4
+ 5250 N^3 g_1^2 g_4^3
+ 31500 N^2 g_1^2 g_4^3
+ 42000 N g_1^2 g_4^3
\right. \right. \nonumber \\
&& \left. \left. ~~~~~~~~~~
+ 656250 N^4 g_2^4 g_4
+ 4375000 N^3 g_2^4 g_4
+ 9625000 N^2 g_2^4 g_4
+ 7000000 N g_2^4 g_4
\right. \right. \nonumber \\
&& \left. \left. ~~~~~~~~~~
+ 5906250 N^3 g_2^2 g_3^2 g_4
+ 16012500 N^2 g_2^2 g_3^2 g_4
+ 8400000 N g_2^2 g_3^2 g_4
\right. \right. \nonumber \\
&& \left. \left. ~~~~~~~~~~
+ 525000 N^2 g_2^2 g_4^3
+ 1050000 N g_2^2 g_4^3
+ 11812500 N^2 g_3^4 g_4
\right. \right. \nonumber \\
&& \left. \left. ~~~~~~~~~~
+ 2362500 N g_3^4 g_4
+ 1968750 N g_3^2 g_4^3
+ 78750 g_4^5
\right]
\Gamma(\fourfifth) \Gamma(\twofifth) \Gamma^{10}(\onefifth)
\right. \nonumber \\
&& \left. ~~~~~~+
\left[ 
21 N^6 g_1^4 g_4
+ 756 N^5 g_1^4 g_4
+ 9492 N^4 g_1^4 g_4
+ 53984 N^3 g_1^4 g_4
\right. \right. \nonumber \\
&& \left. \left. ~~~~~~~~~~
+ 138432 N^2 g_1^4 g_4
+ 125440 N g_1^4 g_4
+ 4900 N^5 g_1^2 g_2^2 g_4
+ 112000 N^4 g_1^2 g_2^2 g_4
\right. \right. \nonumber \\
&& \left. \left. ~~~~~~~~~~
+ 870800 N^3 g_1^2 g_2^2 g_4
+ 2676800 N^2 g_1^2 g_2^2 g_4
+ 2688000 N g_1^2 g_2^2 g_4
\right. \right. \nonumber \\
&& \left. \left. ~~~~~~~~~~
+ 9450 N^4 g_1^2 g_3^2 g_4
+ 94500 N^3 g_1^2 g_3^2 g_4
+ 302400 N^2 g_1^2 g_3^2 g_4
\right. \right. \nonumber \\
&& \left. \left. ~~~~~~~~~~
+ 302400 N g_1^2 g_3^2 g_4
+ 63000 N^4 g_1 g_2^2 g_3 g_4
+ 1008000 N^3 g_1 g_2^2 g_3 g_4
\right. \right. \nonumber \\
&& \left. \left. ~~~~~~~~~~
+ 4284000 N^2 g_1 g_2^2 g_3 g_4
+ 5040000 N g_1 g_2^2 g_3 g_4
+ 42000 N^3 g_1 g_2 g_3 g_4^2
\right. \right. \nonumber \\
&& \left. \left. ~~~~~~~~~~
+ 252000 N^2 g_1 g_2 g_3 g_4^2
+ 336000 N g_1 g_2 g_3 g_4^2
+ 161875 N^4 g_2^4 g_4
\right. \right. \nonumber \\
&& \left. \left. ~~~~~~~~~~
+ 1977500 N^3 g_2^4 g_4
+ 7857500 N^2 g_2^4 g_4
+ 9100000 N g_2^4 g_4
\right. \right. \nonumber \\
&& \left. \left. ~~~~~~~~~~
+ 2100000 N^3 g_2^2 g_3^2 g_4
+ 14700000 N^2 g_2^2 g_3^2 g_4
+ 21000000 N g_2^2 g_3^2 g_4
\right. \right. \nonumber \\
&& \left. \left. ~~~~~~~~~~
+ 43750 N^2 g_2^2 g_4^3
+ 87500 N g_2^2 g_4^3
+ 1575000 N^2 g_2 g_3^2 g_4^2
\right. \right. \nonumber \\
&& \left. \left. ~~~~~~~~~~
+ 3150000 N g_2 g_3^2 g_4^2
+ 2953125 N^2 g_3^4 g_4
+ 9450000 N g_3^4 g_4
\right. \right. \nonumber \\
&& \left. \left. ~~~~~~~~~~
+ 1417500 N g_3^2 g_4^3
+ 91875 g_4^5
\right]
\Gamma^2(\twofifth) \Gamma^{12}(\onefifth)
\right]
\frac{1}{7776000\Gamma(\fourfifth)\Gamma(\twofifth)} \nonumber \\
&& +~ O(g_i^7) ~.
\end{eqnarray}

\sect{Renormalization group functions for the $\Phi^9$ theory with $O(-8)$ 
symmetry.}

For completeness we present renormalization group functions for the $\Phi^9$
structure. In particular we focus on the enhanced case of the $O(N)$ theory
when $N$~$=$~$-$~$8$. The field anomalous dimensions are 
\begin{eqnarray}
\left. \frac{}{} \gamma_\Phi^{\Phi^9}(g_i) \right|_{N=-8} &=&
\left[
- 16 g_1^2
+ 392 g_2^2
- 490 g_3^2
+ 35 g_4^2 \right]
\frac{\Gamma^7(\oneseventh)}{25200} \nonumber \\
&& + \left[
\left[
122880 g_1^4
- 7902720 g_1^2 g_3^2
+ 873600 g_1^2 g_4^2
+ 31610880 g_1 g_2^2 g_3
\right. \right. \nonumber \\
&& \left. \left. ~~~~~
- 19756800 g_1 g_2 g_3 g_4
+ 376320 g_1 g_2 g_4 g_5
- 110638080 g_2^4
\right. \right. \nonumber \\
&& \left. \left. ~~~~~
- 69148800 g_2^2 g_3^2
+ 57953280 g_2^2 g_4^2
- 493920 g_2^2 g_5^2
+ 276595200 g_2 g_3^2 g_4
\right. \right. \nonumber \\
&& \left. \left. ~~~~~
- 4939200 g_2 g_3 g_4 g_5
+ 610814400 g_3^4
- 513676800 g_3^2 g_4^2
+ 2058000 g_3^2 g_5^2
\right. \right. \nonumber \\
&& \left. \left. ~~~~~
+ 2469600 g_3 g_4^2 g_5
+ 49098000 g_4^4
- 323400 g_4^2 g_5^2
\right]
\Gamma^2(\sixseventh) \Gamma(\fiveseventh) \Gamma(\threeseventh) 
\Gamma^3(\oneseventh) 
\right. \nonumber \\
&& \left. ~~~+
\left[ 
36879360 g_1^2 g_2^2
- 1806336 g_1^4
+ 46099200 g_1^2 g_3^2
- 17781120 g_1^2 g_4^2
\right. \right. \nonumber \\
&& \left. \left. ~~~~~~~
+ 82320 g_1^2 g_5^2
+ 1264962048 g_2^4
- 9487215360 g_2^2 g_3^2
+ 1710280320 g_2^2 g_4^2
\right. \right. \nonumber \\
&& \left. \left. ~~~~~~~
- 6050520 g_2^2 g_5^2
+ 13270807200 g_3^4
- 4154690400 g_3^2 g_4^2
+ 12605250 g_3^2 g_5^2
\right. \right. \nonumber \\
&& \left. \left. ~~~~~~~
+ 280917000 g_4^4
- 1260525 g_4^2 g_5^2
\right]
\Gamma(\sixseventh) \Gamma(\fiveseventh) \Gamma(\threeseventh) 
\Gamma(\twoseventh) 
\right. \nonumber \\
&& \left. ~~~+
\left[ 
86016 g_1^4
+ 2107392 g_1^2 g_2^2
- 7902720 g_1^2 g_3^2
+ 63221760 g_1 g_2^2 g_3
\right. \right. \nonumber \\
&& \left. \left. ~~~~~~~
+ 13171200 g_1 g_2 g_3 g_4
+ 3292800 g_1 g_3^2 g_5
- 66382848 g_2^4
+ 92198400 g_2^2 g_3^2
\right. \right. \nonumber \\
&& \left. \left. ~~~~~~~
+ 165957120 g_2^2 g_4^2
- 184396800 g_2 g_3^2 g_4
- 69148800 g_2 g_3 g_4 g_5
\right. \right. \nonumber \\
&& \left. \left. ~~~~~~~
+ 875884800 g_3^4
- 1261965600 g_3^2 g_4^2
+ 4321800 g_3^2 g_5^2
+ 86436000 g_3 g_4^2 g_5
\right. \right. \nonumber \\
&& \left. \left. ~~~~~~~
+ 123891600 g_4^4
- 2881200 g_4^2 g_5^2
\right] 
\Gamma(\sixseventh) \Gamma^2(\threeseventh) \Gamma(\twoseventh) 
\Gamma^2(\oneseventh)
\right. \nonumber \\
&& \left. ~~~+
\left[ 
14676480 g_1^2 g_3^2
- 172032 g_1^4
- 2809856 g_1^2 g_2^2
- 470400 g_1^2 g_4^2
\right. \right. \nonumber \\
&& \left. \left. ~~~~~~~
- 94832640 g_1 g_2^2 g_3
+ 6585600 g_1 g_2 g_3 g_4
+ 878080 g_1 g_2 g_4 g_5
\right. \right. \nonumber \\
&& \left. \left. ~~~~~~~
+ 149976064 g_2^4
+ 176713600 g_2^2 g_3^2
- 49172480 g_2^2 g_4^2
- 384160 g_2^2 g_5^2
\right. \right. \nonumber \\
&& \left. \left. ~~~~~~~
+ 1014182400 g_2 g_3^2 g_4
- 34574400 g_2 g_3 g_4 g_5
- 311169600 g_3^4
\right. \right. \nonumber \\
&& \left. \left. ~~~~~~~
- 1071806400 g_3^2 g_4^2
+ 4802000 g_3^2 g_5^2
+ 28812000 g_3 g_4^2 g_5
+ 123891600 g_4^4
\right. \right. \nonumber \\
&& \left. \left. ~~~~~~~
- 1440600 g_4^2 g_5^2 
\right]
\Gamma^2(\fiveseventh) \Gamma^2(\twoseventh) \Gamma^2(\oneseventh) 
\right] 
\frac{\Gamma^{14}(\oneseventh)}{106686720000 \Gamma(\sixseventh) 
\Gamma(\fiveseventh) \Gamma(\threeseventh) \Gamma(\twoseventh)}
\nonumber \\
&& +~ O(g_i^6) \nonumber \\
\left. \frac{}{} \gamma_\sigma^{\Phi^9}(g_i) \right|_{N=-8} &=&
\left[
128 g_1^2
- 12544 g_2^2
+ 39200 g_3^2
- 7840 g_4^2
+ 35 g_5^2 \right]
\frac{\Gamma^7(\oneseventh)}{201600} \nonumber \\
&& +~
\left[ 
\left[ 
63221760 g_1^2 g_3^2
- 245760 g_1^4
- 11182080 g_1^2 g_4^2
- 252887040 g_1 g_2^2 g_3
\right. \right. \nonumber \\
&& \left. \left. ~~~~~
+ 252887040 g_1 g_2 g_3 g_4
- 7526400 g_1 g_2 g_4 g_5
+ 885104640 g_2^4
\right. \right. \nonumber \\
&& \left. \left. ~~~~~
+ 885104640 g_2^2 g_3^2
- 1159065600 g_2^2 g_4^2
+ 15805440 g_2^2 g_5^2
\right. \right. \nonumber \\
&& \left. \left. ~~~~~
- 5531904000 g_2 g_3^2 g_4
+ 158054400 g_2 g_3 g_4 g_5
- 12216288000 g_3^4
\right. \right. \nonumber \\
&& \left. \left. ~~~~~
+ 16437657600 g_3^2 g_4^2
- 115248000 g_3^2 g_5^2
- 138297600 g_3 g_4^2 g_5
\right. \right. \nonumber \\
&& \left. \left. ~~~~~
- 2749488000 g_4^4
+ 41395200 g_4^2 g_5^2
- 132300 g_5^4
\right]
\Gamma^2(\sixseventh) \Gamma(\fiveseventh) \Gamma(\threeseventh) 
\Gamma^3(\oneseventh) 
\right. \nonumber \\
&& \left. ~~~+
\left[ 
4816896 g_1^4
- 354041856 g_1^2 g_2^2
+ 147517440 g_1^2 g_3^2
+ 136980480 g_1^2 g_4^2
\right. \right. \nonumber \\
&& \left. \left. ~~~~~~~
- 1317120 g_1^2 g_5^2
- 2891341824 g_2^4
+ 79511900160 g_2^2 g_3^2
\right. \right. \nonumber \\
&& \left. \left. ~~~~~~~
- 26331863040 g_2^2 g_4^2
+ 161347200 g_2^2 g_5^2
- 203297472000 g_3^4
\right. \right. \nonumber \\
&& \left. \left. ~~~~~~~
+ 109070707200 g_3^2 g_4^2
- 605052000 g_3^2 g_5^2
- 13391817600 g_4^4
\right. \right. \nonumber \\
&& \left. \left. ~~~~~~~
+ 141178800 g_4^2 g_5^2
- 360150 g_5^4
\right]
\Gamma(\sixseventh) \Gamma(\fiveseventh) \Gamma(\threeseventh) 
\Gamma(\twoseventh) 
\right. \nonumber \\
&& \left. ~~~+
\left[ 
63221760 g_1^2 g_3^2
- 172032 g_1^4
- 9633792 g_1^2 g_2^2
- 505774080 g_1 g_2^2 g_3
\right. \right. \nonumber \\
&& \left. \left. ~~~~~~~
- 168591360 g_1 g_2 g_3 g_4
- 65856000 g_1 g_3^2 g_5
+ 531062784 g_2^4
\right. \right. \nonumber \\
&& \left. \left. ~~~~~~~
- 1180139520 g_2^2 g_3^2
- 3319142400 g_2^2 g_4^2
+ 3687936000 g_2 g_3^2 g_4
\right. \right. \nonumber \\
&& \left. \left. ~~~~~~~
+ 2212761600 g_2 g_3 g_4 g_5
- 17517696000 g_3^4
+ 40382899200 g_3^2 g_4^2
\right. \right. \nonumber \\
&& \left. \left. ~~~~~~~
- 242020800 g_3^2 g_5^2
- 4840416000 g_3 g_4^2 g_5
- 6937929600 g_4^4
\right. \right. \nonumber \\
&& \left. \left. ~~~~~~~
+ 368793600 g_4^2 g_5^2
- 1620675 g_5^4
\right]
\Gamma(\sixseventh) \Gamma^2(\threeseventh) \Gamma(\twoseventh) 
\Gamma^2(\oneseventh) 
\right. \nonumber \\
&& \left. ~~~+
\left[ 
344064 g_1^4
+ 12845056 g_1^2 g_2^2
- 117411840 g_1^2 g_3^2
+ 6021120 g_1^2 g_4^2
\right. \right. \nonumber \\
&& \left. \left. ~~~~~~~
+ 758661120 g_1 g_2^2 g_3
- 84295680 g_1 g_2 g_3 g_4
- 17561600 g_1 g_2 g_4 g_5
\right. \right. \nonumber \\
&& \left. \left. ~~~~~~~
- 1199808512 g_2^4
- 2261934080 g_2^2 g_3^2
+ 983449600 g_2^2 g_4^2
\right. \right. \nonumber \\
&& \left. \left. ~~~~~~~
+ 12293120 g_2^2 g_5^2
- 20283648000 g_2 g_3^2 g_4
+ 1106380800 g_2 g_3 g_4 g_5
\right. \right. \nonumber \\
&& \left. \left. ~~~~~~~
+ 6223392000 g_3^4
+ 34297804800 g_3^2 g_4^2
- 268912000 g_3^2 g_5^2
\right. \right. \nonumber \\
&& \left. \left. ~~~~~~~
- 1613472000 g_3 g_4^2 g_5
- 6937929600 g_4^4
+ 184396800 g_4^2 g_5^2
\right. \right. \nonumber \\
&& \left. \left. ~~~~~~~
- 720300 g_5^4)
\right]
\Gamma^2(\fiveseventh) \Gamma^2(\twoseventh) \Gamma^2(\oneseventh) 
\right]
\frac{\Gamma^{14}(\oneseventh)}{213373440000 \Gamma(\sixseventh) 
\Gamma(\fiveseventh) \Gamma(\threeseventh) \Gamma(\twoseventh)} \nonumber \\
&& +~ O(g_i^6) ~.
\end{eqnarray}
The corresponding $\beta$-functions are
\begin{eqnarray}
\left. \frac{}{} \beta_1^{\Phi^9}(g_i) \right|_{N=-8} &=&
\left[ 
1792 g_2^2
- 128 g_1^2
+ 1120 g_3^2
- 800 g_4^2
+ 5 g_5^2 
\right]
\frac{\Gamma^7(\oneseventh) g_1}{28800} \nonumber \\
&&
+~ \left[
\left[
245760 g_1^4
- 9031680 g_1^2 g_3^2
+ 399360 g_1^2 g_4^2
+ 36126720 g_1 g_2^2 g_3
\right. \right. \nonumber \\
&& \left. \left. ~~~~~
- 9031680 g_1 g_2 g_3 g_4
- 215040 g_1 g_2 g_4 g_5
- 126443520 g_2^4
- 31610880 g_2^2 g_3^2
\right. \right. \nonumber \\
&& \left. \left. ~~~~~
- 33116160 g_2^2 g_4^2
+ 1128960 g_2^2 g_5^2
- 158054400 g_2 g_3^2 g_4
\right. \right. \nonumber \\
&& \left. \left. ~~~~~
+ 11289600 g_2 g_3 g_4 g_5
- 349036800 g_3^4
+ 1174118400 g_3^2 g_4^2
\right. \right. \nonumber \\
&& \left. \left. ~~~~~
- 11760000 g_3^2 g_5^2
- 14112000 g_3 g_4^2 g_5
- 280560000 g_4^4
+ 5174400 g_4^2 g_5^2
\right. \right. \nonumber \\
&& \left. \left. ~~~~~
- 18900 g_5^4
\right]
\Gamma^2(\sixseventh) \Gamma(\fiveseventh) \Gamma(\threeseventh) 
\Gamma^3(\oneseventh)
\right. \nonumber \\
&& \left. ~~~+
\left[ 
33718272 g_1^2 g_2^2
- 3440640 g_1^4
+ 126443520 g_1^2 g_3^2
- 21073920 g_1^2 g_4^2
\right. \right. \nonumber \\
&& \left. \left. ~~~~~~~
+ 2478292992 g_2^4
- 10326220800 g_2^2 g_3^2
+ 147517440 g_2^2 g_4^2
\right. \right. \nonumber \\
&& \left. \left. ~~~~~~~
+ 9219840 g_2^2 g_5^2
+ 1290777600 g_3^4
+ 6085094400 g_3^2 g_4^2
\right. \right. \nonumber \\
&& \left. \left. ~~~~~~~
- 57624000 g_3^2 g_5^2
- 1271020800 g_4^4
+ 17287200 g_4^2 g_5^2
\right. \right. \nonumber \\
&& \left. \left. ~~~~~~~
- 51450 g_5^4
\right]
\Gamma(\sixseventh) \Gamma(\fiveseventh) \Gamma(\threeseventh) 
\Gamma(\twoseventh)
\right. \nonumber \\
&& \left. ~~~+
\left[ 
172032 g_1^4
+ 3440640 g_1^2 g_2^2
- 9031680 g_1^2 g_3^2
+ 72253440 g_1 g_2^2 g_3
\right. \right. \nonumber \\
&& \left. \left. ~~~~~~~
+ 6021120 g_1 g_2 g_3 g_4
- 1881600 g_1 g_3^2 g_5
- 75866112 g_2^4
\right. \right. \nonumber \\
&& \left. \left. ~~~~~~~
+ 42147840 g_2^2 g_3^2
- 94832640 g_2^2 g_4^2
+ 105369600 g_2 g_3^2 g_4
\right. \right. \nonumber \\
&& \left. \left. ~~~~~~~
+ 158054400 g_2 g_3 g_4 g_5
- 500505600 g_3^4
+ 2884492800 g_3^2 g_4^2
\right. \right. \nonumber \\
&& \left. \left. ~~~~~~~
- 24696000 g_3^2 g_5^2
- 493920000 g_3 g_4^2 g_5
- 707952000 g_4^4
\right. \right. \nonumber \\
&& \left. \left. ~~~~~~~
+ 46099200 g_4^2 g_5^2
- 231525 g_5^4
\right]
\Gamma(\sixseventh) \Gamma^2(\threeseventh) \Gamma(\twoseventh) 
\Gamma^2(\oneseventh) 
\right. \nonumber \\
&& \left. ~~~+
\left[ 
16773120 g_1^2 g_3^2
- 344064 g_1^4
- 4587520 g_1^2 g_2^2
- 215040 g_1^2 g_4^2
\right. \right. \nonumber \\
&& \left. \left. ~~~~~~~
- 108380160 g_1 g_2^2 g_3
+ 3010560 g_1 g_2 g_3 g_4
- 501760 g_1 g_2 g_4 g_5
\right. \right. \nonumber \\
&& \left. \left. ~~~~~~~
+ 171401216 g_2^4
+ 80783360 g_2^2 g_3^2
+ 28098560 g_2^2 g_4^2
+ 878080 g_2^2 g_5^2
\right. \right. \nonumber \\
&& \left. \left. ~~~~~~~
- 579532800 g_2 g_3^2 g_4
+ 79027200 g_2 g_3 g_4 g_5
+ 177811200 g_3^4
\right. \right. \nonumber \\
&& \left. \left. ~~~~~~~
+ 2449843200 g_3^2 g_4^2
- 27440000 g_3^2 g_5^2
- 164640000 g_3 g_4^2 g_5
\right. \right. \nonumber \\
&& \left. \left. ~~~~~~~
- 707952000 g_4^4
+ 23049600 g_4^2 g_5^2
\right. \right. \nonumber \\
&& \left. \left. ~~~~~~~
- 102900 g_5^4
\right]
\Gamma^2(\fiveseventh) \Gamma^2(\twoseventh) \Gamma^2(\oneseventh)
\right]
\frac{\Gamma^{14}(\oneseventh) g_1}
{30481920000 \Gamma(\sixseventh) \Gamma(\fiveseventh) \Gamma(\threeseventh)
\Gamma(\twoseventh)} \nonumber \\
&& +~ O(g_i^7) \nonumber \\
\left. \frac{}{} \beta_2^{\Phi^9}(g_i) \right|_{N=-8} &=&
\left[
31360 g_3^2
- 128 g_1^2
- 6272 g_2^2
- 7280 g_4^2
+ 35 g_5^2 
\right]
\frac{\Gamma^7(\oneseventh) g_2}{67200} \nonumber \\
&& +~ 
\left[ 
\left[ 
245760 g_1^4
+ 31610880 g_1^2 g_3^2
- 7687680 g_1^2 g_4^2
- 126443520 g_1 g_2^2 g_3
\right. \right. \nonumber \\
&& \left. \left. ~~~~~
+ 173859840 g_1 g_2 g_3 g_4
- 6021120 g_1 g_2 g_4 g_5
+ 442552320 g_2^4
\right. \right. \nonumber \\
&& \left. \left. ~~~~~
+ 608509440 g_2^2 g_3^2
- 927252480 g_2^2 g_4^2
+ 13829760 g_2^2 g_5^2
\right. \right. \nonumber \\
&& \left. \left. ~~~~~
- 4425523200 g_2 g_3^2 g_4
+ 138297600 g_2 g_3 g_4 g_5
- 9773030400 g_3^4
\right. \right. \nonumber \\
&& \left. \left. ~~~~~
+ 14382950400 g_3^2 g_4^2
- 107016000 g_3^2 g_5^2
- 128419200 g_3 g_4^2 g_5
\right. \right. \nonumber \\
&& \left. \left. ~~~~~
- 2553096000 g_4^4
+ 40101600 g_4^2 g_5^2
- 132300 g_5^4
\right]
\Gamma^2(\sixseventh) \Gamma(\fiveseventh) \Gamma(\threeseventh) 
\Gamma^3(\oneseventh) 
\right. \nonumber \\
&& \left. ~~~+
\left[ 
331914240 g_1^2 g_3^2
- 2408448 g_1^4
- 206524416 g_1^2 g_2^2
+ 65856000 g_1^2 g_4^2
\right. \right. \nonumber \\
&& \left. \left. ~~~~~~~
- 987840 g_1^2 g_5^2
+ 2168506368 g_2^4
+ 41563038720 g_2^2 g_3^2
\right. \right. \nonumber \\
&& \left. \left. ~~~~~~~
- 19490741760 g_2^2 g_4^2
+ 137145120 g_2^2 g_5^2
- 150214243200 g_3^4
\right. \right. \nonumber \\
&& \left. \left. ~~~~~~~
+ 92451945600 g_3^2 g_4^2
- 554631000 g_3^2 g_5^2
- 12268149600 g_4^4
\right. \right. \nonumber \\
&& \left. \left. ~~~~~~~
+ 136136700 g_4^2 g_5^2
- 360150 g_5^4
\right]
\Gamma(\sixseventh) \Gamma(\fiveseventh) \Gamma(\threeseventh) 
\Gamma(\twoseventh) 
\right. \nonumber \\
&& \left. ~~~+
\left[ 
172032 g_1^4
- 1204224 g_1^2 g_2^2
+ 31610880 g_1^2 g_3^2
- 252887040 g_1 g_2^2 g_3
\right. \right. \nonumber \\
&& \left. \left. ~~~~~~~
- 115906560 g_1 g_2 g_3 g_4
- 52684800 g_1 g_3^2 g_5
+ 265531392 g_2^4
\right. \right. \nonumber \\
&& \left. \left. ~~~~~~~
- 811345920 g_2^2 g_3^2
- 2655313920 g_2^2 g_4^2
+ 2950348800 g_2 g_3^2 g_4
\right. \right. \nonumber \\
&& \left. \left. ~~~~~~~
+ 1936166400 g_2 g_3 g_4 g_5
- 14014156800 g_3^4
+ 35335036800 g_3^2 g_4^2
\right. \right. \nonumber \\
&& \left. \left. ~~~~~~~
- 224733600 g_3^2 g_5^2
- 4494672000 g_3 g_4^2 g_5
- 6442363200 g_4^4
\right. \right. \nonumber \\
&& \left. \left. ~~~~~~~
+ 357268800 g_4^2 g_5^2
- 1620675 g_5^4
\right]
\Gamma(\sixseventh) \Gamma^2(\threeseventh) \Gamma(\twoseventh) 
\Gamma^2(\oneseventh) 
\right. \nonumber \\
&& \left. ~~~+
\left[ 
1605632 g_1^2 g_2^2
- 344064 g_1^4
- 58705920 g_1^2 g_3^2
+ 4139520 g_1^2 g_4^2
\right. \right. \nonumber \\
&& \left. \left. ~~~~~~~
+ 379330560 g_1 g_2^2 g_3
- 57953280 g_1 g_2 g_3 g_4
- 14049280 g_1 g_2 g_4 g_5
\right. \right. \nonumber \\
&& \left. \left. ~~~~~~~
- 599904256 g_2^4
- 1555079680 g_2^2 g_3^2
+ 786759680 g_2^2 g_4^2
\right. \right. \nonumber \\
&& \left. \left. ~~~~~~~
+ 10756480 g_2^2 g_5^2
- 16226918400 g_2 g_3^2 g_4
+ 968083200 g_2 g_3 g_4 g_5
\right. \right. \nonumber \\
&& \left. \left. ~~~~~~~
+ 4978713600 g_3^4
+ 30010579200 g_3^2 g_4^2
- 249704000 g_3^2 g_5^2
\right. \right. \nonumber \\
&& \left. \left. ~~~~~~~
- 1498224000 g_3 g_4^2 g_5
- 6442363200 g_4^4
+ 178634400 g_4^2 g_5^2
\right. \right. \nonumber \\
&& \left. \left. ~~~~~~~
- 720300 g_5^4
\right]
\Gamma^2(\fiveseventh) \Gamma^2(\twoseventh) \Gamma^2(\oneseventh)
\right]
\frac{\Gamma^{14}(\oneseventh) g_2}
{71124480000 \Gamma(\sixseventh) \Gamma(\fiveseventh) \Gamma(\threeseventh) 
\Gamma(\twoseventh)} \nonumber \\
&& +~ O(g_i^7) \nonumber \\
\left. \frac{}{} \beta_3^{\Phi^9}(g_i) \right|_{N=-8} &=&
\left[ 
128 g_1^2
- 50176 g_2^2
+ 180320 g_3^2
- 38080 g_4^2
+ 175 g_5^2
\right]
\frac{\Gamma^7(\oneseventh) g_3}{201600} \nonumber \\
&& +~
\left[
\left[
252887040 g_1^2 g_3^2
- 245760 g_1^4
- 48921600 g_1^2 g_4^2
- 1011548160 g_1 g_2^2 g_3
\right. \right. \nonumber \\
&& \left. \left. ~~~~~
+ 1106380800 g_1 g_2 g_3 g_4
- 34621440 g_1 g_2 g_4 g_5
+ 3540418560 g_2^4
\right. \right. \nonumber \\
&& \left. \left. ~~~~~
+ 3872332800 g_2^2 g_3^2
- 5331701760 g_2^2 g_4^2
+ 75075840 g_2^2 g_5^2
\right. \right. \nonumber \\
&& \left. \left. ~~~~~
- 25446758400 g_2 g_3^2 g_4
+ 750758400 g_2 g_3 g_4 g_5
- 56194924800 g_3^4
\right. \right. \nonumber \\
&& \left. \left. ~~~~~
+ 78078873600 g_3^2 g_4^2
- 559776000 g_3^2 g_5^2
- 671731200 g_3 g_4^2 g_5
\right. \right. \nonumber \\
&& \left. \left. ~~~~~
- 13354656000 g_4^4
+ 204388800 g_4^2 g_5^2
- 661500 g_5^4
\right]
\Gamma^2(\sixseventh) \Gamma(\fiveseventh) \Gamma(\threeseventh)
\Gamma^3(\oneseventh) 
\right. \nonumber \\
&& \left. ~~~+
\left[ 
9633792 g_1^4
- 1475174400 g_1^2 g_2^2
+ 1106380800 g_1^2 g_3^2
+ 542653440 g_1^2 g_4^2
\right. \right. \nonumber \\
&& \left. \left. ~~~~~~~
- 5927040 g_1^2 g_5^2
- 4337012736 g_2^4
+ 321661777920 g_2^2 g_3^2
\right. \right. \nonumber \\
&& \left. \left. ~~~~~~~
- 117977072640 g_2^2 g_4^2
+ 758331840 g_2^2 g_5^2
- 910320902400 g_3^4
\right. \right. \nonumber \\
&& \left. \left. ~~~~~~~
+ 512116012800 g_3^2 g_4^2
- 2924418000 g_3^2 g_5^2
- 64711752000 g_4^4
\right. \right. \nonumber \\
&& \left. \left. ~~~~~~~
+ 695809800 g_4^2 g_5^2
- 1800750 g_5^4
\right]
\Gamma(\sixseventh) \Gamma(\fiveseventh) \Gamma(\threeseventh) 
\Gamma(\twoseventh) 
\right. \nonumber \\
&& \left. ~~~+
\left[ 
252887040 g_1^2 g_3^2
- 172032 g_1^4
- 31309824 g_1^2 g_2^2
- 2023096320 g_1 g_2^2 g_3
\right. \right. \nonumber \\
&& \left. \left. ~~~~~~~
- 737587200 g_1 g_2 g_3 g_4
- 302937600 g_1 g_3^2 g_5
+ 2124251136 g_2^4
\right. \right. \nonumber \\
&& \left. \left. ~~~~~~~
- 5163110400 g_2^2 g_3^2
- 15268055040 g_2^2 g_4^2
+ 16964505600 g_2 g_3^2 g_4
\right. \right. \nonumber \\
&& \left. \left. ~~~~~~~
+ 10510617600 g_2 g_3 g_4 g_5
- 80581401600 g_3^4
+ 191818771200 g_3^2 g_4^2
\right. \right. \nonumber \\
&& \left. \left. ~~~~~~~
- 1175529600 g_3^2 g_5^2
- 23510592000 g_3 g_4^2 g_5
- 33698515200 g_4^4
\right. \right. \nonumber \\
&& \left. \left. ~~~~~~~
+ 1820918400 g_4^2 g_5^2
- 8103375 g_5^4
\right]
\Gamma^2(\sixseventh) \Gamma(\threeseventh) \Gamma(\twoseventh) 
\Gamma^2(\oneseventh) 
\right. \nonumber \\
&& \left. ~~~+
\left[ 
344064 g_1^4
+ 41746432 g_1^2 g_2^2
- 469647360 g_1^2 g_3^2
+ 26342400 g_1^2 g_4^2
\right. \right. \nonumber \\
&& \left. \left. ~~~~~~~
+ 3034644480 g_1 g_2^2 g_3
- 368793600 g_1 g_2 g_3 g_4
- 80783360 g_1 g_2 g_4 g_5
\right. \right. \nonumber \\
&& \left. \left. ~~~~~~~
- 4799234048 g_2^4
- 9895961600 g_2^2 g_3^2
+ 4523868160 g_2^2 g_4^2
\right. \right. \nonumber \\
&& \left. \left. ~~~~~~~
+ 58392320 g_2^2 g_5^2
- 93304780800 g_2 g_3^2 g_4
+ 5255308800 g_2 g_3 g_4 g_5
\right. \right. \nonumber \\
&& \left. \left. ~~~~~~~
+ 28627603200 g_3^4
+ 162914572800 g_3^2 g_4^2
- 1306144000 g_3^2 g_5^2
\right. \right. \nonumber \\
&& \left. \left. ~~~~~~~
- 7836864000 g_3 g_4^2 g_5
- 33698515200 g_4^4
+ 910459200 g_4^2 g_5^2
\right. \right. \nonumber \\
&& \left. \left. ~~~~~~~
- 3601500 g_5^4
\right]
\Gamma^2(\fiveseventh) \Gamma^2(\twoseventh) \Gamma^2(\oneseventh) 
\right]
\frac{\Gamma^{14}(\oneseventh) g_3}{213373440000 \Gamma(\sixseventh) 
\Gamma(\fiveseventh) \Gamma(\threeseventh) \Gamma(\twoseventh))}
\nonumber \\
&& +~ O(g_i^7)
\nonumber \\
\left. \frac{}{} \beta_4^{\Phi^9}(g_i) \right|_{N=-8} &=&
\left[
640 g_1^2
- 81536 g_2^2
+ 266560 g_3^2
- 54320 g_4^2
+ 245 g_5^2
\right]
\frac{\Gamma^7(\oneseventh) g_4}{201600} \nonumber \\
&& +~
\left[
\left[
410941440 g_1^2 g_3^2
- 1228800 g_1^4
- 74780160 g_1^2 g_4^2
- 1643765760 g_1 g_2^2 g_3
\right. \right. \nonumber \\
&& \left. \left. ~~~~~
+ 1691182080 g_1 g_2 g_3 g_4
- 51179520 g_1 g_2 g_4 g_5
+ 5753180160 g_2^4
\right. \right. \nonumber \\
&& \left. \left. ~~~~~
+ 5919137280 g_2^2 g_3^2
- 7881646080 g_2^2 g_4^2
+ 108662400 g_2^2 g_5^2
\right. \right. \nonumber \\
&& \left. \left. ~~~~~
- 37616947200 g_2 g_3^2 g_4
+ 1086624000 g_2 g_3 g_4 g_5
- 83070758400 g_3^4
\right. \right. \nonumber \\
&& \left. \left. ~~~~~
+ 113008896000 g_3^2 g_4^2
- 798504000 g_3^2 g_5^2
- 958204800 g_3 g_4^2 g_5
\right. \right. \nonumber \\
&& \left. \left. ~~~~~
- 19050024000 g_4^4
+ 288472800 g_4^2 g_5^2
- 926100 g_5^4
\right]
\Gamma^2(\sixseventh) \Gamma(\fiveseventh) \Gamma(\threeseventh) 
\Gamma^3(\oneseventh)
\right. \nonumber \\
&& \left. ~~~+
\left[ 
26492928 g_1^4
- 2330775552 g_1^2 g_2^2
+ 1217018880 g_1^2 g_3^2
+ 887738880 g_1^2 g_4^2
\right. \right. \nonumber \\
&& \left. \left. ~~~~~~~
- 8890560 g_1^2 g_5^2
- 15179544576 g_2^4
+ 518634439680 g_2^2 g_3^2
\right. \right. \nonumber \\
&& \left. \left. ~~~~~~~
- 177481920000 g_2^2 g_4^2
+ 1105228320 g_2^2 g_5^2
- 1369999075200 g_3^4
\right. \right. \nonumber \\
&& \left. \left. ~~~~~~~
+ 746876188800 g_3^2 g_4^2
- 4184943000 g_3^2 g_5^2
- 92619055200 g_4^4
\right. \right. \nonumber \\
&& \left. \left. ~~~~~~~
+ 983209500 g_4^2 g_5^2
- 2521050 g_5^4
\right]
\Gamma(\sixseventh) \Gamma(\fiveseventh) \Gamma(\threeseventh) 
\Gamma(\twoseventh) 
\right. \nonumber \\
&& \left. ~~~+
\left[ 
410941440 g_1^2 g_3^2
- 860160 g_1^4
- 59006976 g_1^2 g_2^2
- 3287531520 g_1 g_2^2 g_3
\right. \right. \nonumber \\
&& \left. \left. ~~~~~~~
- 1127454720 g_1 g_2 g_3 g_4
- 447820800 g_1 g_3^2 g_5
+ 3451908096 g_2^4
\right. \right. \nonumber \\
&& \left. \left. ~~~~~~~
- 7892183040 g_2^2 g_3^2
- 22570168320 g_2^2 g_4^2
+ 25077964800 g_2 g_3^2 g_4
\right. \right. \nonumber \\
&& \left. \left. ~~~~~~~
+ 15212736000 g_2 g_3 g_4 g_5
- 119120332800 g_3^4
+ 277632432000 g_3^2 g_4^2
\right. \right. \nonumber \\
&& \left. \left. ~~~~~~~
- 1676858400 g_3^2 g_5^2
- 33537168000 g_3 g_4^2 g_5
- 48069940800 g_4^4
\right. \right. \nonumber \\
&& \left. \left. ~~~~~~~
+ 2570030400 g_4^2 g_5^2
- 11344725 g_5^4
\right]
\Gamma^2(\sixseventh) \Gamma(\threeseventh) \Gamma(\twoseventh) 
\Gamma^2(\oneseventh)
\right. \nonumber \\
&& \left. ~~~+
\left[ 
1720320 g_1^4
+ 78675968 g_1^2 g_2^2
- 763176960 g_1^2 g_3^2
+ 40266240 g_1^2 g_4^2
\right. \right. \nonumber \\
&& \left. \left. ~~~~~~~
+ 4931297280 g_1 g_2^2 g_3
- 563727360 g_1 g_2 g_3 g_4
- 119418880 g_1 g_2 g_4 g_5
\right. \right. \nonumber \\
&& \left. \left. ~~~~~~~
- 7798755328 g_2^4
- 15126684160 g_2^2 g_3^2
+ 6687457280 g_2^2 g_4^2
\right. \right. \nonumber \\
&& \left. \left. ~~~~~~~
+ 84515200 g_2^2 g_5^2
- 137928806400 g_2 g_3^2 g_4
+ 7606368000 g_2 g_3 g_4 g_5
\right. \right. \nonumber \\
&& \left. \left. ~~~~~~~
+ 42319065600 g_3^4
+ 235797408000 g_3^2 g_4^2
- 1863176000 g_3^2 g_5^2
\right. \right. \nonumber \\
&& \left. \left. ~~~~~~~
- 11179056000 g_3 g_4^2 g_5
- 48069940800 g_4^4
+ 1285015200 g_4^2 g_5^2
\right. \right. \nonumber \\
&& \left. \left. ~~~~~~~
- 5042100 g_5^4))
\right]
\Gamma^2(\fiveseventh) \Gamma^2(\twoseventh) \Gamma^2(\oneseventh) 
\right]
\frac{\Gamma^{14}(\oneseventh) g_4}{213373440000 \Gamma(\sixseventh) 
\Gamma(\fiveseventh) \Gamma(\threeseventh) \Gamma(\twoseventh)} \nonumber \\
&& +~ O(g_i^7) \nonumber \\
\left. \frac{}{} \beta_5^{\Phi^9}(g_i) \right|_{N=-8} &=&
\left[
128 g_1^2
- 12544 g_2^2
+ 39200 g_3^2
- 7840 g_4^2
+ 35 g_5^2 
\right]
\frac{\Gamma^7(\oneseventh) g_5}{22400} \nonumber \\
&& +~
\left[
\left[
63221760 g_1^2 g_3^2
- 245760 g_1^4
- 11182080 g_1^2 g_4^2
- 252887040 g_1 g_2^2 g_3
\right. \right. \nonumber \\
&& \left. \left. ~~~~~
+ 252887040 g_1 g_2 g_3 g_4
- 7526400 g_1 g_2 g_4 g_5
+ 885104640 g_2^4
\right. \right. \nonumber \\
&& \left. \left. ~~~~~
+ 885104640 g_2^2 g_3^2
- 1159065600 g_2^2 g_4^2
+ 15805440 g_2^2 g_5^2
\right. \right. \nonumber \\
&& \left. \left. ~~~~~
- 5531904000 g_2 g_3^2 g_4
+ 158054400 g_2 g_3 g_4 g_5
- 12216288000 g_3^4
\right. \right. \nonumber \\
&& \left. \left. ~~~~~
+ 16437657600 g_3^2 g_4^2
- 115248000 g_3^2 g_5^2
- 138297600 g_3 g_4^2 g_5
\right. \right. \nonumber \\
&& \left. \left. ~~~~~
- 2749488000 g_4^4
+ 41395200 g_4^2 g_5^2
- 132300 g_5^4
\right]
\Gamma^2(\sixseventh) \Gamma(\fiveseventh) \Gamma(\threeseventh) 
\Gamma^3(\oneseventh)
\right. \nonumber \\
&& \left. ~~~+
\left[ 
4816896 g_1^4
- 354041856 g_1^2 g_2^2
+ 147517440 g_1^2 g_3^2
+ 136980480 g_1^2 g_4^2
\right. \right. \nonumber \\
&& \left. \left. ~~~~~~~
- 1317120 g_1^2 g_5^2
- 2891341824 g_2^4
+ 79511900160 g_2^2 g_3^2
\right. \right. \nonumber \\
&& \left. \left. ~~~~~~~
- 26331863040 g_2^2 g_4^2
+ 161347200 g_2^2 g_5^2
- 203297472000 g_3^4
\right. \right. \nonumber \\
&& \left. \left. ~~~~~~~
+ 109070707200 g_3^2 g_4^2
- 605052000 g_3^2 g_5^2
- 13391817600 g_4^4
\right. \right. \nonumber \\
&& \left. \left. ~~~~~~~
+ 141178800 g_4^2 g_5^2
- 360150 g_5^4
\right]
\Gamma(\sixseventh) \Gamma(\fiveseventh) \Gamma(\threeseventh) 
\Gamma(\twoseventh) 
\right. \nonumber \\
&& \left. ~~~+
\left[ 
63221760 g_1^2 g_3^2
- 172032 g_1^4
- 9633792 g_1^2 g_2^2
- 505774080 g_1 g_2^2 g_3
\right. \right. \nonumber \\
&& \left. \left. ~~~~~~~
- 168591360 g_1 g_2 g_3 g_4
- 65856000 g_1 g_3^2 g_5
+ 531062784 g_2^4
\right. \right. \nonumber \\
&& \left. \left. ~~~~~~~
- 1180139520 g_2^2 g_3^2
- 3319142400 g_2^2 g_4^2
+ 3687936000 g_2 g_3^2 g_4
\right. \right. \nonumber \\
&& \left. \left. ~~~~~~~
+ 2212761600 g_2 g_3 g_4 g_5
- 17517696000 g_3^4
+ 40382899200 g_3^2 g_4^2
\right. \right. \nonumber \\
&& \left. \left. ~~~~~~~
- 242020800 g_3^2 g_5^2
- 4840416000 g_3 g_4^2 g_5
- 6937929600 g_4^4
\right. \right. \nonumber \\
&& \left. \left. ~~~~~~~
+ 368793600 g_4^2 g_5^2
- 1620675 g_5^4
\right]
\Gamma(\sixseventh) \Gamma^2(\threeseventh) \Gamma(\twoseventh) 
\Gamma^2(\oneseventh)
\right. \nonumber \\
&& \left. ~~~+
\left[ 
+ 344064 g_1^4
+ 12845056 g_1^2 g_2^2
- 117411840 g_1^2 g_3^2
+ 6021120 g_1^2 g_4^2
\right. \right. \nonumber \\
&& \left. \left. ~~~~~~~
+ 758661120 g_1 g_2^2 g_3
- 84295680 g_1 g_2 g_3 g_4
- 17561600 g_1 g_2 g_4 g_5
\right. \right. \nonumber \\
&& \left. \left. ~~~~~~~
- 1199808512 g_2^4
- 2261934080 g_2^2 g_3^2
+ 983449600 g_2^2 g_4^2
\right. \right. \nonumber \\
&& \left. \left. ~~~~~~~
+ 12293120 g_2^2 g_5^2
- 20283648000 g_2 g_3^2 g_4
+ 1106380800 g_2 g_3 g_4 g_5
\right. \right. \nonumber \\
&& \left. \left. ~~~~~~~
+ 6223392000 g_3^4
+ 34297804800 g_3^2 g_4^2
- 268912000 g_3^2 g_5^2
\right. \right. \nonumber \\
&& \left. \left. ~~~~~~~
- 1613472000 g_3 g_4^2 g_5
- 6937929600 g_4^4
+ 184396800 g_4^2 g_5^2
\right. \right. \nonumber \\
&& \left. \left. ~~~~~~~
- 720300 g_5^4))
\right]
\Gamma^2(\fiveseventh) \Gamma^2(\twoseventh) \Gamma^2(\oneseventh) 
\right]
\frac{\Gamma^{14}(\oneseventh) g_5}{23708160000 \Gamma(\sixseventh) 
\Gamma(\fiveseventh) \Gamma(\threeseventh) \Gamma(\twoseventh)} \nonumber \\
&& +~ O(g_i^7) ~.
\end{eqnarray}


\begin{thebibliography}{99}
\bibitem{1} S.-S. Lee, Phys. Rev. {\bf B76} (2007), 075103.
%%CITATION = COND-MAT/0611658;%%
\bibitem{2} B. Roy, V. Juri\u{c}i\'{c} \& I.F. Herbut, Phys. Rev. {\bf B87}
(2013), 041401(R).
%%CITATION = ARXIV:1608.02560;%%
\bibitem{3} L. Fei, S. Giombi, I.R. Klebanov \& G. Tarnopolsky, Prog. Theor.
Exp. Phys. (2016) 12C105.
%%CITATION = ARXIV:1607.05316;%%
\bibitem{4} D. Gross \& A. Neveu, Phys. Rev. {\bf D10} (1974), 3235.
%%CITATION = PHRVA,D10,3235;%%
\bibitem{5} J. Zinn-Justin, Nucl. Phys. {\bf B367} (1991), 105.
%%CITATION = NUPHA,B367,105;%%
\bibitem{6} T. Grover, D.N. Sheng \& A. Vishwanath, Science {\bf 344} (2014),
280.
%%CITATION = ARXIV:1301.7449;%%
\bibitem{7} N. Zerf, C.-H. Lin \& J. Maciejko, Phys. Rev. {\bf B94} (2016),
205106.
%%CITATION = ARXIV:1605.09423;%%
\bibitem{8} L.N. Mihaila, N. Zerf, B. Ihrig, I.F. Herbut \& M.M. Scherer,
Phys. Rev. {\bf B96} (2017), 165133.
%%CITATION = ARXIV:1703.08801;%%
\bibitem{9} N. Zerf, L.N. Mihaila, P. Marquard, I.F. Herbut \& M.M. Scherer,
Phys. Rev. {\bf D96} (2017), 096010.
%%CITATION = ARXIV:1709.05057;%%
\bibitem{10} B. Ihrig, L.N. Mihaila \& M.M. Scherer, Phys. Rev. {\bf B98}
(2018), 125109.
%%CITATION = ARXIV:1806.04977;%%
\bibitem{11} N. Bobev, S. El-Showk, D. Maz\'{a}\u{c} \& M.F. Paulos, Phys.
Rev. Lett. {\bf 115} (2015), 051601.
%%CITATION = ARXIV:1502.04124;%%
\bibitem{12} M. Baggio, N. Bobev, S.M. Chester, E. Lauria \& S.S. Pufu, JHEP
{\bf 02} (2018), 062.
%%CITATION = ARXIV:1712.02698;%%
\bibitem{13} J. Wess \& B. Zumino, Phys. Lett. {\bf B49} (1974), 52.
%%CITATION = PHLTA,B49,52;%%
\bibitem{14} J.A. Gracey, Phys. Rev. {\bf D105} (2022), 025004.
%%CITATION = ARXIV:2108.13133;%%
\bibitem{15} P.K. Townsend \& P. van Nieuwenhuizen, Phys. Rev. {\bf D20}
(1979), 1832.
%%CITATION = PHRVA,D20,1832;%%
\bibitem{16} L. Abbott \& M.T. Grisaru, Nucl. Phys. {\bf B169} (1980), 415.
%%CITATION = NUPHA,B169,415;%%
\bibitem{17} A. Sen \& M.K. Sundaresan, Phys. Lett. {\bf B101} (1981), 61.
%%CITATION = PHLTA,B101,61;%%
\bibitem{18} L.V. Avdeev, S.G. Gorishny, A.Yu. Kamenshchik \& S.A. Larin, Phys.
Lett. {\bf B117} (1982), 321.
%%CITATION = PHLTA,B117,321;%%
\bibitem{19} L. Janssen \& I.F. Herbut, Phys. Rev. {\bf B89} (2014), 205403.
%%CITATION = ARXIV:1402.6277;%%
\bibitem{20} L. Fei, S. Giombi \& I.R. Klebanov, Phys. Rev. {\bf D90} (2014),
025018.
%%CITATION = ARXIV:1404.1094;%%
\bibitem{21} I.F. Herbut \& L. Janssen, Phys. Rev. {\bf D93} (2016), 085005.
%%CITATION = ARXIV:1510.05691;%%
\bibitem{22} D. Roscher \& I.F. Herbut, Phys. Rev. {\bf D97} (2018), 116019.
%%CITATION = ARXIV:1805.01480;%%
\bibitem{23} J.A. Gracey, I.F. Herbut \& D. Roscher, Phys. Rev. {\bf D98}
(2018), 096014.
%%CITATION = ARXIV:1810.05721;%%
\bibitem{24} I.R. Klebanov, Phys. Rev. Lett. {\bf 128} (2022), 061601.
%%CITATION = ARXIV:2111.12648;%%
\bibitem{25} M. Blume, Phys. Rev. {\bf 141} (1966), 517.
%%CITATION = PHRVA,141,517;%%
\bibitem{26} W. Capel, Physica {\bf 32} (1966), 966.
%%CITATION = PHYSA,32,966;%%
\bibitem{27} L. Zambelli \& O. Zanusso, Phys. Rev. {\bf D95} (2017), 085001.
%%CITATION = ARXIV:1612.08739;%%
\bibitem{28} A. Codello, N. Defenu \& G. D'Odorico, Phys. Rev. {\bf D91}
(2015), 105003.
%%CITATION = ARXIV:1410.3308;%%
\bibitem{29} A. Codello, M. Safari, G.P. Vacca \& O. Zanusso, Phys. Rev.
{\bf D96} (2017), 081701.
%%CITATION = ARXIV:1706.06887;%%
\bibitem{30} R. Ben Al\`{\i} Zinati \& A. Codello, J. Stat. Mech. {\bf 1801}
(2018), 013206.
%%CITATION = ARXIV:1707.03410;%%
\bibitem{31} K. Fujikawa \& W. Lang, Nucl. Phys. {\bf B88} (1975), 61.
%%CITATION = NUPHA,B88,61;%%
\bibitem{32} A. Codello, M. Safari, G.P. Vacca \& O. Zanusso, JHEP {\bf 1704}
(2017), 127.
%%CITATION = ARXIV:1703.04830;%%
\bibitem{33} J.A. Gracey, Eur. Phys. J. {\bf C80} (2020), 604.
%%CITATION = ARXIV:1703.09685;%%
\bibitem{34} J.A.M. Vermaseren, math-ph/0010025.
%%CITATION = MATH-PH/0010025;%%
\bibitem{35} M. Tentyukov \& J.A.M. Vermaseren, Comput. Phys. Commun. {\bf 181}
(2010), 1419.
%%CITATION = HEP-PH 0702279;%% 
\bibitem{36} P. Nogueira, J. Comput. Phys. {\bf 105} (1993), 279.
%%CITATION = JCTPA,105,279;%%
\bibitem{37} J.S. Hager, J. Phys. {\bf A35} (2002), 2703.
%%CITATION = JPAGA,A35,2703;%%
\bibitem{38} J.C. Collins \& J.A.M. Vermaseren, arXiv:1606.01177 [cs.OH].
%%CITATION = ARXIV:1606.01177;%%
\end{thebibliography}
\end{document}